%% file: neurips_arxiv.tex
\documentclass{article}

\usepackage[preprint]{neurips_2026}
\usepackage{graphicx}
\usepackage{amsmath}
\usepackage{multirow}
\usepackage{float}
\usepackage{algorithm}
\usepackage{algpseudocode}
\usepackage{amsmath,amssymb}
\usepackage[table]{xcolor}
\usepackage{colortbl}
\usepackage{subcaption}
\usepackage{makecell}

\usepackage[utf8]{inputenc} 
\usepackage[T1]{fontenc}    
\usepackage{hyperref}       
\usepackage{url}            
\usepackage{booktabs}       
\usepackage{amsfonts}       
\usepackage{nicefrac}       
\usepackage{microtype}      
\usepackage{xcolor}         

\title{\textbf{TIGER-FG}: \textbf{T}ext-Guided \textbf{I}mplicit Fine-Grained \textbf{G}rounding for \textbf{E}-commerce \textbf{R}etrieval}

\author{%
  {\small Xinyu Sun \thanks{Equal contribution.}} \\[-0.15em]
  {\small Kuaishou Technology} \\[-0.15em]
  {\small Beijing, China} \\[-0.15em]
  {\scriptsize\texttt{22221189@zju.edu.cn}} \\
  \And
  {\small Huangyu Dai\footnotemark[1]} \\[-0.15em]
  {\small Kuaishou Technology} \\[-0.15em]
  {\small Beijing, China} \\[-0.15em]
  {\scriptsize\texttt{11931034@zju.edu.cn}} \\
  \And
  {\small Lingtao Mao\footnotemark[1]} \\[-0.15em]
  {\small Kuaishou Technology} \\[-0.15em]
  {\small Beijing, China} \\[-0.15em]
  {\scriptsize\texttt{mltzju@163.com}} \\
  \And
  {\small Zexin Zheng} \\[-0.15em]
  {\small Kuaishou Technology} \\[-0.15em]
  {\small Beijing, China} \\[-0.15em]
  {\scriptsize\texttt{zhengzx25@mail2.sysu.edu.cn}} \\
  \AND
  {\small Zihan Liang} \\[-0.15em]
  {\small Kuaishou Technology} \\[-0.15em]
  {\small Beijing, China} \\[-0.15em]
  {\scriptsize\texttt{liangzih@seas.upenn.edu}} \\
  \And
  {\small Ben Chen\thanks{Corresponding author.}} \\[-0.15em]
  {\small Kuaishou Technology} \\[-0.15em]
  {\small Beijing, China} \\[-0.15em]
  {\scriptsize\texttt{benchen4395@gmail.com}} \\
  \And
  {\small Chenyi Lei} \\[-0.15em]
  {\small Kuaishou Technology} \\[-0.15em]
  {\small Beijing, China} \\[-0.15em]
  {\scriptsize\texttt{leichy@mail.ustc.edu.cn}} \\
  \And
  {\small Wenwu Ou} \\[-0.15em]
  {\small Kuaishou Technology} \\[-0.15em]
  {\small Beijing, China} \\[-0.15em]
  {\scriptsize\texttt{ouwenweu@gmail.com}}
}

\begin{document}

\maketitle

\input{tex_files/00.abstract}

\input{tex_files/01.introduction}
\input{tex_files/02.related_works}
\input{tex_files/03.method}
\input{tex_files/04.experiments}
\input{tex_files/05.conclusion}

\bibliographystyle{plainnat}
\bibliography{neurips_reference}

\newpage
\appendix
\input{tex_files/F.limitation}

\input{tex_files/A.related_work_appendix.tex}

\input{tex_files/B.data_construction.tex}

\input{tex_files/C.visualization_recall.tex}

\input{tex_files/D.ablation_appendix.tex}

\input{tex_files/E.train_appendix}

\newpage

\end{document}

%% file: tex_files/00.abstract.tex
\begin{abstract}
E-commerce image search often takes a cropped image as the query, while each candidate is represented by full item images and structured text. This image-to-multimodal retrieval setting presents two asymmetries: a \emph{modality disparity} -- a visual query must match image--text items, and a \emph{granularity disparity}  -- a cropped query must be compared with full images containing background context and possible distractors. Detection-based pipelines handle the granularity disparity through explicit localization but incur extra cost and error propagation, whereas CLIP-style encoders avoid detection, but are vulnerable to backgrounds or irrelevant items. To address these limitations, we propose \textbf{TIGER-FG}, a \textbf{t}ext-guided \textbf{i}mplicit fine-grained \textbf{g}rounding framework for image-to-multimodal \textbf{e}-commerce \textbf{r}etrieval. TIGER-FG uses item text as semantic guidance to produce target-focused item representations without object detection for retrieval. We further introduce dual distillation objectives that preserve target-region spatial consistency and query--item similarity structure, yielding more stable and discriminative multimodal representations. In addition, we construct \textbf{ECom-RF-IMMR}, a realistic benchmark suite with a 10M-pair training set and two evaluation benchmarks covering standard and cluttered item layouts. TIGER-FG improves Recall@1 over the strongest baseline by 6.1 and 34.4 percentage points on the two evaluation benchmarks, respectively, with only 85.7M query-side parameters and 256-dim embeddings. Results on public e-commerce benchmarks further demonstrate its generalization to noisy and one-to-many retrieval scenarios. Code and data will be released.
\end{abstract}

%% file: tex_files/01.introduction.tex
\section{Introduction}
\vspace{-6pt}
\label{sec:intro}

In e-commerce image search, users often start with an image containing a product of interest. Practical systems first localize the target product, typically through an upstream detector that proposes candidate regions for user selection, and then use the resulting cropped region as the visual query. However, each candidate is represented by full item images with structured text such as title, category, and attributes, where the image often contains background context, multiple products, or other distractors. This setting induces the image-to-multimodal retrieval (IMMR) task~\citep{cheng2023category}, where a cropped visual query must be matched against scene-level multimodal item candidates.

\begin{figure}[htbp]
    \vspace{-6pt}
    \centering
    \includegraphics[width=0.9\linewidth]{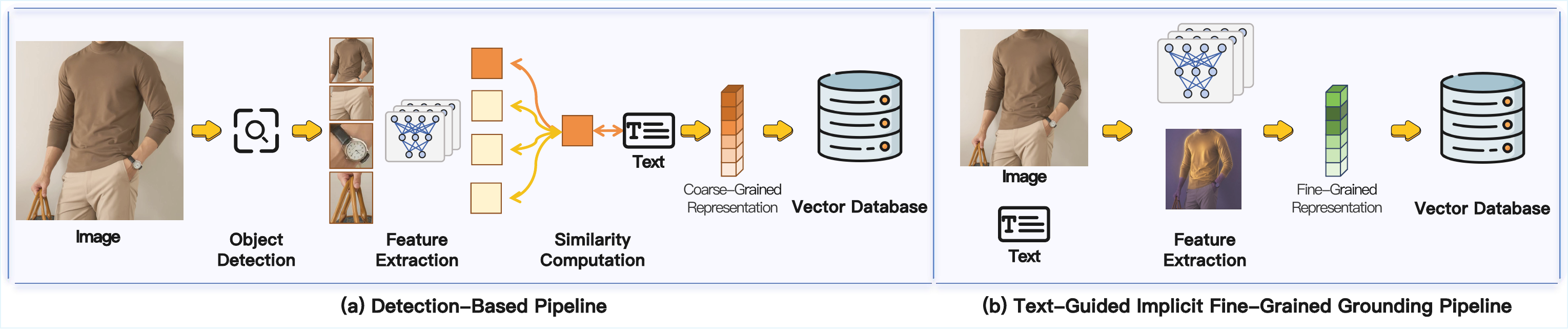}
    \caption{\textbf{Pipeline comparison for IMMR.}
    (a) Detection-based methods localize candidate regions from the full item image and select region features using text.
(b) TIGER-FG encodes the full image--text item into a target-focused representation without object detection.}
    \label{fig:pipeline_comparison}
    \vspace{-6pt}
\end{figure}

This setting exposes two challenges that are difficult for existing retrieval methods to handle. The first is the \emph{modality disparity}, where a visual query must be represented in a space compatible with item candidates described by both images and structured text. This differs from homogeneous image--image retrieval, where the query and candidate share the same modality. The second is the \emph{granularity disparity}, where a cropped query must be compared with full item images. As a result, global image-level representations can be dominated by visually salient but query-irrelevant content.
Existing approaches address these disparities only partially. One practical paradigm, illustrated in Figure~\ref{fig:pipeline_comparison}a, uses explicit object detection before item encoding. Given a full item image, such methods first detect candidate item boxes, crop the corresponding regions, encode each cropped region, and then compare region embeddings with structured item text to retain the most text-compatible representation for indexing~\citep{cheng2024yolo, nan2025unidgf}. A related grounding-based variant, not shown in the figure, replaces detector-based candidate region generation with text-conditioned grounding models such as GroundingDINO~\citep{liu2024grounding}. It takes the full item image and item text as input, grounds text-relevant regions, and then uses the selected region features for item representation. These explicit-region pipelines can mitigate the granularity disparity, but they introduce a multi-stage indexing process and make the stored representation dependent on detection or grounding quality and region selection. These issues become more pronounced in e-commerce data, where item titles and categories are long, structured, and attribute-dense, and adapting generic detection or grounding models usually requires costly in-domain annotation. Another paradigm adopts dual encoders built on vision--language pretraining, including CLIP~\citep{radford2021learning, chinese-clip}, BLIP~\citep{li2022blip, li2023blip}, ALIGN~\citep{jia2021scaling}, and recent MLLM-based embedders~\citep{qwen3vlembedding, zhang2024gme}. These models support efficient ANN retrieval by encoding queries and candidates independently, but they mainly produce global image--text representations and often fail to focus on the query-relevant region when candidate images contain salient backgrounds, multiple objects, or other distractors. Although IMMR has been formalized as an item retrieval problem~\citep{cheng2023category}, learning fine-grained multimodal item representations without explicit detection remains underexplored.

We observe that e-commerce item candidates naturally provide structured text, including titles, categories, and attributes, which often specifies the target item and its discriminative properties. Such text can serve as semantic guidance for learning target-focused item representations from full item images, without relying on explicit detection. Based on this observation, we propose \textbf{TIGER-FG}, a text-guided implicit fine-grained grounding framework for IMMR, as illustrated in Figure~\ref{fig:pipeline_comparison}b. TIGER-FG directly produces target-focused item embeddings from full image--text candidates without box prediction during indexing or retrieval. It mitigates the modality and granularity gaps while retaining the efficiency of a dual-encoder retrieval architecture. We further use distillation objectives to preserve target-region structure and query--item similarity relationships. 

Experiments on \textbf{ECom-RF-IMMR} show that TIGER-FG consistently improves image-to-multimodal item retrieval. On \textbf{ECom-RF-IMMR-Normal}, TIGER-FG achieves 80.1 Recall@1, outperforming the strongest baseline by 6.1 points with compact 256-dimensional embeddings. On \textbf{ECom-RF-IMMR-Mosaic}, which introduces multi-item candidate images and stronger cross-item interference, TIGER-FG reaches 75.2 Recall@1 and improves over the strongest baseline by 34.4 points. Results on \textbf{LookBench}~\citep{gao2026lookbench} further show that TIGER-FG transfers to noisy and one-to-many item retrieval settings. Together with ablation and qualitative analyses, these results show that structured item text, clutter-aware training, and distillation provide effective guidance for detection-free fine-grained grounding in IMMR. Our contributions are summarized as follows:







\noindent(1) \textbf{Text-guided implicit fine-grained grounding for IMMR.}
We propose a detection-free item encoder that uses structured item text as semantic guidance for visual token interaction. Instead of extracting boxes from full images, the encoder directly produces target-focused representations from item candidates. This design reduces deployment cost and improves robustness when candidate images contain background context, multiple products, or other distractors.

\noindent(2) \textbf{Distillation-enhanced multimodal item representation learning.}
We introduce two complementary distillation objectives for multimodal item representation learning. Spatial-relational distillation aligns target-region spatial consistency, while similarity-distribution distillation preserves the global query--item similarity structure. These objectives jointly constrain fine-grained visual structure and global retrieval behavior, improving representation stability and discriminability in IMMR.

\noindent(3) \textbf{A large-scale benchmark suite for IMMR.}
We construct \textbf{ECom-RF-IMMR}, a realistic benchmark suite with a 10M-pair training set and two evaluation splits, a standard split and a cluttered mosaic split with cross-category distractors.
The evaluation splits provide structured item text and region-level annotations for assessing retrieval under clean and cluttered item layouts.
Experiments on ECom-RF-IMMR and two public benchmarks demonstrate the effectiveness of our method.

%% file: tex_files/02.related_works.tex
\vspace{-6pt}
\section{Related Work}
\label{sec:related}
\vspace{-6pt}

Vision--language retrieval has been widely studied with dual-encoder architectures such as CLIP~\citep{radford2021learning, chinese-clip}  and ALIGN~\citep{jia2021scaling}, and with cross-attention models such as ALBEF~\citep{li2021align}. Subsequent work targets unified multimodal retrieval~\citep{wei2024uniir, liang2025uniecs} and MLLM-based embeddings~\citep{zhang2024gme, qwen3vlembedding}. In parallel, self-supervised models such as DINO~\citep{oquab2023dinov2,  darcet2023vision, simeoni2025dinov3} learn strongly object-centric representations, and recent CLIP variants---DeCLIP~\citep{wang2025declip} and SmartCLIP~\citep{xie2025smartclip}---explicitly improve region-level alignment. However, these methods all target symmetric image--text matching under global supervision and do not address the region-to-scene asymmetry that defines IMMR, where a cropped query is matched against full-scene multimodal candidates. A separate line of detection-based retrieval pipelines instead localizes products explicitly before matching, which we discussed in Section~\ref{sec:intro} and extend in Appendix~\ref{app:related}.

%% file: tex_files/03.method.tex
\vspace{-6pt}
\section{Methodology}
\vspace{-6pt}
\textbf{Problem definition.}
We formalize image-to-multimodal retrieval (IMMR) as follows. The candidate set $\mathcal{G} = \{(\mathbf{I}^{\mathrm{p}}_j, \mathbf{T}^{\mathrm{p}}_j)\}_{j=1}^{|\mathcal{G}|}$ consists of multimodal item entries. 
Each entry pairs a full item image
$\mathbf{I}^{p}_j \in \mathbb{R}^{H_p \times W_p \times 3}$
with a structured item text $\mathbf{T}^{p}_j$, such as title, category, and attributes.
The query is a cropped visual region specified by a box $\mathbf{b}^{q}$ in a source image
$\mathbf{I}^{q} \in \mathbb{R}^{H \times W \times 3}$, which depicts the target product instance.
Given $(\mathbf{I}^\mathrm{q}, \mathbf{b}^\mathrm{q})$ and a relevance indicator
$y_j \in \{0, 1\}$, the task is defined as
\begin{equation}
    \Psi_{\mathrm{IMMR}}(\mathbf{I}^\mathrm{q}, \mathbf{b}^\mathrm{q})
    = \{(\mathbf{I}^\mathrm{p}_j, \mathbf{T}^\mathrm{p}_j) \in \mathcal{G} \mid y_j = 1\}.
\end{equation}
This task is characterized by two disparities between the visual query and the multimodal item candidates.
The \emph{modality disparity} arises because the query is purely visual,
whereas each candidate entry is represented by both image appearance and structured text semantics.
The \emph{granularity disparity} arises because the query captures a local item region while each candidate image shows a full item scene that may include background context, multiple products, or other distractors.
We use the item-side box $\mathbf{b}^{\mathrm{p}}_j$ only as auxiliary supervision during training (\S\ref{sec:item}), and at indexing and retrieval time the encoder consumes only $(\mathbf{I}^{\mathrm{p}}_j, \mathbf{T}^{\mathrm{p}}_j)$.

\begin{figure}[htbp]
    \centering
    \includegraphics[width=1\linewidth]{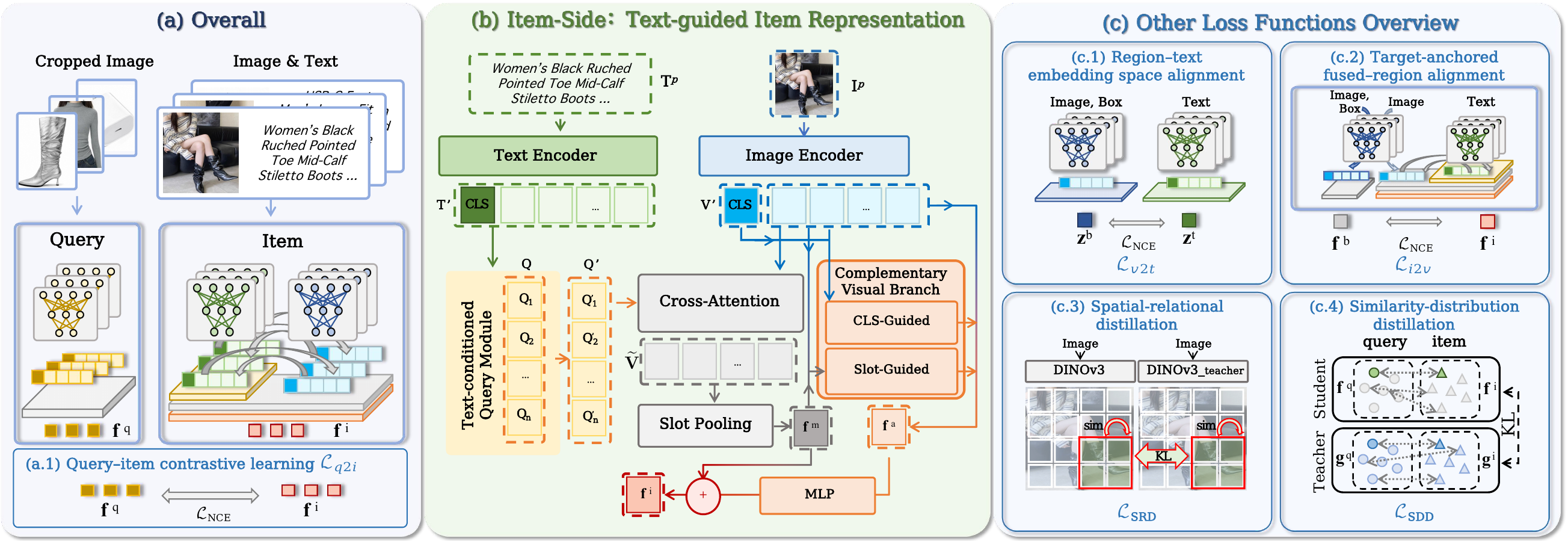}
    \caption{\textbf{Overview of the proposed TIGER-FG framework.}
    \textbf{(a)} Dual-encoder retrieval architecture.
    \textbf{(b)} Text-guided item representation learning.
    \textbf{(c)} Joint training objectives for item representation and query--item alignment.}
    \label{fig:architecture}
    \vspace{-6pt}
\end{figure}

\textbf{Framework overview (Fig.~\ref{fig:architecture}).}
TIGER-FG addresses the two disparities with an asymmetric dual-encoder architecture. The query encoder maps the cropped query region into a visual representation, while the item encoder uses the full item image and structured text to produce a target-focused multimodal representation. Structured text guides item-side visual token interaction, enabling target-focused encoding without explicit box prediction during retrieval.
We next describe the text-guided item representation
(\S\ref{sec:repr}), item-side regularization objectives
(\S\ref{sec:item}), dual-encoder training with similarity-distribution
distillation (\S\ref{sec:dual}), and dataset construction
(\S\ref{sec:dataset}).

\subsection{Text-Guided Item Representation}
\label{sec:repr}

\vspace{-6pt}
For item-side encoding, given a multimodal item $(\mathbf{I}^{\mathrm{p}}, \mathbf{T}^{\mathrm{p}}) \in \mathcal{G}$, a DINOv3 ViT encodes $\mathbf{I}^{\mathrm{p}}$ into visual tokens $\mathbf{V} \in \mathbb{R}^{N_v \times C_v}$, and a BERT-based text encoder encodes $\mathbf{T}^{\mathrm{p}}$ into textual tokens $\mathbf{T} \in \mathbb{R}^{N_t \times C_t}$. Two linear layers project them into a shared $C_u$-dimensional space, giving $\mathbf{V}' \in \mathbb{R}^{N_v \times C_u}$ and $\mathbf{T}' \in \mathbb{R}^{N_t \times C_u}$. For notational simplicity, let $\mathbf{c}' = \mathbf{V}'_{[\mathrm{CLS}]} \in \mathbb{R}^{C_u}$ denote the visual class token, and let $\mathrm{sim}(\cdot,\cdot)$ denote cosine similarity. We omit the item index $j$ in this subsection since the same forward pass applies to all items.

\noindent\textbf{Text-guided cross-attention.}
Following Fig.~\ref{fig:architecture}(b), we maintain a set of $N_q$ learnable query tokens 
$\mathbf{Q} \in \mathbb{R}^{N_q \times C_u}$, similar to the learnable queries in DETR~\citep{carion2020detr} and BLIP-2~\citep{li2023blip}.
These query tokens are updated through two cross-attention steps.
They first attend to the textual tokens to form text-conditioned queries, which then attend to the visual tokens,
\begin{equation}
\mathbf{Q}' = \mathrm{CA}(\mathbf{Q}, \mathbf{T}'),
\qquad
\widetilde{\mathbf{V}} = \mathrm{CA}(\mathbf{Q}', \mathbf{V}'),
\end{equation}
where $\mathrm{CA}(\mathbf{X}, \mathbf{Y})$ denotes cross-attention with $\mathbf{X}$ as queries and $\mathbf{Y}$ as keys and values. Through this two-step interaction, each slot in $\widetilde{\mathbf{V}}$ carries text-conditioned visual evidence. We aggregate the slots with an MLP-based weighting and produce a single text-guided item feature $\mathbf{f}^{\mathrm{m}} \in \mathbb{R}^{C_u}$,
\begin{equation}
\mathbf{f}^{\mathrm{m}} = \sum_{k=1}^{N_q} \pi_k\, \widetilde{\mathbf{V}}_k,
\qquad
\boldsymbol{\pi} = \mathrm{softmax}\bigl(\mathrm{MLP}_{\pi}(\widetilde{\mathbf{V}})\bigr).
\end{equation}

\noindent\textbf{Complementary visual branch.}
While the slot module captures text-relevant semantics, it may underweight visual cues that are not explicitly mentioned in the text. We therefore add a parallel branch (the CLS-Guided path in Fig.~\ref{fig:architecture}(b)) that pools the visual patch tokens $\mathbf{V}'$ using two complementary anchors, the text-guided feature $\mathbf{f}^{\mathrm{m}}$ and the visual class token $\mathbf{c}'$,
\begin{equation}
\mathbf{s}
=
\lambda_m \,\mathrm{norm}\bigl(\mathrm{sim}(\mathbf{f}^{\mathrm{m}}, \mathbf{V}')\bigr)
+
\lambda_c \,\mathrm{norm}\bigl(\mathrm{sim}(\mathbf{c}', \mathbf{V}')\bigr),
\qquad
\lambda_m + \lambda_c = 1,
\end{equation}
where $\mathrm{norm}(\cdot)$ denotes $\ell_2$ normalization over the $N_v$ tokens, putting the two similarity vectors on the same scale before fusion. The patches are then pooled by a temperature-controlled softmax over $\mathbf{s}$,
\begin{equation}
\mathbf{f}^{\mathrm{a}}
=
\sum_{k=1}^{N_v}
\frac{\exp(s_k/\tau_p)}
{\sum_{\ell=1}^{N_v}\exp(s_\ell/\tau_p)}
\,\mathbf{V}'_k.
\end{equation}
A residual MLP combines the semantic and appearance features into the final item embedding,
\begin{equation}
\mathbf{f}^{\mathrm{i}}
=
\mathbf{f}^{\mathrm{m}}
+
\mathrm{MLP}(\mathbf{f}^{\mathrm{a}})
\in \mathbb{R}^{C_u}.
\end{equation}
This embedding is used by the item-side regularizers in \S\ref{sec:item} and the query--item training in \S\ref{sec:dual}.

\subsection{Item-Side Regularization}
\label{sec:item}
Given the item embedding defined above, we further regularize the item encoder with two cross-modal alignment objectives and one self-supervised distillation, all defined over a batch of size $B$. To avoid repetition we abbreviate the InfoNCE loss as
\vspace{-1pt}
\begin{equation}
\mathcal{L}_{\mathrm{NCE}}(\mathbf{a}\!\rightarrow\!\mathbf{b};\, \tau)
= -\frac{1}{B}\sum_{j=1}^{B} \log
\frac{\exp(\mathrm{sim}(\mathbf{a}_j, \mathbf{b}_j)/\tau)}
{\sum_{\ell=1}^{B} \exp(\mathrm{sim}(\mathbf{a}_j, \mathbf{b}_\ell)/\tau)}.
\end{equation}

\vspace{-6pt}
\noindent\textbf{Region--text embedding space alignment ($\mathcal{L}_{v2t}$).}
Following Fig.~\ref{fig:architecture}(c.1), we directly align the cropped target region with its text in a shared contrastive space. By aligning the cropped target region with its structured text before fusion, this objective explicitly establishes region--text correspondence in the shared alignment space. The subsequent text-guided cross-attention therefore starts from compatible visual and textual embeddings, rather than learning this correspondence only through the final fused item representation.


Let $\mathbf{I}^{\mathrm{p}}_{j,\mathrm{box}}$ denote the target region of the $j$-th item and $\mathbf{T}^{\mathrm{p}}_j$ its structured text.
Before dual-encoder training, we perform CLIP-style image--text contrastive pretraining on our constructed \textbf{ECom-RF-IMMR-10M} training set, described in \S\ref{sec:dataset}, with details provided in \S\ref{sec:experiments}.
We then encode $\mathbf{I}^{\mathrm{p}}_{j,\mathrm{box}}$ and $\mathbf{T}^{\mathrm{p}}_j$, and project their class-token features into a $C_a$-dimensional alignment space using the pretrained heads $\mathrm{Proj}_v$ and $\mathrm{Proj}_t$.
This alignment space is separate from the $C_u$ space used in \S\ref{sec:repr},
\begin{equation}
\mathbf{c}_j^{\mathrm{b}}=\mathrm{CLS}(\mathrm{Enc}^{\mathrm{p}}_v(\mathbf{I}^{\mathrm{p}}_{j,\mathrm{box}})),\;
\mathbf{z}_j^{\mathrm{b}}=\mathrm{norm}(\mathrm{Proj}_v(\mathbf{c}_j^{\mathrm{b}})),\;
\mathbf{z}_j^{\mathrm{t}}=\mathrm{norm}(\mathrm{Proj}_t(\mathrm{CLS}(\mathrm{Enc}_t(\mathbf{T}^{\mathrm{p}}_j)))).
\end{equation}
with $\mathbf{z}_j^{\mathrm{b}}, \mathbf{z}_j^{\mathrm{t}} \in \mathbb{R}^{C_a}$ and $\mathrm{norm}(\cdot)$ denoting $\ell_2$ normalization. We use a bidirectional InfoNCE loss,
\begin{equation}
\mathcal{L}_{v2t} = \tfrac{1}{2}\bigl[
\mathcal{L}_{\mathrm{NCE}}(\mathbf{z}^{\mathrm{b}}\!\to\!\mathbf{z}^{\mathrm{t}}; \tau_{v2t})
+ \mathcal{L}_{\mathrm{NCE}}(\mathbf{z}^{\mathrm{t}}\!\to\!\mathbf{z}^{\mathrm{b}}; \tau_{t2v})
\bigr].
\end{equation}

\noindent\textbf{Target-anchored fused--region alignment ($\mathcal{L}_{i2v}$).}
As shown in Fig.~\ref{fig:architecture}(c.2), we use the target-region feature $\mathbf{f}_j^{\mathrm{b}}$ as a visual anchor to keep the fused item embedding focused on the target object.
The item-side branch maps the item image and structured text to
$\mathbf{f}_j^{\mathrm{i}} = \mathrm{Fuse}(\mathbf{I}^{\mathrm{p}}_j, \mathbf{T}^{\mathrm{p}}_j)$.
The visual anchor reuses the box CLS $\mathbf{c}^{\mathrm{b}}_j$ from $\mathcal{L}_{v2t}$ and applies a separate detached projection,
$\mathbf{f}_j^{\mathrm{b}} = \mathrm{sg}\!\left[\mathrm{norm}(\mathbf{W}_r\mathbf{c}_j^{\mathrm{b}})\right]$,
where $\mathbf{W}_r \in \mathbb{R}^{C_u \times C_v}$.
The positive term uses an in-batch InfoNCE objective to pull the fused embedding toward this anchor, defined as $\mathcal{L}_{i2v}^{\mathrm{pos}} = \mathcal{L}_{\mathrm{NCE}}(\mathbf{f}^{\mathrm{i}}\!\to\!\mathbf{f}^{\mathrm{b}}; \tau_{i2v})$. For the negative term, we draw $K$ texts $\widetilde{\mathbf{T}}^{\mathrm{p}}_{j,k}$ from items whose primary category differs from that of the $j$-th item, and re-fuse each with the original item image to obtain $\mathbf{f}_{j,k}^{\mathrm{i,neg}} = \mathrm{norm}(\mathrm{Fuse}(\mathbf{I}^{\mathrm{p}}_j, \widetilde{\mathbf{T}}^{\mathrm{p}}_{j,k}))$. We then apply a softplus penalty to the similarity between each mismatched embedding and the visual anchor,
\begin{equation}
\mathcal{L}_{i2v}^{\mathrm{hard}} = \frac{1}{B K} \sum_{j=1}^{B}\sum_{k=1}^{K} \mathrm{softplus}\bigl(\mathrm{sim}(\mathbf{f}_{j,k}^{\mathrm{i,neg}}, \mathbf{f}_j^{\mathrm{b}})\bigr),
\qquad
\mathcal{L}_{i2v} = \mathcal{L}_{i2v}^{\mathrm{pos}} + \lambda_h\, \mathcal{L}_{i2v}^{\mathrm{hard}}.
\end{equation}
The cross-category constraint avoids false negatives that arise when a same-category title is visually compatible with the target. The negative term is an additive penalty rather than an in-batch denominator, which avoids competing with the InfoNCE positive on the same partition function.

\noindent\textbf{Spatial-relational distillation ($\mathcal{L}_{\mathrm{SRD}}$).}
Following Fig.~\ref{fig:architecture}(c.3), we distill patch-level spatial relations from a frozen DINOv3 ViT-B/16 teacher into the item-side visual encoder. This encourages the patch-level spatial structure to remain consistent with that of the pre-trained self-supervised backbone during multimodal training. 
Let $\mathbf{F}_j^{\mathrm{s}}, \mathbf{F}_j^{\mathrm{t}} \in \mathbb{R}^{H_f \times W_f \times C_v}$ denote the spatially reshaped patch tokens of the $j$-th item from the DINOv3-based encoder on the item-side as the student encoder and the frozen DINOv3 teacher encoder, respectively, with $H_f W_f = N_v$.
Using ROI Align, which resamples the item-box region into a fixed-size feature grid, we obtain
$\mathbf{R}_j^{\{\mathrm{s},\mathrm{t}\}} 
= \mathrm{ROIAlign}(\mathbf{F}_j^{\{\mathrm{s},\mathrm{t}\}}, \mathbf{b}_j^{\mathrm{p}})$.
We then flatten each $\mathbf{R}_j^{\{\mathrm{s},\mathrm{t}\}}$ into $h_r w_r$ patch features and form
$\mathbf{A}_j^{\mathrm{s}} = \mathrm{sim}(\mathbf{R}_j^{\mathrm{s}}, \mathbf{R}_j^{\mathrm{s}})$ and
$\mathbf{A}_j^{\mathrm{t}} = \mathrm{sim}(\mathbf{R}_j^{\mathrm{t}}, \mathbf{R}_j^{\mathrm{t}})$,
both in $\mathbb{R}^{h_r w_r \times h_r w_r}$.
After converting each row into a distribution with a row-wise softmax, the student similarity structure is aligned to the teacher with KL divergence,
\begin{equation}
\mathcal{L}_{\mathrm{SRD}} = \frac{1}{B}\sum_{j=1}^{B} D_{\mathrm{KL}}\!\bigl(\mathrm{softmax}(\mathbf{A}_j^{\mathrm{t}}) \,\|\, \mathrm{softmax}(\mathbf{A}_j^{\mathrm{s}})\bigr).
\end{equation}
Restricting the supervision to the box region keeps the signal on the target object and avoids transferring teacher noise from cluttered backgrounds.
 
The full item-side objective is
\begin{equation}
\mathcal{L}_{\mathrm{item}} = \lambda_{v2t}\mathcal{L}_{v2t} + \lambda_{i2v}\mathcal{L}_{i2v} + \lambda_{\mathrm{SRD}}\mathcal{L}_{\mathrm{SRD}}.
\end{equation}

\subsection{Dual-Encoder Training}
\label{sec:dual}
The query encoder is a DINOv3 ViT with the same architecture as the item-side visual encoder but trained with independent parameters, taking the cropped query region as input and producing the projected \texttt{[CLS]} feature
$\mathbf{f}_j^{\mathrm{q}} = \mathrm{MLP}_q(\mathbf{c}_j^{\mathrm{q}}) \in \mathbb{R}^{C_u}$.



\noindent\textbf{Query--item contrastive learning ($\mathcal{L}_{q2i}$).}
For the query-to-item direction, we augment the in-batch denominator with the same image--text mismatch hard negatives $\{\mathbf{f}_{j,k}^{\mathrm{i,neg}}\}_{k=1}^{K}$ used in $\mathcal{L}_{i2v}$,
\begin{equation}
\mathcal{L}_{q2i}^{\mathrm{hard}}
=
-\frac{1}{B}
\sum_{j=1}^{B}
\log
\frac{
s(\mathbf{f}_j^{\mathrm{q}},\mathbf{f}_j^{\mathrm{i}})
}{
\sum_{\ell=1}^{B}
s(\mathbf{f}_j^{\mathrm{q}},\mathbf{f}_{\ell}^{\mathrm{i}})
+
\sum_{k=1}^{K}
s(\mathbf{f}_j^{\mathrm{q}},\mathbf{f}_{j,k}^{\mathrm{i,neg}})
},
\end{equation}
where $s(\mathbf{a},\mathbf{b}) = \exp(\mathrm{sim}(\mathbf{a},\mathbf{b})/\tau_{q2i})$. The reverse direction uses a standard InfoNCE term, and the two are averaged,
\begin{equation}
\mathcal{L}_{q2i}
=
\tfrac{1}{2}
\left[
\mathcal{L}_{q2i}^{\mathrm{hard}}
+
\mathcal{L}_{\mathrm{NCE}}(\mathbf{f}^{\mathrm{i}}\!\to\!\mathbf{f}^{\mathrm{q}}; \tau_{i2q})
\right].
\end{equation}

\noindent\textbf{Similarity-distribution distillation ($\mathcal{L}_{\mathrm{SDD}}$).}
To regularize the discriminative structure of item representations, we align query-to-item similarity distributions between the student encoder and a frozen MoCo-pretrained image-to-image retrieval ViT, which also appears as a baseline in our experiments. Applying the teacher to the cropped query and item regions yields region-level features $\mathbf{g}_j^{\mathrm{q}}$ and $\{\mathbf{g}_k^{\mathrm{i}}\}_{k=1}^{B}$. We define the student and teacher distributions as
\begin{equation}
\mathbf{p}_j^{\mathrm{s}}
=
\mathrm{softmax}
\bigl(
\bigl[\mathrm{sim}(\mathbf{f}_k^{\mathrm{i}}, \mathbf{f}_j^{\mathrm{q}})\bigr]_{k=1}^{B}
/ \tau_{\mathrm{stu}}
\bigr),
\quad
\mathbf{p}_j^{\mathrm{t}}
=
\mathrm{softmax}
\bigl(
\bigl[\mathrm{sim}(\mathbf{g}_k^{\mathrm{i}}, \mathbf{g}_j^{\mathrm{q}})\bigr]_{k=1}^{B}
/ \tau_{\mathrm{tea}}
\bigr),
\vspace{-6pt}
\end{equation}
and align them with KL divergence,
\begin{equation}
\mathcal{L}_{\mathrm{SDD}}
=
\frac{1}{B}
\sum_{j=1}^{B}
D_{\mathrm{KL}}
\bigl(
\mathbf{p}_j^{\mathrm{t}} \,\|\, \mathbf{p}_j^{\mathrm{s}}
\bigr).
\end{equation}
Distilling from the image-to-image teacher transfers relative visual similarity structure, complementing the multimodal and text-guided item representations learned by the student.



\vspace{-6pt}

\paragraph{Joint optimization objective.}
We combine two cross-branch losses as
$
\mathcal{L}_{\mathrm{dual}}
=
\lambda_{q2i}\mathcal{L}_{q2i}
+
\lambda_{\mathrm{SDD}}\mathcal{L}_{\mathrm{SDD}},
$
and integrate dual-encoder loss with the item-side loss to form the final training objective,
\begin{equation}
\mathcal{L}
=
\lambda_{\mathrm{dual}}\, \mathcal{L}_{\mathrm{dual}}
+
\lambda_{\mathrm{item}}\, \mathcal{L}_{\mathrm{item}}.
\end{equation}

\subsection{Dataset Construction}
\label{sec:dataset}
We construct an e-commerce dataset suite for IMMR, including a large-scale training set \textbf{ECom-RF-IMMR-10M} and two evaluation benchmarks, \textbf{ECom-RF-IMMR-Normal} and \textbf{ECom-RF-IMMR-Mosaic}. Training pairs are mined from two complementary sources, pairs of main and auxiliary item images and user image-search click logs. For these main--auxiliary pairs, no user-drawn query boxes exist. We therefore treat the auxiliary view as the query image and use GroundingDINO followed by CLIP text--image alignment filtering to localize the target product on it as a pseudo query box. For click logs, query boxes are directly obtained from user-drawn regions of interest. Item-side boxes are then generated by matching each query crop against GroundingDINO proposals on the item image using a production image-search embedding model. We further apply Standard Product Unit (SPU) level deduplication and category-balanced sampling to obtain 10M training pairs.
 
The Normal evaluation set is constructed from held-out click logs using the same mining pipeline, followed by VLM-assisted and human auditing for quality control. To counteract the center bias commonly observed in e-commerce imagery and evaluate robustness under cluttered multi-item layouts, \textbf{ECom-RF-IMMR-Mosaic} re-synthesizes the candidate image for each Normal sample. The query image, query box, item title, and item category are kept unchanged, while the candidate image is rebuilt by compositing the ground-truth target crop with cross-category distractor crops on a randomly sampled background. This process creates cluttered scenes with spatial ambiguity and provides a controlled benchmark for evaluating text-guided item grounding. All datasets follow the same schema, $(\mathbf{I}^{\mathrm{q}},\mathbf{b}^{\mathrm{q}},\mathbf{I}^{\mathrm{p}},\mathbf{b}^{\mathrm{p}}, \mathbf{T}^{\mathrm{p}},\mathbf{c}^{\mathrm{p}})$. Full construction details are provided in Appendix~\ref{app:dataset}.

%% file: tex_files/04.experiments.tex
\vspace{-6pt}
\section{Experiments}
\label{sec:experiments}
\vspace{-6pt}
\textbf{Implementation details.}
TIGER-FG is trained on \textbf{ECom-RF-IMMR-10M} with a $1{:}1$ mixture of original and Mosaic-augmented samples by default.
Both query and item visual encoders are initialized from DINOv3 ViT-B/16~\citep{simeoni2025dinov3} at $224\!\times\!224$ resolution and trained as independent copies.
The text encoder is initialized from the text branch of Chinese-CLIP ViT-B/16~\citep{chinese-clip} and further pretrained on \textbf{ECom-RF-IMMR-10M} with CLIP-style image--text contrastive learning.
The unified embedding dimension is $C_u=256$, the item branch uses $N_q=8$ learnable query tokens, and we use $K=1$ mismatched-text hard negative per sample.
We train for 10 epochs with AdamW, a learning rate of $2\!\times\!10^{-5}$, and batch size 256.
For fair comparison, all baseline models are fine-tuned on \textbf{ECom-RF-IMMR-10M} using only original samples, without Mosaic-augmented mixing.
Full hyperparameters and training details are provided in Appendix~\ref{app:impl}.

\noindent\textbf{Training hard-example augmentation.}
We construct hard training signals with two mechanisms that affect the training objective in different ways.
\emph{Image--text mismatch} explicitly increases the number of negative item representations.
For each item image $\mathbf{I}_j^{\mathrm{p}}$, we sample $K$ mismatched texts from cross-category items and fuse them with the same image to obtain hard negative item embeddings.
These negatives are added to both $\mathcal{L}_{i2v}$ and $\mathcal{L}_{q2i}$, encouraging the model to suppress visually plausible but semantically incorrect image--text pairs.
\emph{Mosaic augmentation} does not add extra negative entries to the loss.
Instead, it re-synthesizes the candidate image of an original query--item pair into a cluttered multi-item scene.
Specifically, we composite the target item with several cross-category distractor items on a sampled background, while keeping the query region and its matched item text unchanged.
To strengthen in-batch competition, we place the items used in the same Mosaic composition into the same mini-batch.
Thus, the model observes multiple candidate entries with nearly identical visual content but different item texts and different matched queries.
These samples serve as hard in-batch competitors and force the item encoder to rely on structured text to focus on the correct item, rather than exploiting shared background, layout, or center bias.

\textbf{Datasets and evaluation.}
We evaluate on two in-domain benchmarks, \textbf{ECom-RF-IMMR-Normal} and \textbf{ECom-RF-IMMR-Mosaic}, as described in \S\ref{sec:dataset}.
To evaluate generalization, we further adapt two public e-commerce benchmarks, \textbf{eSSPR}~\citep{chen2023unified} and \textbf{LookBench}, to the IMMR setting, where a cropped image query is used to retrieve image--text item candidates.Results are aggregated across all LookBench subsets, and additional adaptation details are provided in Appendix~\ref{app:public_benchmarks}.

\textbf{Evaluation metrics.}
We evaluate retrieval performance using standard retrieval metrics. For \textbf{ECom-RF-IMMR-Normal}, \textbf{ECom-RF-IMMR-Mosaic}, and \textbf{eSSPR}, we report Recall@K, MRR@K, and NDCG@K with $K \in \{1, 4, 10\}$. For \textbf{LookBench}, we additionally report HitRate@K.

\input{tables/table4_1_compare_total}

\subsection{Multimodal Retrieval Performance Comparison}

\vspace{-6pt}
We compare TIGER-FG with three groups of baselines, including CLIP/BLIP-based vision--language models, large-scale multimodal embedding models, and image retrieval models.
Following UniIR~\citep{wei2024uniir}, SF and FF denote score-level fusion and feature-level fusion variants for CLIP/BLIP-based models.
Results on our constructed benchmarks are reported in Table~\ref{tab:4_1_comparisons}, and results on public benchmarks are reported in Table~\ref{tab:4_2_comparisons}.

On \textbf{ECom-RF-IMMR-Normal}, TIGER-FG achieves the best performance across Recall, MRR, and NDCG.
The gain over strong multimodal embedding baselines shows that text-guided item representation improves the alignment between cropped visual queries and image--text item candidates, even when candidate images are relatively clean.
TIGER-FG-RAW also outperforms all baselines on this split, indicating that the proposed architecture and objectives already provide strong item-level alignment without Mosaic training.
The advantage becomes much larger on \textbf{ECom-RF-IMMR-Mosaic}.
This split introduces dense object co-occurrence and stronger visual distractors, causing a substantial degradation for existing models.
For example, Qwen3-VL-Emb drops from 74.0 to 40.8 in Recall@1, while TIGER-FG only decreases from 80.1 to 75.2.
TIGER-FG-RAW also drops sharply to 47.8 Recall@1, showing that clutter-aware training is critical for robustness under multi-item ambiguity.
These results suggest that model scale and generic multimodal pretraining are insufficient for IMMR, where the item representation must bridge both modality and granularity disparities.

\input{tables/table4_2_compare}

The same trend is observed on public benchmarks in Table~\ref{tab:4_2_comparisons}.
On \textbf{eSSPR}, where images usually contain a single clean item, Recall is already close across methods.
TIGER-FG obtains comparable Recall while achieving the best MRR and NDCG, suggesting that relevant items are placed earlier in the retrieved list.
On \textbf{LookBench}, which contains noisy candidates and one-to-many relevance, TIGER-FG achieves the best performance across HitRate, Recall, MRR, and NDCG.
This confirms that the learned text-guided item representation generalizes beyond our constructed benchmark and remains effective under ambiguous candidate pools.

\begin{figure}[htbp]
    \vspace{-6pt}
    \centering
    \includegraphics[width=1\linewidth]{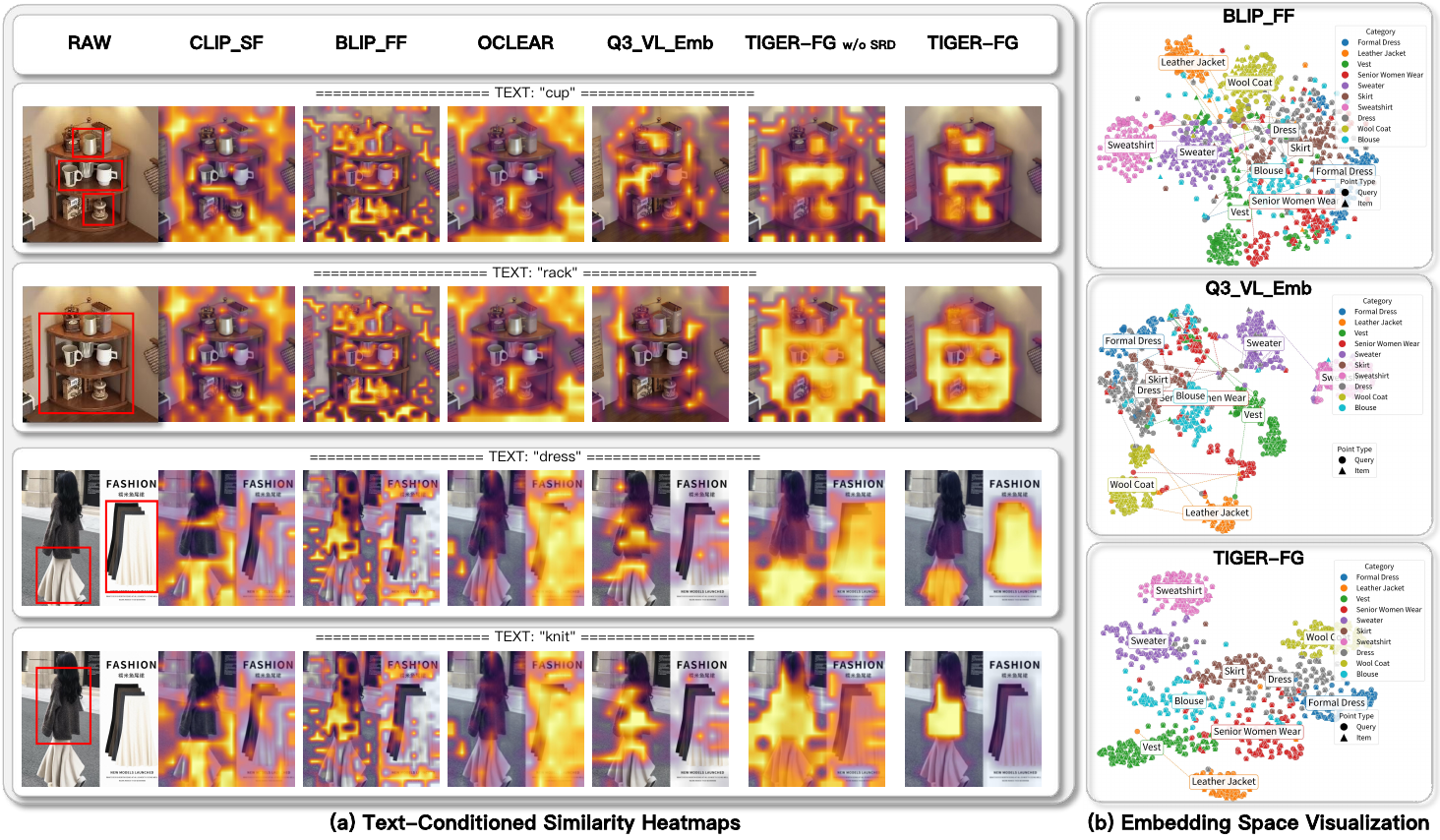}
    \caption{\textbf{Heatmap and embedding visualizations.}
\textbf{(a)} Text-conditioned heatmaps show that TIGER-FG focuses on query-relevant regions.
\textbf{(b)} Compared with Qwen3-VL-Emb, TIGER-FG forms tighter category clusters and closer query--item alignment.}
    \label{fig:figure3}
    \vspace{-6pt}
\end{figure}

Qualitative results are shown in Figure~\ref{fig:figure3}.
In Figure~\ref{fig:figure3}(a), given the same candidate image, TIGER-FG produces concentrated and semantically consistent responses under different text queries.
It distinguishes cup from rack in the first example and category information such as dress from attribute information such as knit in the second.
Removing SRD leads to more diffuse responses, indicating that spatial-relational distillation helps preserve localized visual structure.
In Figure~\ref{fig:figure3}(b), TIGER-FG places query features close to matched item features and forms compact category clusters with clear separation.
Although Qwen3-VL-Emb also shows separable clusters, it contains more scattered outliers and query--item mismatches, suggesting weaker fine-grained alignment for IMMR.

\vspace{-6pt}
\subsection{Ablation Study}
\vspace{-6pt}
We conduct ablations on \textbf{ECom-RF-IMMR-Mosaic}, with results summarized in Table~\ref{tab:4_3_ablation}.
Here CVB refers to the complementary visual branch, SRD to spatial-relational distillation, and SDD to similarity-distribution distillation.
The largest drop comes from TIGER-FG-RAW, which uses the same architecture and objectives but is trained only on original samples without Mosaic augmentation.
Its Recall@1 decreases from 75.2 to 47.8, showing that clutter-aware training is essential for handling multi-item ambiguity.
Replacing DINOv3 with a CLIP-based backbone also causes a clear degradation, confirming the importance of region-sensitive visual representations for IMMR.
Removing CVB leads to only a small drop, indicating that the text-guided branch captures the main semantic signal while CVB provides complementary appearance cues.
For distillation, removing SDD substantially reduces performance, suggesting that query--item similarity structure from the image-to-image teacher improves retrieval discrimination.
In contrast, removing SRD has little effect on ranking metrics, indicating that SRD mainly acts as a grounding-oriented regularizer rather than the primary driver of retrieval accuracy.

\vspace{-6pt}
\subsection{Modality Comparison}
\label{sec:modality_comparison}

We analyze query-side and item-side modality configurations on
\textbf{ECom-RF-IMMR-Normal}, with results reported in
Table~\ref{tab:4_4_modalities}. The query is represented by a cropped region,
while the item side uses the corresponding image, its associated text, or both.
Single-modality settings show clear limitations. The DINOv3 visual baseline
achieves 76.7 Recall@1, indicating that region-level visual representations
provide a strong retrieval signal. In contrast, using only textual information
drops Recall@1 to 30.1, showing that item titles lack the fine-grained visual
grounding required for accurate retrieval. Our model with only item images is
lower than the DINOv3 baseline, since it is optimized for multimodal alignment
rather than unimodal matching. Adding textual information on the item side
consistently improves performance. Combining visual and textual signals gives
the best results, reaching 80.6 and 80.1 Recall@1 under different visual
configurations, and surpassing all single-modality baselines. These results
show that text-guided interaction helps align localized query regions with
item candidates through both visual evidence and textual semantics.

\input{tables/table4_34}

%% file: tables/table4_1_compare_total.tex
\begin{table*}[htbp]

    \centering
    \scriptsize
    \setlength{\tabcolsep}{1.7pt}
    \renewcommand\arraystretch{1}
    \caption{\textbf{Comparison with representative retrieval baselines on the constructed e-commerce benchmarks.}
Results are reported on \textbf{ECom-RF-IMMR-Normal} and \textbf{ECom-RF-IMMR-Mosaic}.
Abbreviations of large multimodal embedding models are as follows:
BGE-L (\textit{BGE-VL-Large}),
GME-2B (\textit{gme-Qwen2-VL-2B-Instruct}),
Qwen3-VL-Emb (\textit{Qwen3-VL-Embedding-2B}),
and Ops-MM-Emb (\textit{Ops-MM-embedding-v1-2B}).
TIGER-FG-RAW uses the same architecture and objectives as TIGER-FG but is trained only on original samples without Mosaic augmentation, while TIGER-FG uses a $1\!:\!1$ mixture of original and Mosaic samples.
All values are in \%.}
    \label{tab:4_1_comparisons}
    \begin{tabular}{l c c ccc cc cc ccc cc cc}
    \toprule
    \multirow{3}{*}{Method}
        & \multirow{3}{*}{\makecell{Query\\Param}}
        & \multirow{3}{*}{Dim}
        & \multicolumn{7}{c}{ECom-RF-IMMR-Normal}
        & \multicolumn{7}{c}{ECom-RF-IMMR-Mosaic} \\
    \cmidrule(lr){4-10} \cmidrule(lr){11-17}
        & & & \multicolumn{3}{c}{Recall} & \multicolumn{2}{c}{MRR} & \multicolumn{2}{c}{NDCG}
            & \multicolumn{3}{c}{Recall} & \multicolumn{2}{c}{MRR} & \multicolumn{2}{c}{NDCG} \\
    \cmidrule(lr){4-6}\cmidrule(lr){7-8}\cmidrule(lr){9-10}
    \cmidrule(lr){11-13}\cmidrule(lr){14-15}\cmidrule(lr){16-17}
        & & & @1 & @4 & @10 & @4 & @10 & @4 & @10
              & @1 & @4 & @10 & @4 & @10 & @4 & @10 \\
    \midrule
    \multicolumn{17}{l}{\textit{(a) CLIP/BLIP-based Vision-Language Models}} \\
    CLIP-SF~\citep{wei2024uniir}    & 188.3M & 512
        & 66.9 & 85.7 & 92.5 & 74.8 & 75.8 & 77.6 & 79.9
        & 30.0 & 49.4 & 61.7 & 37.8 & 39.6 & 40.7 & 44.9 \\
    CLIP-FF~\citep{wei2024uniir}    & 188.3M & 512
        & 37.2 & 59.3 & 72.0 & 46.0 & 48.0 & 49.4 & 53.7
        &  8.1 & 18.1 & 27.6 & 12.0 & 13.4 &  1.5 & 16.7 \\
    BLIP-SF~\citep{wei2024uniir}    & 195.4M & 768
        & 67.8 & 84.8 & 91.0 & 74.9 & 75.9 & 77.4 & 79.6
        & 14.9 & 25.8 & 33.6 & 19.3 & 20.4 & 20.9 & 23.6 \\
    BLIP-FF~\citep{wei2024uniir}    & 223.7M & 768
        & 68.8 & 86.0 & 92.2 & 76.0 & 77.0 & 78.6 & 80.7
        & 16.1 & 27.7 & 35.8 & 20.7 & 22.0 & 22.5 & 25.2 \\
    BGE-L~\citep{zhou2024megapairs} & 0.4B & 768
        & 69.4 & 87.0 & 93.1 & 76.8 & 77.8 & 79.4 & 81.5
        & 19.3 & 31.6 & 39.7 & 24.3 & 25.4 & 26.1 & 28.8 \\
    \midrule
    \multicolumn{17}{l}{\textit{(b) Large-scale Multimodal Embedding Models}} \\
    GME-2B~\citep{zhang2024gme}       & 2.2B & 1536
        & 54.9 & 77.1 & 86.4 & 64.1 & 65.5 & 67.4 & 70.6
        & 30.4 & 50.5 & 63.1 & 38.4 & 40.3 & 41.5 & 45.7 \\
    Qwen3-VL-Emb~\citep{qwen3vlembedding} & 2.1B & 2048
        & 74.0 & 90.0 & 94.6 & 80.8 & 81.5 & 83.2 & 84.8
        & 40.8 & 59.7 & 69.1 & 48.5 & 49.9 & 51.3 & 54.5 \\
    Ops-MM-Emb~\citep{ops_mm_embedding}   & 2.2B & 1536
        & 72.7 & 90.3 & 95.4 & 80.1 & 81.0 & 82.7 & 84.5
        & 31.9 & 51.8 & 63.6 & 39.9 & 41.7 & 42.9 & 46.9 \\
    \midrule
    \multicolumn{17}{l}{\textit{(c) Image Retrieval Models}} \\
    UniEcs~\citep{liang2025uniecs}  & 198.0M & 256
        & 41.4 & 62.1 & 73.7 & 49.8 & 51.5 & 52.9 & 56.8
        &  3.4 &  7.3 & 10.9 &  4.9 &  5.5 &  5.5 &  6.7 \\
    OClear~\citep{cheng2023category}  & 189.2M & 256
        & 62.7 & 83.0 & 90.8 & 71.1 & 72.3 & 74.1 & 76.8
        & 14.3 & 24.7 & 31.9 & 18.4 & 19.5 & 20.0 & 22.4 \\
    \rowcolor{gray!12}
    \textbf{TIGER-FG-RAW} & \textbf{85.7M} & 256
        & \underline{79.4} & \underline{94.8} & \underline{98.2} & \underline{86.0} & \underline{86.5} & \underline{88.2} & \underline{89.4}
        & \underline{47.8} & \underline{64.0} & \underline{70.7} & \underline{54.5} & \underline{55.6} & \underline{56.9} & \underline{59.2} \\
    \rowcolor{gray!12}
    \textbf{TIGER-FG} & \textbf{85.7M} & 256
        & \textbf{80.1} & \textbf{95.3} & \textbf{98.4} & \textbf{86.7} & \textbf{87.2} & \textbf{88.9} & \textbf{90.0}
        & \textbf{75.2} & \textbf{93.5} & \textbf{97.8} & \textbf{83.0} & \textbf{83.7} & \textbf{85.7} & \textbf{87.2} \\
    \bottomrule
    \end{tabular}
    \vspace{-6pt}
    
\end{table*}

%% file: tables/table4_2_compare.tex
\begin{table*}[htbp]
    \vspace{-3pt}
    \centering
    \scriptsize
    \setlength{\tabcolsep}{2pt}
    \renewcommand\arraystretch{1.1}
    \caption{\textbf{Comparison with representative baselines on public benchmarks.}
        Results are reported on \textbf{eSSPR} and \textbf{LookBench}.}
    \label{tab:4_2_comparisons}
    \begin{tabular}{l ccc cc cc ccc ccc cc cc}
    \toprule
    \multirow{3}{*}{Method}
        & \multicolumn{7}{c}{eSSPR}
        & \multicolumn{10}{c}{LookBench} \\
    \cmidrule(lr){2-8} \cmidrule(lr){9-18}
        & \multicolumn{3}{c}{Recall} & \multicolumn{2}{c}{MRR} & \multicolumn{2}{c}{NDCG}
        & \multicolumn{3}{c}{HitRate} & \multicolumn{3}{c}{Recall} & \multicolumn{2}{c}{MRR} & \multicolumn{2}{c}{NDCG} \\
    \cmidrule(lr){2-4}\cmidrule(lr){5-6}\cmidrule(lr){7-8}
    \cmidrule(lr){9-11}\cmidrule(lr){12-14}\cmidrule(lr){15-16}\cmidrule(lr){17-18}
        & @1 & @4 & @10 & @4 & @10 & @4 & @10
        & @1 & @4 & @10 & @1 & @4 & @10 & @4 & @10 & @4 & @10 \\
    \midrule
    \multicolumn{18}{l}{\textit{(a) CLIP/BLIP-based Vision-Language Models}} \\
    CLIP-SF~\citep{wei2024uniir}
        & 24.8 & 97.5 & \underline{98.6} & \underline{60.6} & \underline{60.8} & \underline{70.2} & \underline{70.6}
        & \underline{36.3} & \underline{55.4} & \underline{65.5} & \underline{16.9} & \underline{32.4} & \underline{43.5} & \underline{44.0} & \underline{45.6} & \underline{35.4} & \underline{38.1} \\
    BLIP-FF~\citep{wei2024uniir}
        & 20.2 & \textbf{97.7} & \textbf{98.9} & 58.6 & 58.8 & 68.8 & 69.2
        & 30.5 & 47.4 & 57.2 & 14.9 & 28.9 & 38.4 & 37.3 & 38.8 & 30.6 & 33.2 \\
    \midrule
    \multicolumn{18}{l}{\textit{(b) Large-scale Multimodal Embedding Models}} \\
    Qwen3-VL-Emb~\citep{qwen3vlembedding}
        & \textbf{28.8} & 88.8 & 93.4 & 57.5 & 58.2 & 65.7 & 67.2
        & 33.1 & 51.0 & 59.2 & 14.8 & 29.0 & 38.6 & 40.3 & 41.5 & 32.2 & 34.1 \\
    \midrule
    \multicolumn{18}{l}{\textit{(c) Image Retrieval Models}} \\
    UniEcs~\citep{liang2025uniecs}
        & 25.5 & 93.3 & 95.9 & 58.6 & 59.1 & 67.7 & 68.6
        & 28.9 & 45.6 & 55.7 & 12.3 & 24.7 & 34.7 & 35.6 & 37.2 & 27.8 & 30.1 \\
    OClear~\citep{cheng2023category}
        & 25.1 & 96.5 & 97.8 & 60.2 & 60.4 & 69.7 & 70.1
        & 29.4 & 46.4 & 57.5 & 14.0 & 27.9 & 38.2 & 36.1 & 37.8 & 29.7 & 32.6 \\
    \rowcolor{gray!12}
    \textbf{TIGER-FG}
        & \underline{26.4} & \underline{97.1} & \underline{98.6} & \textbf{61.3} & \textbf{61.6} & \textbf{70.7} & \textbf{71.2}
        & \textbf{39.8} & \textbf{61.3} & \textbf{71.0} & \textbf{18.5} & \textbf{38.8} & \textbf{52.2} & \textbf{48.6} & \textbf{50.0} & \textbf{41.5} & \textbf{45.0} \\
    \bottomrule
    \end{tabular}
    \vspace{-3pt}
\end{table*}

%% file: tables/table4_34.tex
\begin{table}[htbp]
\vspace{-6pt}

\centering
\scriptsize
\renewcommand\arraystretch{1}

\begin{minipage}{0.46\linewidth}
\centering
\caption{\textbf{Ablation Studies.}
}
\label{tab:4_3_ablation}
\setlength\tabcolsep{2pt}
\begin{tabular}{l ccc cc cc}
\toprule
\multirow{3}{*}{Method}
    & \multicolumn{7}{c}{ECom-RF-IMMR-Mosaic} \\
\cmidrule(lr){2-8}
    & \multicolumn{3}{c}{Recall} & \multicolumn{2}{c}{MRR} & \multicolumn{2}{c}{NDCG} \\
\cmidrule(lr){2-4}\cmidrule(lr){5-6}\cmidrule(lr){7-8}
    & @1 & @4 & @10 & @4 & @10 & @4 & @10 \\
\midrule
\rowcolor{gray!12}
\textbf{TIGER-FG} & \underline{75.2} & \textbf{93.5} & \textbf{97.8} & \underline{83.0} & \underline{83.7} & \textbf{85.7} & \textbf{87.2} \\
\midrule
CLIP-backbone      & 64.6 & 85.8 & 93.1 & 73.5 & 74.6 & 76.6 & 79.1 \\
w/o CVB  & 74.8 & \underline{93.3} & \textbf{97.8} & 82.7 & 83.4 & \underline{85.4} & \underline{87.0} \\
w/o SRD   & \textbf{75.3} & \textbf{93.5} & \textbf{97.8} & \textbf{83.1} & \textbf{83.8} & \textbf{85.7} & \textbf{87.2} \\
w/o SDD & 69.9 & 91.1 & \underline{96.8} & 78.8 & 79.7 & 81.9 & 83.9 \\
TIGER-FG-RAW & {47.8} & {64.0} & {70.7} & {54.5} & {55.6} & {56.9} & {59.2} \\

\bottomrule
\end{tabular}
\end{minipage}%
\hfill
\begin{minipage}{0.52\linewidth}
\centering
\caption{\textbf{Modalities comparison.}}
\label{tab:4_4_modalities}
\setlength\tabcolsep{2pt}
\begin{tabular}{l ccc cc cc}
\toprule
\multirow{3}{*}{Method}
    & \multicolumn{7}{c}{ECom-RF-IMMR-Normal} \\
\cmidrule(lr){2-8}
    & \multicolumn{3}{c}{Recall} & \multicolumn{2}{c}{MRR} & \multicolumn{2}{c}{NDCG} \\
\cmidrule(lr){2-4}\cmidrule(lr){5-6}\cmidrule(lr){7-8}
    & @1 & @4 & @10 & @4 & @10 & @4 & @10 \\
\midrule
DINOv3 (C.$\to$C.)       & 76.7 & 93.8 & 97.9 & 84.0 & 84.6 & 86.5 & 87.9 \\
DINOv3 (C.$\to$I.)       & 72.6 & 90.4 & 95.3 & 80.1 & 80.9 & 82.8 & 84.5 \\
Eco-CLIP (C.$\to$T.)     & 30.1 & 57.1 & 73.3 & 40.8 & 43.2 & 44.9 & 50.4 \\
TIGER-FG (C.$\to$C.)         & 27.4 & 45.2 & 59.2 & 34.4 & 36.5 & 37.2 & 41.9 \\
TIGER-FG (C.$\to$I.)         & 26.0 & 43.2 & 56.9 & 32.8 & 34.8 & 35.4 & 40.0 \\
TIGER-FG (C.$\to$C.\!+\!T.)  & \textbf{80.6} & \textbf{95.7} & \textbf{98.7} & \textbf{87.1} & \textbf{87.6} & \textbf{89.3} & \textbf{90.4} \\
\rowcolor{gray!12}
\textbf{TIGER-FG (C.$\to$I.\!+\!T.)} & \underline{80.1} & \underline{95.3} & \underline{98.4} & \underline{86.7} & \underline{87.2} & \underline{88.9} & \underline{90.0} \\
\bottomrule
\end{tabular}
\end{minipage}
\vspace{-6pt}
\end{table}

%% file: tex_files/05.conclusion.tex
\vspace{-6pt}
\section{Conclusion}
\vspace{-6pt}

We study image-to-multimodal item retrieval (IMMR), where a cropped visual query is matched against item candidates represented by full images and structured text.
This setting introduces modality and granularity disparities that are not well addressed by standard image--text retrieval models.
We propose TIGER-FG, a text-guided retrieval framework that learns target-focused item representations.
We further construct \textbf{ECom-RF-IMMR}, a large-scale benchmark suite for training and evaluating IMMR under clean and cluttered mosaic item scenes.
Experiments on both in-domain and public e-commerce benchmarks show consistent improvements over strong retrieval baselines, especially in multi-item, noisy and ne-to-many scenarios.
Future work will explore more robust grounding under noisy, sparse, or weakly aligned supervision.
We provide a brief discussion of scope and practical limitations in Appendix~\ref{app:limitations}.

%% file: tex_files/F.limitation.tex
\section{Limitations}
\label{app:limitations}

TIGER-FG is designed for image-to-multimodal item retrieval. In this work, we instantiate and evaluate it in the e-commerce setting, where each candidate is typically represented by full item images paired with structured item text. This setting provides a practical testbed for studying cross-modal and granularity disparities in fine-grained item retrieval. Further evaluation may consider item entries with different text fields, category systems, and language styles.

Our benchmarks focus on item-level retrieval under clean and cluttered visual layouts. They are intended to provide controlled and reproducible evaluation of retrieval performance, rather than to simulate all factors in online serving, such as user personalization or live traffic dynamics.

Finally, our training recipe uses item boxes only for auxiliary supervision and dataset construction. These boxes are not required during indexing or retrieval, where TIGER-FG directly encodes full image--text item candidates. Exploring weaker or automatically checked signals may further simplify data construction.

%% file: tex_files/A.related_work_appendix.tex
\section{Extended Related Work}
\label{app:related}

\subsection{Vision--language representation learning}

Vision--language retrieval learns a shared embedding space for visual and
textual inputs. Large-scale contrastive models such as
ALIGN~\citep{jia2021scaling} and CLIP~\citep{radford2021learning, chinese-clip}
show strong transferability across multimodal retrieval tasks. Subsequent work
improves cross-modal alignment with stronger interaction or training signals:
ALBEF~\citep{li2021align} introduces cross-attention, BLIP~\citep{li2022blip}
uses bootstrapped captioning and filtering, BLIP-2~\citep{li2023blip}
connects frozen image encoders and language models with a lightweight querying
transformer, and FILIP~\citep{yao2022filip} strengthens fine-grained alignment
with token-level late interaction. More recent studies, including
UniIR~\citep{wei2024uniir} and UniECS~\citep{liang2025uniecs}, further explore
unified embedding spaces for heterogeneous retrieval.

These methods provide strong general-purpose retrieval foundations, but they do
not directly address the asymmetric matching problem in e-commerce IMMR. In this
setting, and in related multimodal item-retrieval studies~\citep{zhan2021product1m,
dong2022m5product,yu2022commercemm,chen2023unified,jin2023eclip}, a cropped
visual query must be matched to the corresponding item region inside a full
image--text candidate. This requires item-level discrimination and fine-grained
visual grounding beyond coarse image--text relevance~\citep{chen2023rethinking}.

\subsection{Region-aware visual representation learning}

Another line of work studies local and object-level visual representations.
Self-supervised models in the DINO family~\citep{zhang2022dino, oquab2023dinov2,
darcet2023vision, simeoni2025dinov3} often produce spatially meaningful patch
features, making them useful priors for region-aware representation learning
without dense supervision. Recent CLIP-based variants, such as
DeCLIP~\citep{wang2025declip} and SmartCLIP~\citep{xie2025smartclip}, further
improve local semantic consistency and region-level alignment in
vision--language encoders. In parallel, e-commerce pretraining and retrieval
benchmarks emphasize instance-level and fine-grained item
discrimination~\citep{zhan2021product1m,dong2022m5product,jin2023eclip}.

These works are closely related to fine-grained representation learning, but
most of them are developed for generic vision--language settings or for
symmetric item matching. Our e-commerce setting has a different retrieval
asymmetry: the query is already object-centric, whereas the candidate image may
contain multiple objects. We therefore use DINO-style object-centric features as
a frozen teacher for spatial-relational distillation, together with structured
item text as task-specific guidance for implicit region
selection~\citep{yu2022commercemm,chen2023unified}.

\subsection{Grounding-based retrieval pipelines}

Industrial e-commerce retrieval often handles the granularity disparity through
an explicit detection or grounding stage. Grounded vision--language models such
as MDETR~\citep{kamath2021mdetr} and GLIP~\citep{li2022glip} learn
phrase-region or text-region alignment from detection-style supervision.
Open-vocabulary detectors and grounding models, including
YOLO-World~\citep{cheng2024yolo} and Grounding DINO~\citep{liu2024grounding},
can localize candidate objects from textual prompts. UniDGF~\citep{nan2025unidgf}
further combines detection with fine-grained recognition, but still relies on
explicit object localization.

This pipeline design introduces additional computation and can propagate
localization errors to region filtering and post-hoc matching. It also faces
domain-transfer challenges in e-commerce, where item titles, categories, and
attributes are often structured, attribute-dense, and domain-specific. TIGER-FG
removes this explicit stage by using structured item text as a semantic cue for
soft, patch-level region selection inside the retrieval encoder.

\subsection{MLLM-based multimodal embeddings}

Recent MLLM-based embedding models extend retrieval to more flexible multimodal
inputs, including images, text, and their combinations. Models such as
GME~\citep{zhang2024gme}, MM-Embed~\citep{lin2024mm},
Ops-MM-Embedding~\citep{ops_mm_embedding}, jina-embeddings~\citep{gunther2025jina},
and Qwen3-VL-Embedding~\citep{qwen3vlembedding} show strong retrieval ability
under heterogeneous multimodal inputs. They typically build on large multimodal
backbones and are trained with broad mixtures of single-modal, cross-modal, and
fused-modal data, which improves generalization across retrieval tasks.

Compared with earlier dual-encoder retrieval models, MLLM-based embedders offer
stronger generality and support more flexible input formats. Their large
backbones and broad objectives make them closer to general-purpose multimodal
retrieval systems than lightweight domain-specific item encoders. We include
them as representative large-scale multimodal embedding baselines, while keeping
TIGER-FG compatible with offline item encoding and standard vector retrieval
systems~\citep{douze2024faiss}.

%% file: tex_files/B.data_construction.tex
\makeatletter
\newenvironment{breakablealgorithm}
  {\begin{center}
     \refstepcounter{algorithm}
     \hrule height.8pt depth0pt \kern2pt
     \renewcommand{\caption}[2][\relax]{
       {\raggedright\textbf{\ALG@name~\thealgorithm} ##2\par}
       \ifx\relax##1\relax
         \addcontentsline{loa}{algorithm}{\protect\numberline{\thealgorithm}##2}
       \else
         \addcontentsline{loa}{algorithm}{\protect\numberline{\thealgorithm}##1}
       \fi
       \kern2pt\hrule\kern2pt
     }
  }
  {\kern2pt\hrule\relax
   \end{center}}
\makeatother
 
\section{Dataset Construction Details}
\label{app:dataset}

\begin{figure}[htbp]
    \centering
    \includegraphics[width=1\linewidth]{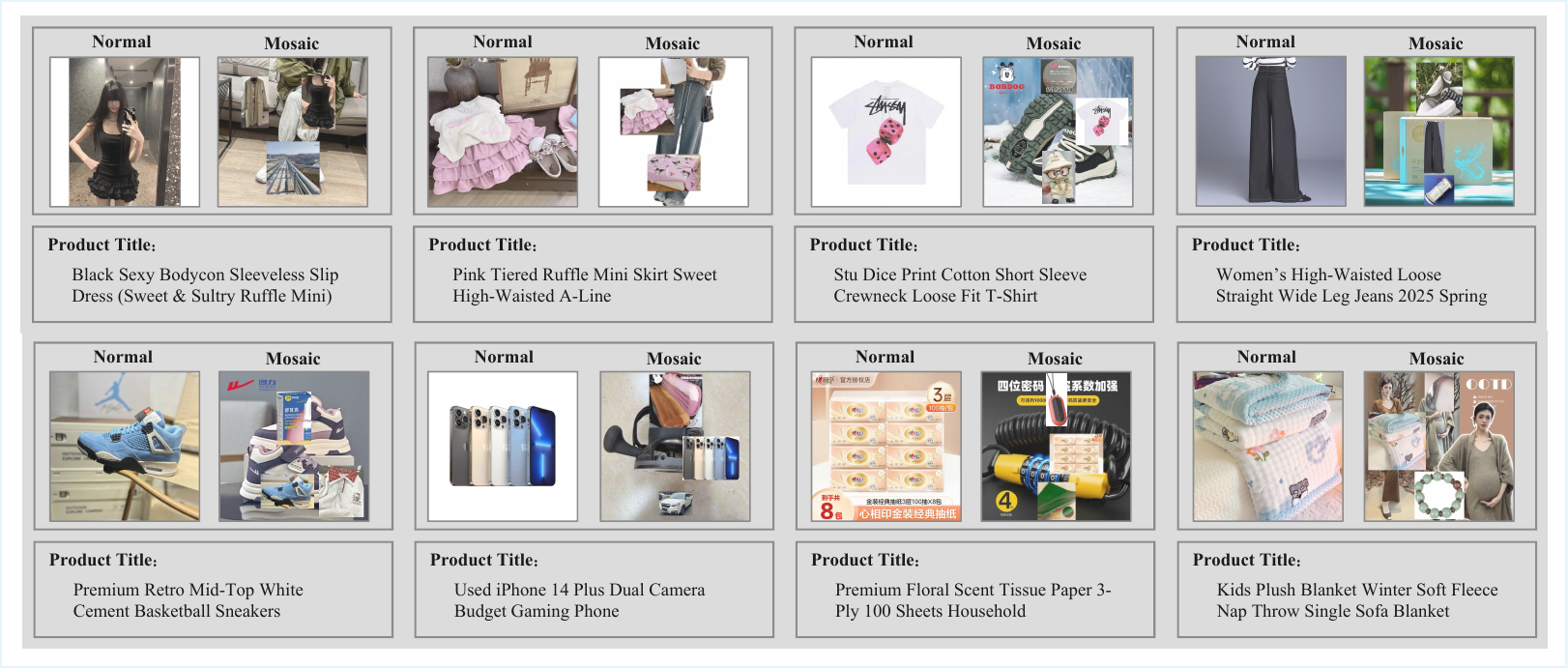}
    \caption{\textbf{Paired examples from ECom-RF-IMMR-Normal and ECom-RF-IMMR-Mosaic.} Each cell shows, for the same item text, the original Normal item image (left) and its Mosaic re-synthesis (right). The Mosaic image keeps the Normal target crop pixel-verbatim and pastes it, together with cross-category distractors, onto a random background at random scale and location.}
    \label{fig:mosaic_showcase}
\end{figure}

We construct the ECom-RF-IMMR suite---the training set \textbf{ECom-RF-IMMR-10M} together with the evaluation benchmarks \textbf{ECom-RF-IMMR-Normal} and \textbf{ECom-RF-IMMR-Mosaic}---under a single schema $(\mathbf{I}^{\mathrm{q}},\mathbf{b}^{\mathrm{q}},\mathbf{I}^{\mathrm{p}},\mathbf{b}^{\mathrm{p}}, \mathbf{T}^{\mathrm{p}},\mathbf{c}^{\mathrm{p}})$,with two shared subroutines doing most of the work: \textsc{CandidateMining} extracts the query-side region, and \textsc{ItemBoxPairing} localizes its counterpart on the item side. Algorithm~\ref{alg:ecom_rf_immr} gives the full pipeline; the rest of this section walks through the parts that need elaboration.
 
\paragraph{Query-side sources.} Query tuples come from (i) product catalogs, where each product ships a main view $\mathbf{I}^{\mathrm{p}}_{\text{main}}$ and an auxiliary view $\mathbf{I}^{\mathrm{p}}_{\text{sub}}$, and (ii) user image-search click logs, which natively contain user-drawn query boxes $\mathbf{b}^{\mathrm{q}}$. For the product source we first drop perceptually identical main/sub pairs via a perceptual hash, then treat the auxiliary view as the query and use GroundingDINO conditioned on the title $\mathbf{T}^{\mathrm{p}}$ to propose a region, keeping only those whose CLIP text--image similarity exceeds a category-aware threshold $\tau_{\text{clip}}$ (typically in the range of 0.25--0.35 in our e-commerce setting). This yields a localized $\mathbf{b}^{\mathrm{q}}$ whose content is consistent with the title.
 
\paragraph{Item-side box assignment.} Item images frequently depict full scenes or multi-object compositions, so we do not assume a trivial one-to-one correspondence. For every item image $\mathbf{I}^{\mathrm{p}}$, we generate open-vocabulary proposals via GroundingDINO, embed both the query crop and each proposal through an online representation model $\mathcal{E}$, and select the proposal $\mathbf{b}^{\mathrm{p\star}}$ whose embedding is closest to that of the query crop. We then filter by this query--item cosine similarity $s$: pairs with $s>\tau_{\text{high}}{=}0.97$ are discarded as the two crops are nearly identical, and pairs with $s<\tau_{\text{low}}{=}0.80$ are discarded as the two crops are unlikely to depict the same product.

\paragraph{Training-set post-processing.} We deduplicate at the SPU level: if multiple pairs share the same item SPU, we keep only one and drop the rest. This ensures distinct pairs in a batch correspond to distinct products, so they can serve as valid in-batch negatives for contrastive learning. After category-balanced sampling, we obtain 10M pairs as \textbf{ECom-RF-IMMR-10M}.
 
\paragraph{Normal evaluation set.} \textbf{ECom-RF-IMMR-Normal} is built mainly from held-out user click logs, where the query is a user-uploaded photo and the item is the product the user clicked on. After the same mining pipeline, every candidate is passed through a verifier $\mathcal{V}$ (VLM pre-filtering followed by human auditing) to remove residual noise from automated mining. We then apply category-balanced sampling to obtain 100K final pairs as \textbf{ECom-RF-IMMR-Normal}.
 
\begin{figure}[htbp]
    \centering
    \includegraphics[width=1\linewidth]{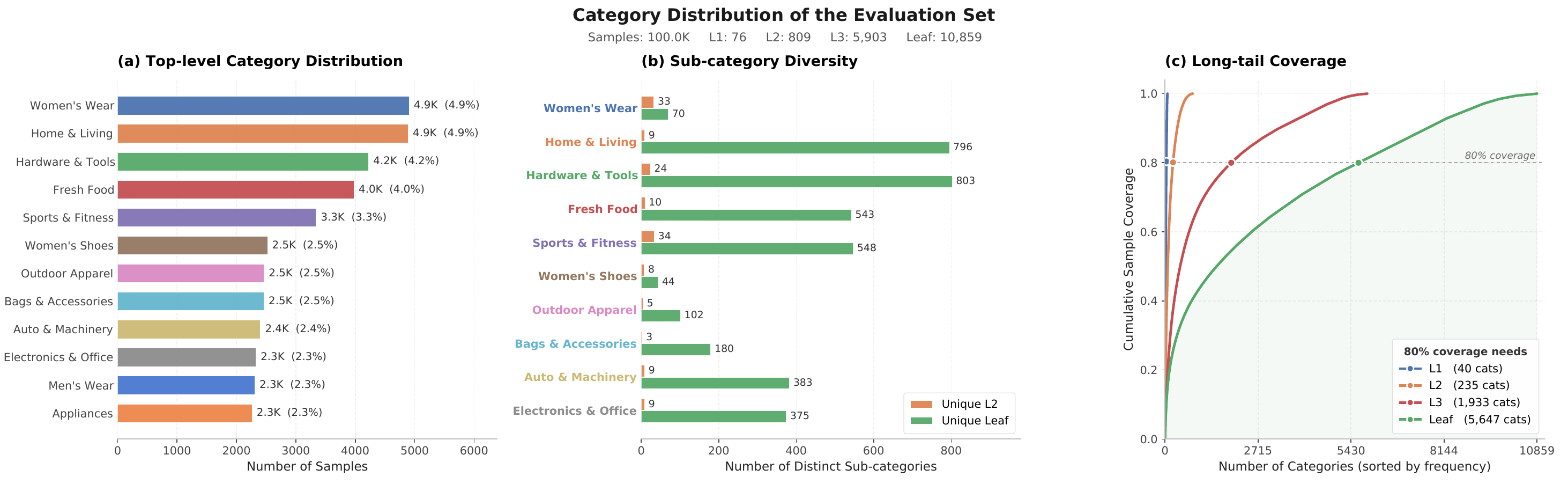}
    \caption{\textbf{Category distribution of ECom-RF-IMMR-Normal.}
    (a) Top-level (L1) distribution showing a balanced mix of verticals, with no single L1 exceeding $\sim$5\% of samples.
    (b) Within-L1 diversity, measured by the number of distinct L2 and leaf categories under each of the top L1 verticals.
    (c) Long-tail cumulative coverage: reaching 80\% of samples requires 40 L1, 235 L2, 1{,}933 L3, and 5{,}647 leaf categories, highlighting the benchmark's long-tail nature.
    Overall statistics: 100K samples over 76 L1 / 809 L2 / 5{,}903 L3 / 10{,}859 leaf categories.}
    \label{fig:dataset_category}
\end{figure}

\paragraph{Category coverage and diversity.} A natural concern is whether category-balanced sampling produces a benchmark that is genuinely diverse, or one that simply over-represents a few large verticals. To answer this, we analyze the taxonomy of \textbf{ECom-RF-IMMR-Normal} along three axes (Figure~\ref{fig:dataset_category}). Each sample carries a four-level category path $\mathbf{c}^{\mathrm{p}}{=}(c_{1},c_{2},c_{3},c_{\text{leaf}})$, with 100K samples spanning 76 L1, 809 L2, 5{,}903 L3, and 10{,}859 leaf categories in total. Panel (a) shows the L1 distribution: no single L1 exceeds 5\%, and the top verticals span apparel, home, hardware, fresh food, electronics, and appliances---going well beyond the fashion- or general-object-centric setups of most existing benchmarks. Panel (b) shows within-L1 diversity: L2 counts stay in the single or low double digits across most verticals while leaf counts often run into the hundreds (e.g., \emph{Home \& Living} has 9 L2 and 796 leaves; \emph{Hardware \& Tools} has 24 L2 and 803 leaves), indicating that the bulk of the fine-grained variation lives at the leaf level. Panel (c) makes the long tail explicit: covering 80\% of samples requires only 40 of 76 L1 but 5{,}647 of 10{,}859 leaves---roughly half of L1 versus over half of leaves, showing that L1 is close to uniform while the leaf distribution is heavily long-tailed, so the benchmark genuinely stresses long-tail retrieval.

\paragraph{Mosaic evaluation set.} \textbf{ECom-RF-IMMR-Mosaic} is derived entirely from \textbf{ECom-RF-IMMR-Normal}: the query image $\mathbf{I}^{\mathrm{q}}$, the query box $\mathbf{b}^{\mathrm{q}}$, and the item title/category $(\mathbf{T}^{\mathrm{p}},\mathbf{c}^{\mathrm{p}})$ are all kept verbatim, and only the item image and its box are re-synthesized. We build two pools directly from Normal: a \emph{background pool} $\mathcal{I}_{\text{bg}}$, consisting of the full item images, and a \emph{distractor pool} $\mathcal{I}_{\text{dist}}$, consisting of the item target crops together with their categories. For each Normal tuple with target crop $\mathbf{I}_{\text{tgt}}{=}\operatorname{crop}(\mathbf{I}^{\mathrm{p}},\mathbf{b}^{\mathrm{p}})$, we sample (i) a background $\mathbf{I}_{\text{bg}}\sim\mathcal{I}_{\text{bg}}$ drawn from a different Normal sample whose SPU and category both differ from those of the current tuple, to prevent leakage from the background itself, and (ii) up to $k{=}4$ distractor crops from $\mathcal{I}_{\text{dist}}$ whose categories differ from $\mathbf{c}^{\mathrm{p}}$. The target and distractors are then pasted onto the background at random scales and positions, with placements rejected and resampled to keep overlaps between objects to a minimum. Each Normal tuple yields exactly one Mosaic tuple, giving a final set of 100K pairs. Because the target crop is copied verbatim, pixel-level ground truth is preserved; randomizing the surrounding context with cross-category distractors introduces controlled multi-object interference and removes the center bias of e-commerce imagery, forcing models to ground the query and title to the correct object rather than to positional or saliency priors.
 
\paragraph{Qualitative examples of Normal vs.\ Mosaic.} Figure~\ref{fig:mosaic_showcase} juxtaposes item images of eight representative samples from \textbf{ECom-RF-IMMR-Normal} and their re-synthesized counterparts in \textbf{ECom-RF-IMMR-Mosaic}, while the query image, query box, and item title are held fixed. In Normal, the target product typically dominates the frame and is roughly centered, so a naïve global-matching model can often succeed by relying on the dominant object or positional priors. In contrast, Mosaic embeds the same target crop---unaltered in pixels---into a heterogeneous scene that also contains one or more cross-category distractors (e.g., a handbag next to a mini skirt, a sneaker next to a T-shirt, or a hairdryer next to a tissue-paper pack), at varying scales and spatial locations. This explicitly breaks the center bias and forces retrieval to rely on the item title $\mathbf{T}^{\mathrm{p}}$ to disambiguate the target.

\vspace{2mm}  
\begin{breakablealgorithm}
\caption{Unified Data Construction Pipeline for the ECom-RF-IMMR Benchmarks}
\label{alg:ecom_rf_immr}
\small
\begin{algorithmic}[1]
\Require Product catalog $\mathcal{P}$, where each $p\in\mathcal{P}$ provides
$(\mathbf{I}^{\mathrm{p}}_{\text{main}},\,\mathbf{I}^{\mathrm{p}}_{\text{sub}},\,\mathbf{T}^{\mathrm{p}},\,\mathbf{c}^{\mathrm{p}})$;
user image-search click log $\mathcal{L}$ with tuples $(\mathbf{I}^{\mathrm{q}},\mathbf{b}^{\mathrm{q}},\mathbf{I}^{\mathrm{p}},\mathbf{T}^{\mathrm{p}},\mathbf{c}^{\mathrm{p}})$;
GroundingDINO detector $\mathcal{G}$; CLIP model $\mathcal{C}=(\mathcal{C}_v,\mathcal{C}_t)$;
online representation model $\mathcal{E}$; SPU taxonomy $\mathcal{S}$;
VLM/human verifier $\mathcal{V}$;
thresholds $\tau_{\text{clip}},\,\tau_{\text{low}}{=}0.80,\,\tau_{\text{high}}{=}0.97$; number of distractors $k$
\Ensure Training set $\mathcal{D}_{\text{train}}$ and evaluation sets $\mathcal{D}_{\text{eval}}^{\text{N}},\,\mathcal{D}_{\text{eval}}^{\text{M}}$; all tuples share the schema $(\mathbf{I}^{\mathrm{q}},\mathbf{b}^{\mathrm{q}},\mathbf{I}^{\mathrm{p}},\mathbf{b}^{\mathrm{p}},\mathbf{T}^{\mathrm{p}},\mathbf{c}^{\mathrm{p}})$
\Statex

\Procedure{CandidateMining}{$\mathcal{P},\mathcal{L}$}\Comment{query-side mining from two heterogeneous sources}
    \State $\mathcal{Q}\gets\emptyset$
    \Statex \hspace{1.2em}\textit{\# Source 1: product main / sub image pairs ($\mathbf{b}^{\mathrm{q}}$ must be inferred)}
    \For{each product $\mathrm{p}\in\mathcal{P}$}
        \If{$\textsc{PHash}(\mathbf{I}^{\mathrm{p}}_{\text{main}})=\textsc{PHash}(\mathbf{I}^{\mathrm{p}}_{\text{sub}})$} \textbf{continue}
            \Comment{drop visually identical main/sub}
        \EndIf
        \State $\mathbf{I}^{\mathrm{q}}\gets\mathbf{I}^{\mathrm{p}}_{\text{sub}},\ \ \mathbf{I}^{\mathrm{p}}\gets\mathbf{I}^{\mathrm{p}}_{\text{main}}$
            \Comment{treat sub-image as query, main image as item}
        \State $\mathbf{b}^{\mathrm{q}}\gets\mathcal{G}(\mathbf{I}^{\mathrm{q}};\,\mathbf{T}^{\mathrm{p}})$ \Comment{title-grounded box proposal}
        \If{$\cos\!\big(\mathcal{C}_v(\operatorname{crop}(\mathbf{I}^{\mathrm{q}},\mathbf{b}^{\mathrm{q}})),\,\mathcal{C}_t(\mathbf{T}^{\mathrm{p}})\big)<\tau_{\text{clip}}$}
            \State \textbf{continue} \Comment{CLIP text--image alignment filter}
        \EndIf
        \State $\mathcal{Q}\gets\mathcal{Q}\cup\{(\mathbf{I}^{\mathrm{q}},\mathbf{b}^{\mathrm{q}},\mathbf{I}^{\mathrm{p}},\mathbf{T}^{\mathrm{p}},\mathbf{c}^{\mathrm{p}})\}$
    \EndFor
    \Statex \hspace{1.2em}\textit{\# Source 2: click log ($\mathbf{b}^{\mathrm{q}}$ is the user-drawn crop from production traffic)}
    \For{each $(\mathbf{I}^{\mathrm{q}},\mathbf{b}^{\mathrm{q}},\mathbf{I}^{\mathrm{p}},\mathbf{T}^{\mathrm{p}},\mathbf{c}^{\mathrm{p}})\in\mathcal{L}$}
        \State $\mathcal{Q}\gets\mathcal{Q}\cup\{(\mathbf{I}^{\mathrm{q}},\mathbf{b}^{\mathrm{q}},\mathbf{I}^{\mathrm{p}},\mathbf{T}^{\mathrm{p}},\mathbf{c}^{\mathrm{p}})\}$
    \EndFor
    \State \Return $\mathcal{Q}$
\EndProcedure
\Statex

\Procedure{ItemBoxPairing}{$\mathcal{Q}$}\Comment{lift item images to item crops via embedding similarity}
    \State $\mathcal{D}\gets\emptyset$
    \For{each $(\mathbf{I}^{\mathrm{q}},\mathbf{b}^{\mathrm{q}},\mathbf{I}^{\mathrm{p}},\mathbf{T}^{\mathrm{p}},\mathbf{c}^{\mathrm{p}})\in\mathcal{Q}$}
        \State $\mathcal{B}^{\mathrm{p}}\gets\mathcal{G}(\mathbf{I}^{\mathrm{p}})$ \Comment{open-vocabulary proposals on the item image}
        \State $\mathbf{e}^{\mathrm{q}}\gets\mathcal{E}(\operatorname{crop}(\mathbf{I}^{\mathrm{q}},\mathbf{b}^{\mathrm{q}}))$
        \State $\mathbf{b}^{\mathrm{p\star}}\gets\displaystyle\arg\max_{\mathbf{b}\in\mathcal{B}^{\mathrm{p}}}\,\cos\!\big(\mathbf{e}^{\mathrm{q}},\,\mathcal{E}(\operatorname{crop}(\mathbf{I}^{\mathrm{p}},\mathbf{b}))\big)$
        \State $s\gets\cos\!\big(\mathbf{e}^{\mathrm{q}},\,\mathcal{E}(\operatorname{crop}(\mathbf{I}^{\mathrm{p}},\mathbf{b}^{\mathrm{p\star}}))\big)$
        \If{$s>\tau_{\text{high}}$} \textbf{continue} \Comment{near-duplicate crop: leakage / trivial match}
        \ElsIf{$s<\tau_{\text{low}}$} \textbf{continue} \Comment{likely not the same instance}
        \EndIf
        \State $\mathcal{D}\gets\mathcal{D}\cup\{(\mathbf{I}^{\mathrm{q}},\mathbf{b}^{\mathrm{q}},\mathbf{I}^{\mathrm{p}},\mathbf{b}^{\mathrm{p\star}},\mathbf{T}^{\mathrm{p}},\mathbf{c}^{\mathrm{p}})\}$
    \EndFor
    \State \Return $\mathcal{D}$
\EndProcedure
\Statex

\Statex \textbf{Stage 1 --- Training set {\normalfont (ECom-RF-IMMR-10M)}}
\State $\mathcal{Q}\gets\textsc{CandidateMining}(\mathcal{P},\mathcal{L})$
\State $\mathcal{D}\gets\textsc{ItemBoxPairing}(\mathcal{Q})$
\State $\mathcal{D}_{\text{train}}\gets\textsc{SPUCoarseDedup}(\mathcal{D},\,\mathcal{S})$
\Statex \hspace{1.2em}\Comment{remove pairs sharing the same SPU so that distinct pairs serve as valid in-batch negatives}
\State $\mathcal{D}_{\text{train}}\gets\textsc{CategoryBalancedSampling}(\mathcal{D}_{\text{train}})$
\Statex \hspace{1.2em}\Comment{re-balance category distribution to avoid long-tail dominance}
\Statex

\Statex \textbf{Stage 2 --- ECom-RF-IMMR-Normal evaluation set}
\State $\mathcal{D}_{\text{eval}}^{\text{N}}\gets\emptyset$
\For{each $x=(\mathbf{I}^{\mathrm{q}},\mathbf{b}^{\mathrm{q}},\mathbf{I}^{\mathrm{p}},\mathbf{b}^{\mathrm{p}},\mathbf{T}^{\mathrm{p}},\mathbf{c}^{\mathrm{p}})\in\mathcal{D}$}
    \Comment{$\mathcal{D}$ biased toward click-log pairs, augmented by main/sub pairs}
    \If{$\mathcal{V}(x)=\textsc{valid}$} \Comment{VLM pre-filter followed by human audit}
        \State $\mathcal{D}_{\text{eval}}^{\text{N}}\gets\mathcal{D}_{\text{eval}}^{\text{N}}\cup\{x\}$
    \EndIf
\EndFor
\Statex

\Statex \textbf{Stage 3 --- ECom-RF-IMMR-Mosaic evaluation set {\normalfont (derived from Normal)}}
\State $\mathcal{I}_{\text{bg}}\gets\{\mathbf{I}^{\mathrm{p}}\mid(\cdot,\cdot,\mathbf{I}^{\mathrm{p}},\cdot,\cdot,\cdot)\in\mathcal{D}_{\text{eval}}^{\text{N}}\}$
    \Comment{background pool: full item images from Normal}
\State $\mathcal{I}_{\text{dist}}\gets\{(\operatorname{crop}(\mathbf{I}^{\mathrm{p}},\mathbf{b}^{\mathrm{p}}),\,\mathbf{c}^{\mathrm{p}})\mid(\cdot,\cdot,\mathbf{I}^{\mathrm{p}},\mathbf{b}^{\mathrm{p}},\cdot,\mathbf{c}^{\mathrm{p}})\in\mathcal{D}_{\text{eval}}^{\text{N}}\}$
\Statex \hspace{1.2em}\Comment{distractor pool: target crops with categories (used for cross-category sampling)}
\State $\mathcal{D}_{\text{eval}}^{\text{M}}\gets\emptyset$
\For{each $(\mathbf{I}^{\mathrm{q}},\mathbf{b}^{\mathrm{q}},\mathbf{I}^{\mathrm{p}},\mathbf{b}^{\mathrm{p}},\mathbf{T}^{\mathrm{p}},\mathbf{c}^{\mathrm{p}})\in\mathcal{D}_{\text{eval}}^{\text{N}}$}
    \Statex \hspace{1.2em}\Comment{keep $(\mathbf{I}^{\mathrm{q}},\mathbf{b}^{\mathrm{q}},\mathbf{T}^{\mathrm{p}},\mathbf{c}^{\mathrm{p}})$ unchanged from Normal}
    \State Sample background $\mathbf{I}_{\text{bg}}\sim\mathcal{I}_{\text{bg}}$
    \State Sample $k$ distractors $\{(\mathbf{I}_{d_i},\mathbf{c}_{d_i})\}_{i=1}^{k}\sim\mathcal{I}_{\text{dist}}$ s.t.\ $\mathbf{c}_{d_i}\neq\mathbf{c}^{\mathrm{p}},\ \forall i$
        \Comment{cross-category distractors only}
    \State $\mathbf{I}_{\text{tgt}}\gets\operatorname{crop}(\mathbf{I}^{\mathrm{p}},\mathbf{b}^{\mathrm{p}})$
        \Comment{ground-truth target crop preserved verbatim}
    \State $(\tilde{\mathbf{I}}^{\mathrm{p}},\,\tilde{\mathbf{b}}^{\mathrm{p}})\gets\textsc{Compose}\!\big(\mathbf{I}_{\text{bg}},\,\mathbf{I}_{\text{tgt}},\,\{\mathbf{I}_{d_i}\}_{i=1}^{k}\big)$
    \Statex \hspace{1.2em}\Comment{paste target and distractors on the background at random scale/location; $\tilde{\mathbf{b}}^{\mathrm{p}}$ is the target bbox in $\tilde{\mathbf{I}}^{\mathrm{p}}$}
    \State $\mathcal{D}_{\text{eval}}^{\text{M}}\gets\mathcal{D}_{\text{eval}}^{\text{M}}\cup\{(\mathbf{I}^{\mathrm{q}},\mathbf{b}^{\mathrm{q}},\tilde{\mathbf{I}}^{\mathrm{p}},\tilde{\mathbf{b}}^{\mathrm{p}},\mathbf{T}^{\mathrm{p}},\mathbf{c}^{\mathrm{p}})\}$
\EndFor
\Statex

\State \Return $\mathcal{D}_{\text{train}},\ \mathcal{D}_{\text{eval}}^{\text{N}},\ \mathcal{D}_{\text{eval}}^{\text{M}}$
\end{algorithmic}
\end{breakablealgorithm}

%% file: tex_files/C.visualization_recall.tex
\section{Qualitative Retrieval Comparison}
\label{app:recall_vis}

\begin{figure}[htbp]
    \centering
    \includegraphics[width=1\linewidth]{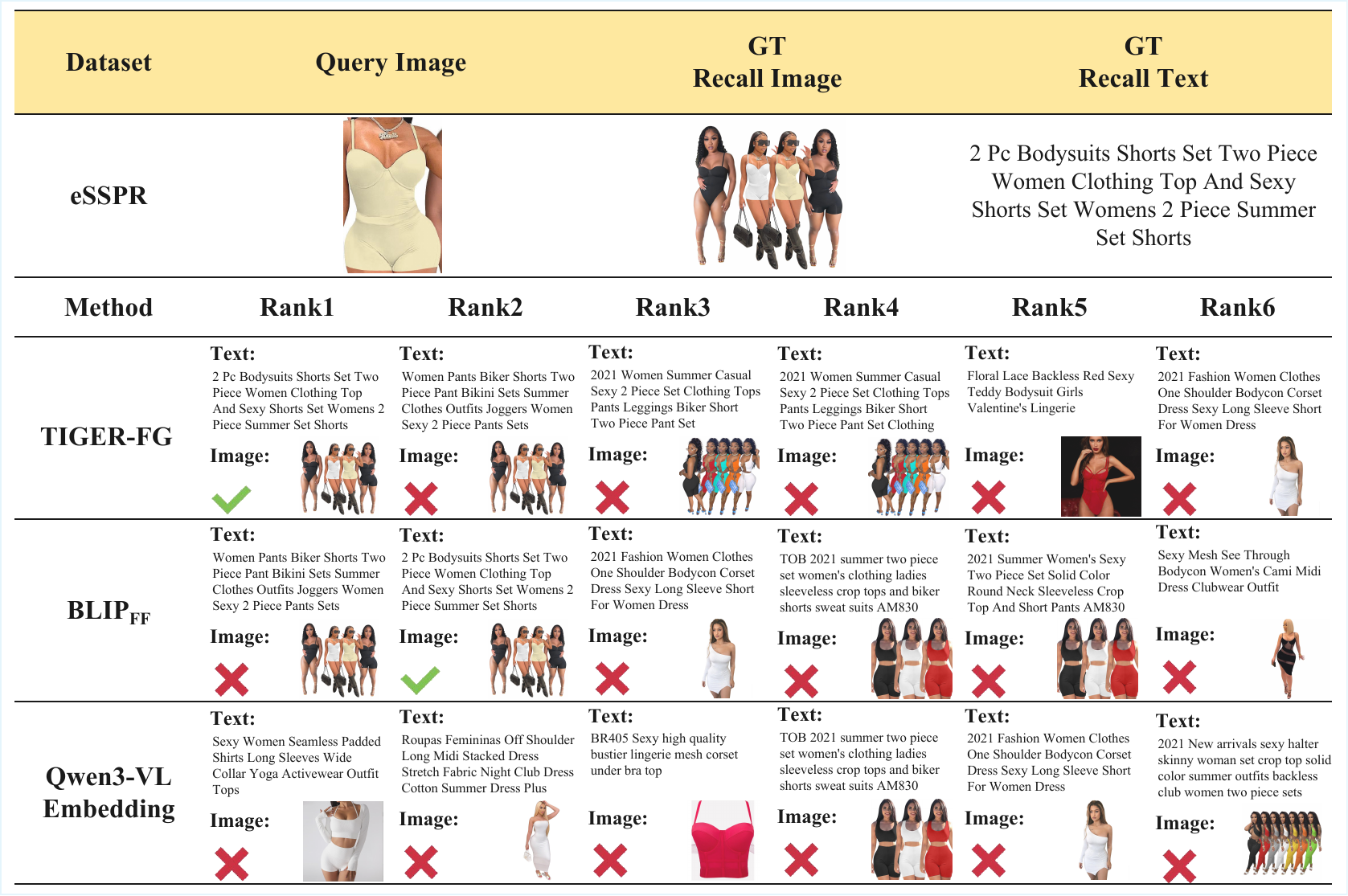}
    \caption{\textbf{Top-6 retrieval results on eSSPR} for the query ``2 Pc Bodysuits Shorts Set \dots''. TIGER-FG hits the ground truth at rank 1; BLIP$_{\text{FF}}$ hits at rank 2, with rank 1 being an image-matching but title-mismatching candidate (biker shorts); Qwen3-VL-Embedding returns semantically related but incorrect items (yoga tops, bodycon dresses, lingerie) throughout the top-6.}
    \label{fig:recall_esspr}
\end{figure}

\begin{figure}[htbp]
    \centering
    \includegraphics[width=1\linewidth]{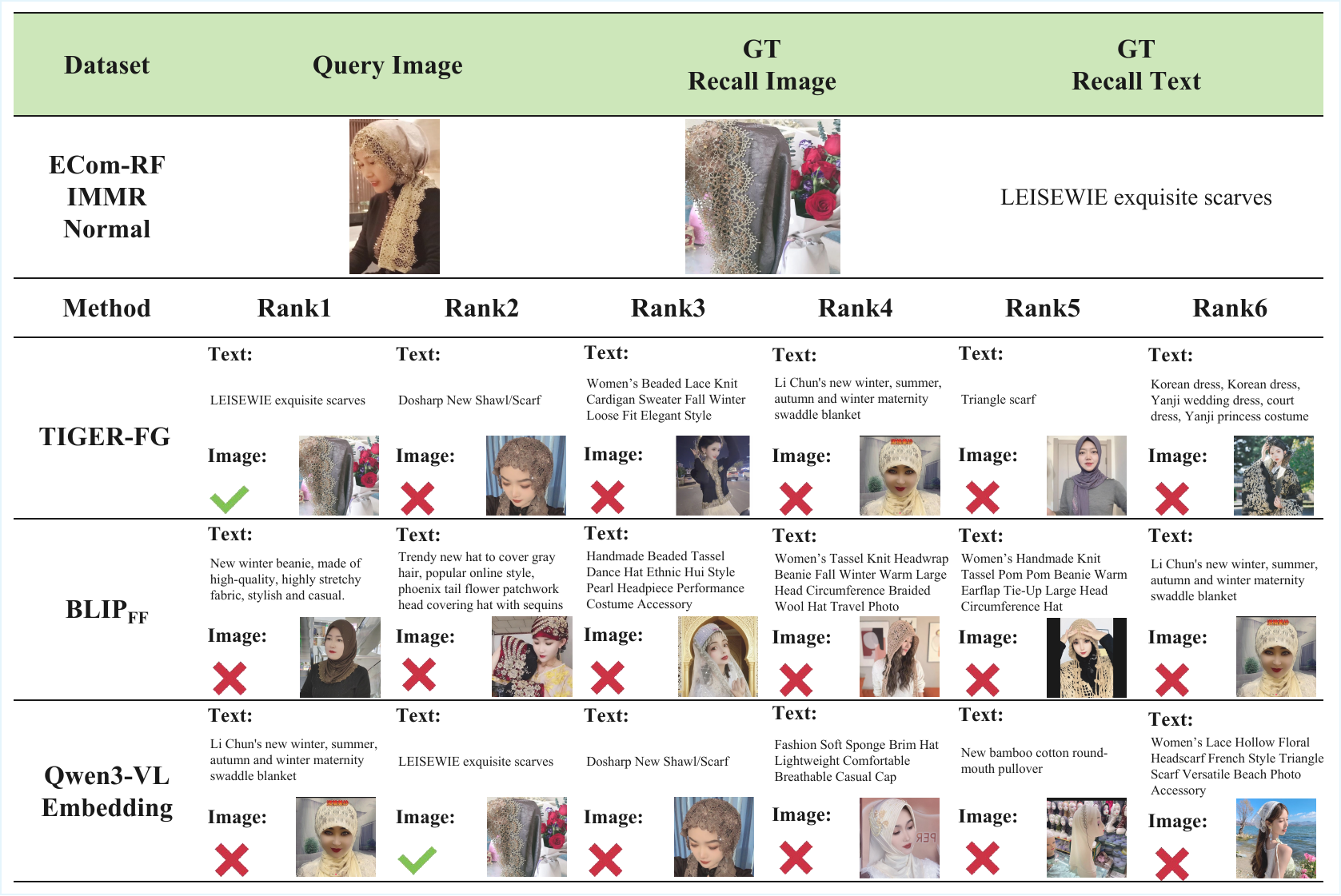}
    \caption{\textbf{Top-6 retrieval results on ECom-RF-IMMR-Normal} for the query ``LEISEWIE exquisite scarves''. TIGER-FG ranks the ground truth first; BLIP$_{\text{FF}}$ reads the worn scarf as headwear and returns hats throughout the top-6; Qwen3-VL-Embedding hits the target only at rank 2, with an unrelated maternity-blanket item at rank 1.}
    \label{fig:recall_ours}
\end{figure}

To complement the quantitative results in Section~\ref{sec:experiments}, we visualize the top-6 retrieved candidates of three representative methods---our \textbf{TIGER-FG}, the strongest VLM-based baseline \textbf{BLIP$_{\text{FF}}$}, and the strongest MLLM-based embedder \textbf{Qwen3-VL-Embedding}---on one query drawn from the public \textbf{eSSPR} benchmark (Figure~\ref{fig:recall_esspr}) and one from our \textbf{ECom-RF-IMMR-Normal} (Figure~\ref{fig:recall_ours}). For each method we show the retrieved item title and image at ranks 1--6, with the ground-truth candidate marked by a green check and incorrect candidates by a red cross.

\paragraph{eSSPR (Figure~\ref{fig:recall_esspr}).} The query is a cropped image of a beige one-piece bodysuit, and the ground truth is a two-piece bodysuit--shorts set whose item image shows four models jointly wearing dresses, bodysuits, and shorts---a cluttered candidate that global matching tends to mishandle. \textbf{TIGER-FG} places the ground truth at rank 1, with several semantically coherent two-piece sets and bodysuits filling out the rest of the top-6. \textbf{BLIP$_{\text{FF}}$} retrieves the ground truth at rank 2; its rank-1 candidate is image-matching but title-mismatching (a biker-shorts set), exactly the failure mode that text-guided grounding is meant to fix. \textbf{Qwen3-VL-Embedding} misses the ground truth in the top-6 entirely, drifting into adjacent categories such as yoga activewear, bodycon dresses, and lingerie---a sign that a global MLLM embedding is insufficient when the query crop shares low-level visual cues with many garment categories.

\paragraph{ECom-RF-IMMR-Normal (Figure~\ref{fig:recall_ours}).} The query shows a woman with a lace fabric draped over her head---without an accompanying title, the user could equally be searching for a scarf or for a piece of clothing/headwear, and there is no way to tell from the image alone. The ground truth is in fact the same scarf product, but its item image shows the scarf laid flat next to a flower vase---a viewpoint that shares almost no low-level cues with the query, so coarse global matching alone has little chance of bridging the two. \textbf{TIGER-FG} still ranks the ground truth first, and its remaining top-6 spans both scarves/shawls and clothing items---a reasonable spread given the inherent ambiguity of the query, and a sign that the model is actually using the title to ground the search rather than collapsing to one visual interpretation. \textbf{BLIP$_{\text{FF}}$} commits to the headwear reading and returns six hat/beanie/headwrap products in a row, never reaching the target. \textbf{Qwen3-VL-Embedding} retrieves the ground truth at rank 2, but its rank-1 result is an unrelated maternity swaddle blanket---the kind of category drift that follows from squeezing a multimodal candidate into a single global embedding.

%% file: tex_files/D.ablation_appendix.tex
\section{Extended Ablation}
\label{app:ablation}
 
To complement the subtractive ablation in the main paper, we conduct an additive study that builds the model progressively from a plain dual-encoder baseline. The study serves two purposes. First, it quantifies the contribution of each component under the order in which it is introduced into TIGER-FG. Second, it examines how progressively adding fine-grained region supervision and Mosaic-augmented training data changes both the quantitative retrieval results and the qualitative localization patterns. All experiments are conducted on \textbf{ECom-RF-IMMR-Mosaic} under the same training setup as in Section~\ref{sec:experiments}.

\begin{table}[ht]
    \centering
    \small
    \caption{\textbf{Additive ablation on ECom-RF-IMMR-Mosaic.} 
    Starting from a plain dual-encoder, we progressively enable components of TIGER-FG. 
    \emph{Data}: ``$1$'' uses raw samples only; ``$1{+}4$'' mixes Mosaic-augmented samples at a $1{:}1$ ratio. 
    \emph{Config} is the cumulative set of components, abbreviated as:
    \textbf{S}=slot- and CLS-guided cross-attention; 
    \textbf{B}=item-side box supervision; 
    \textbf{R}=ROI-Align region alignment; 
    \textbf{H}=mismatched-text hard negatives; 
    \textbf{D}=spatial-relational \& similarity-distribution distillation;
    \textbf{T}=image--text contrastive regularizer.
    All values are in \%.}
    \label{tab:additive_ablation}
    \setlength\tabcolsep{6pt}
    \renewcommand\arraystretch{1.25}
    \begin{tabular}{l c l ccc cc cc}
    \toprule
    \multirow{2}{*}{Recall fusion} & \multirow{2}{*}{Data} & \multirow{2}{*}{Config}
        & \multicolumn{3}{c}{Recall} & \multicolumn{2}{c}{MRR} & \multicolumn{2}{c}{NDCG} \\
    \cmidrule(lr){4-6} \cmidrule(lr){7-8} \cmidrule(lr){9-10}
        &  &  & @1 & @4 & @10 & @4 & @10 & @4 & @10 \\
    \midrule
    \multicolumn{10}{l}{\textit{(a) Backbone only}} \\
    CLIP        & $1$ & --              & 46.80 & 65.24 & 73.35 & 54.40 & 55.64 & 57.15 & 59.92 \\
    DINOv3      & $1$ & --              & 49.74 & 67.83 & 75.42 & 57.21 & 58.38 & 59.91 & 62.50 \\
    DINOv3-Slot & $1$ & S               & 49.94 & 68.01 & 75.77 & 57.39 & 58.58 & 60.08 & 62.74 \\
    \midrule
    \multicolumn{10}{l}{\textit{(b) + region supervision on raw data}} \\
    DINOv3-Slot & $1$ & S+B             & 50.11 & 68.09 & 75.87 & 57.53 & 58.72 & 60.21 & 62.87 \\
    DINOv3-Slot & $1$ & S+B+R           & 46.54 & 64.86 & 73.16 & 54.07 & 55.34 & 56.81 & 59.65 \\
    DINOv3-Slot & $1$ & S+B+R+H         & 46.70 & 64.62 & 72.66 & 54.06 & 55.30 & 56.74 & 59.50 \\
    \midrule
    \multicolumn{10}{l}{\textit{(c) + Mosaic augmentation and alignment objectives}} \\
    DINOv3-Slot & $1{+}4$ & S+B+R+H         & 71.30 & 91.55 & 97.04 & 79.91 & 80.78 & 82.87 & 84.78 \\
    DINOv3-Slot & $1{+}4$ & S+B+R+H+D       & 74.93 & 93.20 & 97.72 & 82.70 & 83.42 & 85.38 & 86.95 \\
    \rowcolor{gray!12}
    DINOv3-Slot & $1{+}4$ & S+B+R+H+D+T     & \textbf{75.22} & \textbf{93.47} & \textbf{97.85} & \textbf{83.02} & \textbf{83.71} & \textbf{85.68} & \textbf{87.21} \\
    \bottomrule
    \end{tabular}
\end{table}

Table~\ref{tab:additive_ablation} is organized into three blocks that follow the additive construction of TIGER-FG. Block \emph{(a)} starts from the plain dual-encoder and updates the item-side encoder by replacing the visual backbone and then introducing the text-guided fusion design shown in Figure~\ref{fig:architecture}, corresponding to \textbf{S}. Block \emph{(b)} keeps the training data fixed to raw samples only and progressively adds the main item-side constraints, including target-anchored fused–region alignment \textbf{B}, spatial-relational distillation \textbf{R}, and mismatched-text hard negatives \textbf{H}. Block \emph{(c)} then switches the training data from $1$ to the raw+Mosaic setting $1{+}4$, and further introduces similarity-distribution distillation \textbf{D} and the image--text contrastive objective \textbf{T}. This organization makes it easier to track how the retrieval metrics evolve as the model, supervision, and training data are introduced step by step.
 
\begin{figure}[htbp]
    \centering
    \begin{minipage}{0.95\linewidth}
    \centering
    \captionsetup[subfigure]{font=small}
    \begin{minipage}[b]{0.48\linewidth}
        \centering
        \begin{subfigure}[b]{\linewidth}
            \centering
            \includegraphics[width=\linewidth]{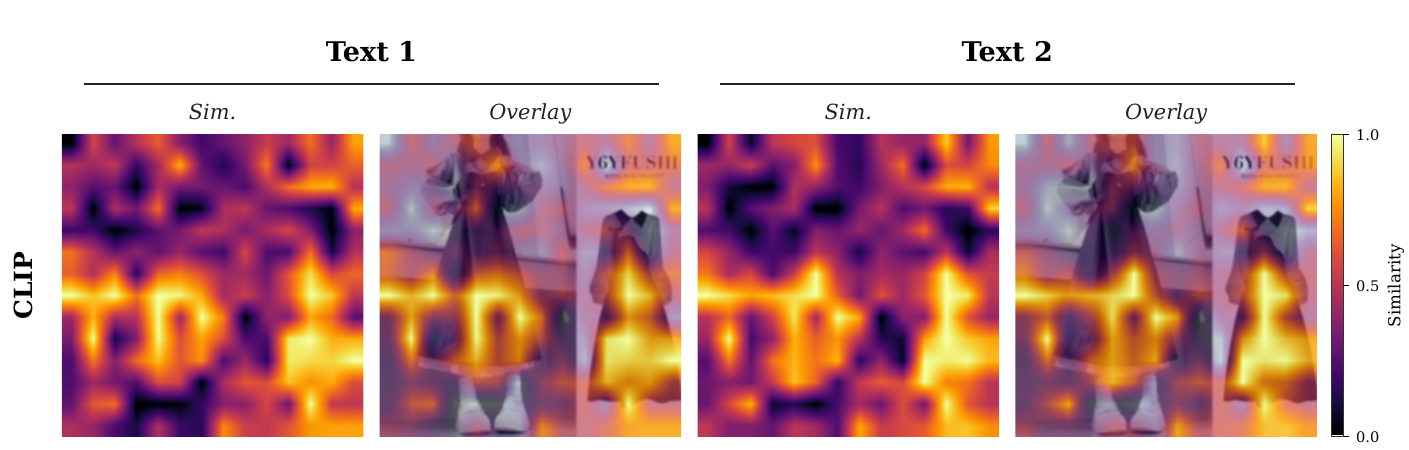}
            \caption{CLIP}
            \label{fig:case1_a}
        \end{subfigure}
 
        \vspace{1mm}
 
        \begin{subfigure}[b]{\linewidth}
            \centering
            \includegraphics[width=\linewidth]{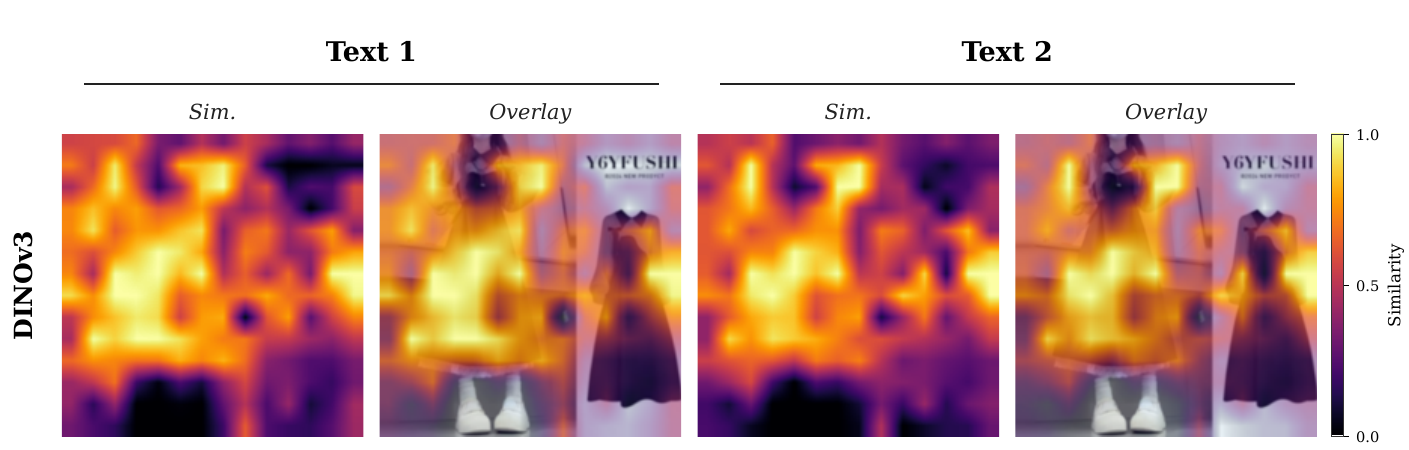}
            \caption{DINOv3}
            \label{fig:case1_b}
        \end{subfigure}
    \end{minipage}
    \hfill
    \begin{subfigure}[b]{0.48\linewidth}
        \centering
        \includegraphics[width=\linewidth]{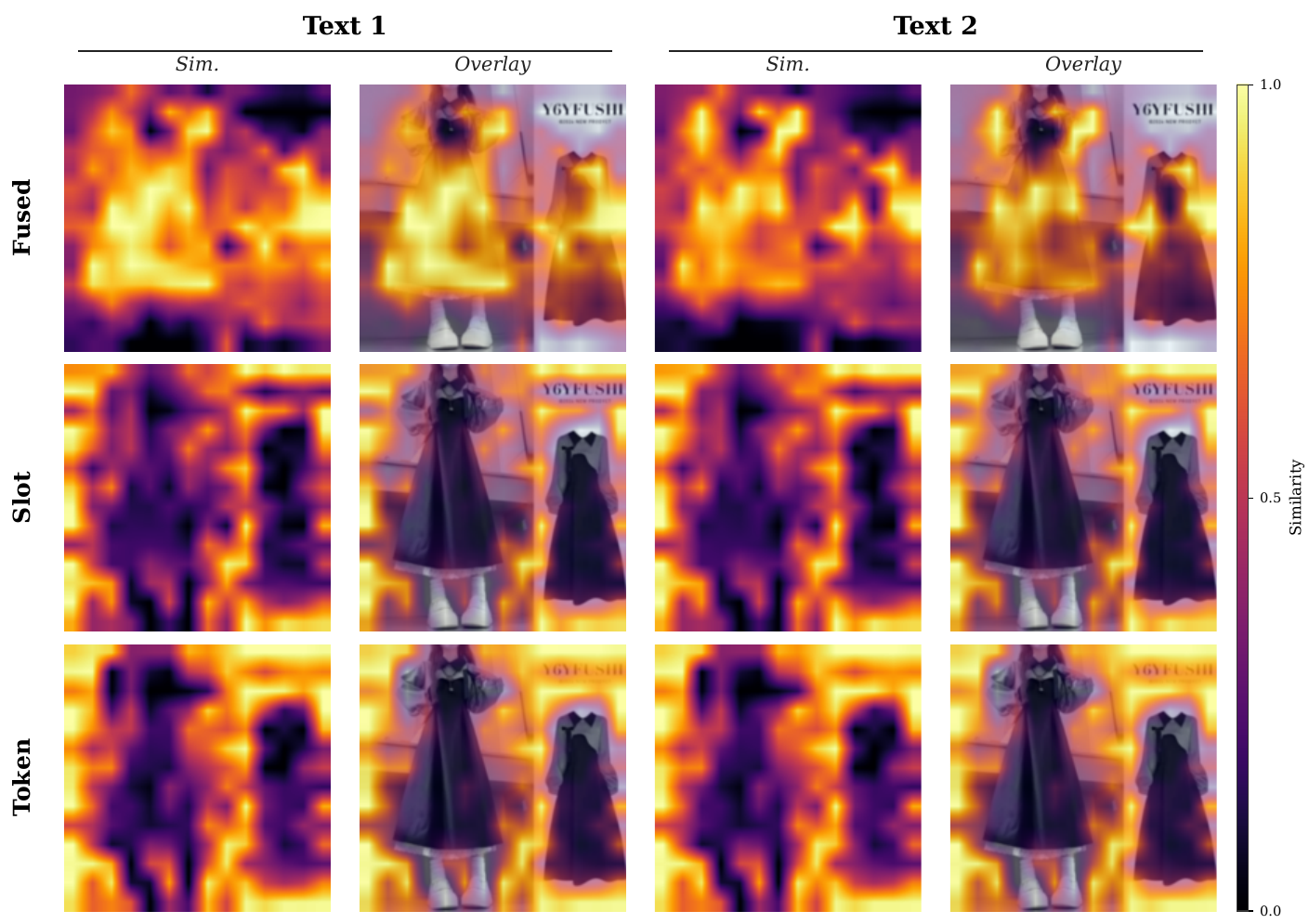}
        \caption{+\,S}
        \label{fig:case1_c}
    \end{subfigure}
 
    \vspace{1mm}
 
    \begin{subfigure}[b]{0.48\linewidth}
        \centering
        \includegraphics[width=\linewidth]{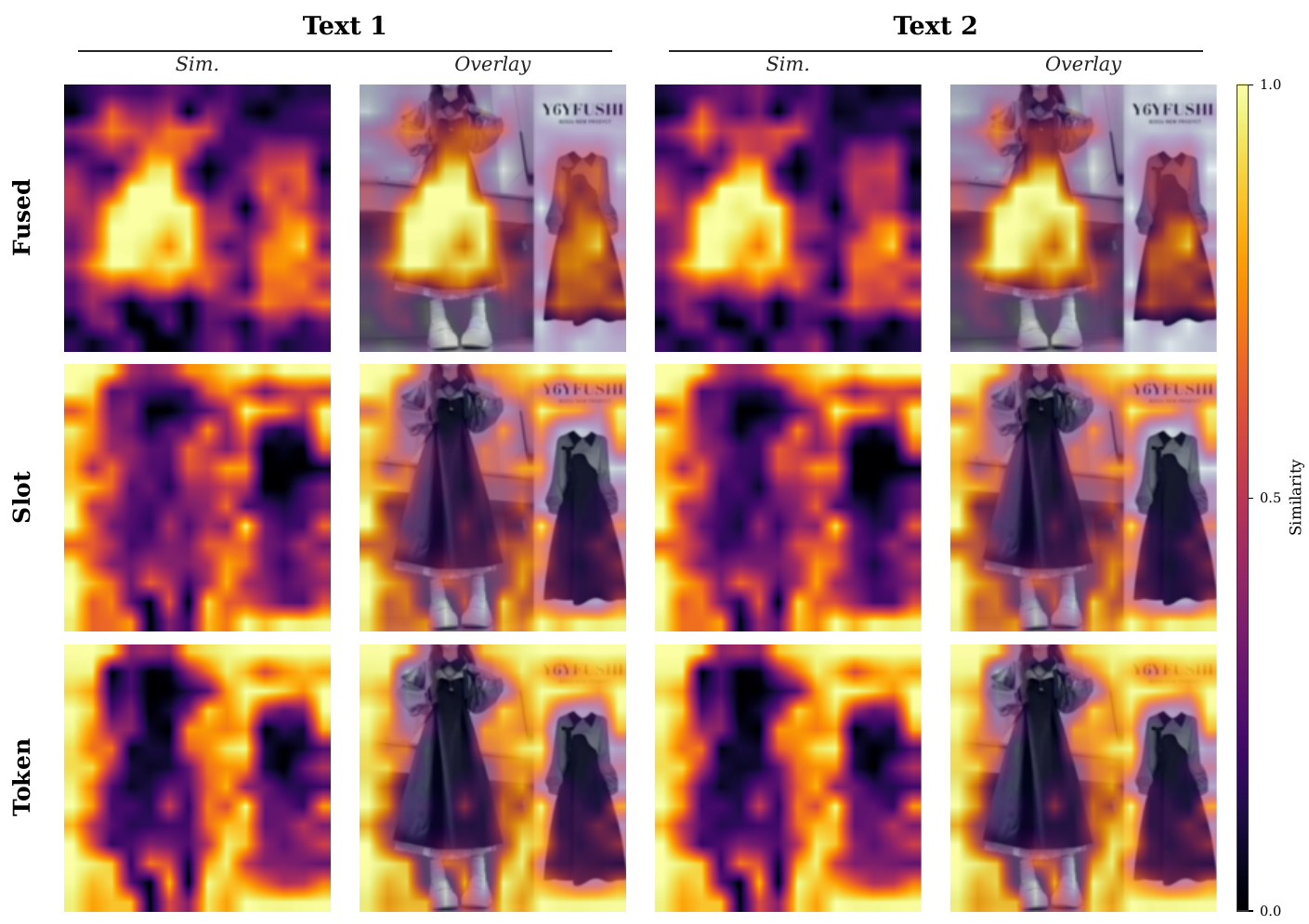}
        \caption{+\,B}
        \label{fig:case1_d}
    \end{subfigure}
    \hfill
    \begin{subfigure}[b]{0.48\linewidth}
        \centering
        \includegraphics[width=\linewidth]{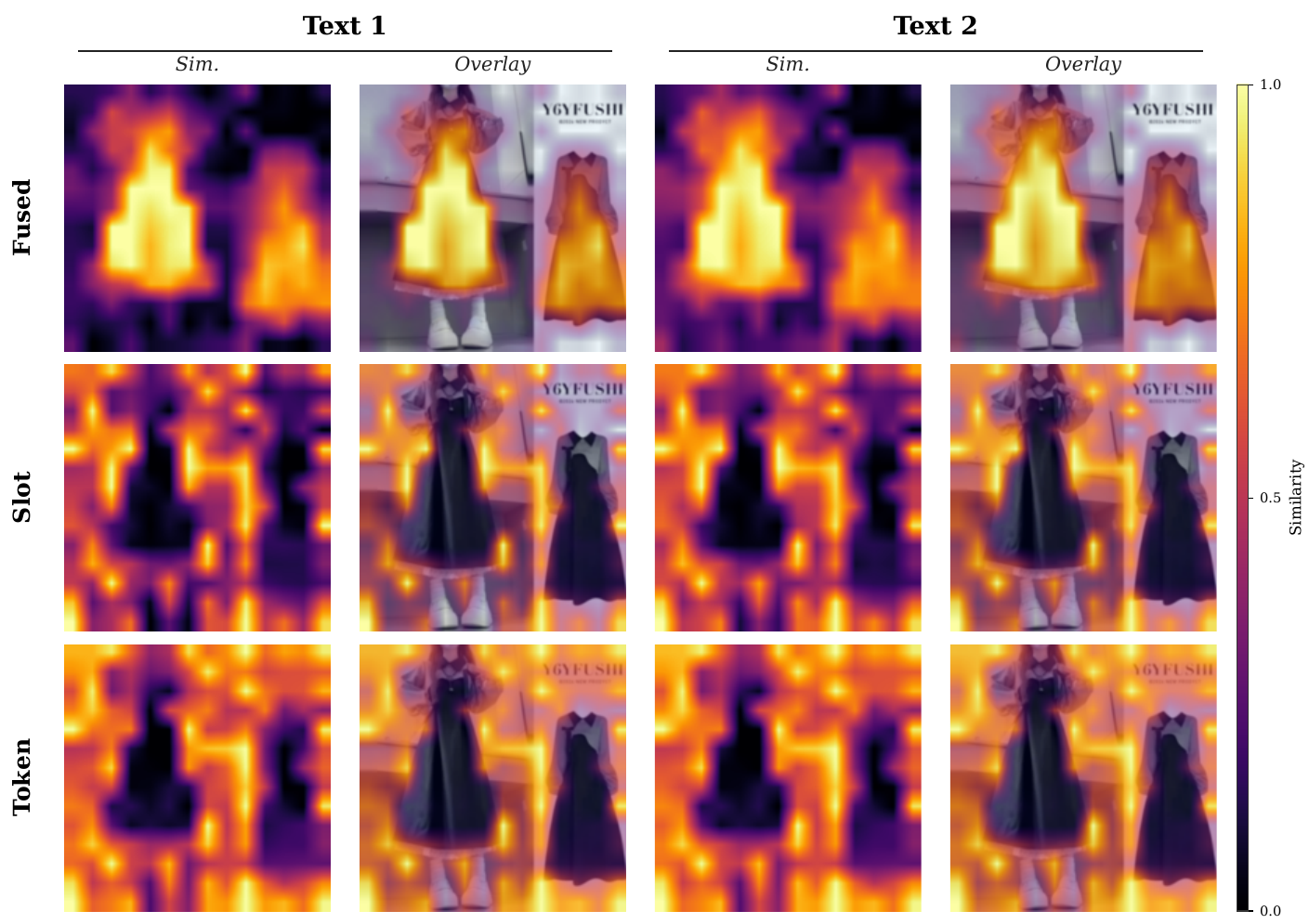}
        \caption{+\,R}
        \label{fig:case1_e}
    \end{subfigure}
 
    \vspace{1mm}
 
    \begin{subfigure}[b]{0.48\linewidth}
        \centering
        \includegraphics[width=\linewidth]{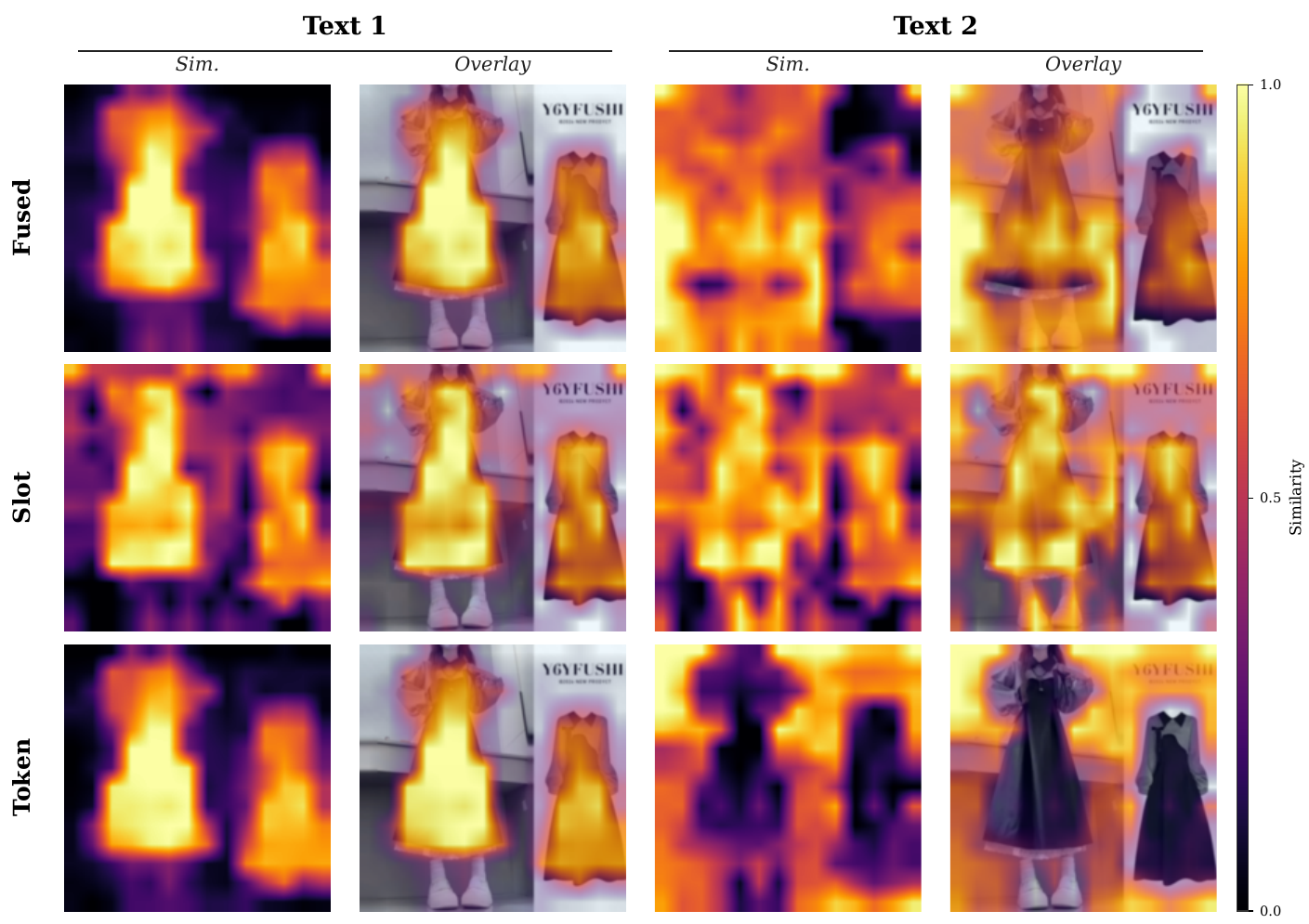}
        \caption{+\,H}
        \label{fig:case1_f}
    \end{subfigure}
    \hfill
    \begin{subfigure}[b]{0.48\linewidth}
        \centering
        \includegraphics[width=\linewidth]{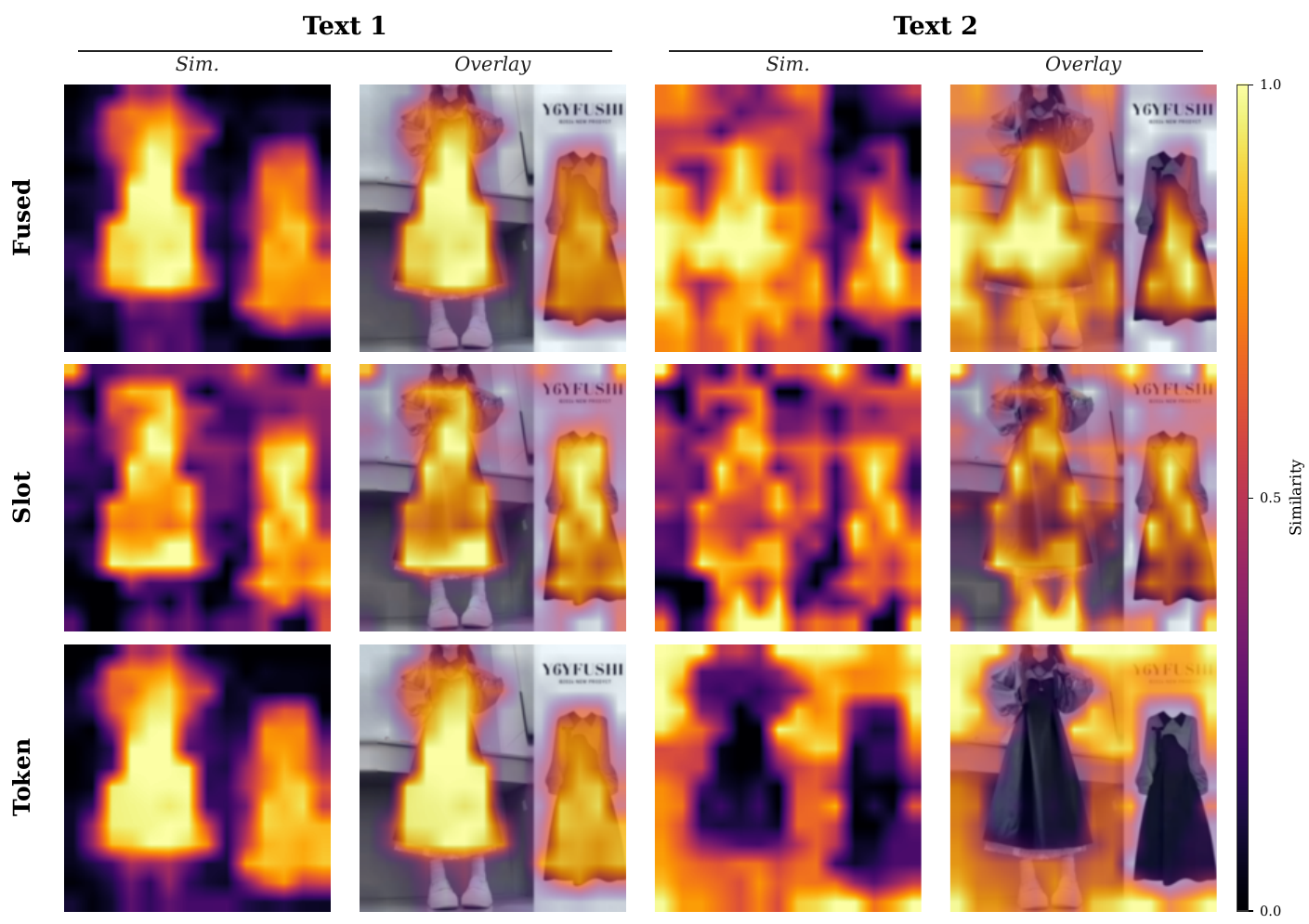}
        \caption{+\,Mosaic}
        \label{fig:case1_g}
    \end{subfigure}
 
    \vspace{1mm}
 
    \begin{subfigure}[b]{0.48\linewidth}
        \centering
        \includegraphics[width=\linewidth]{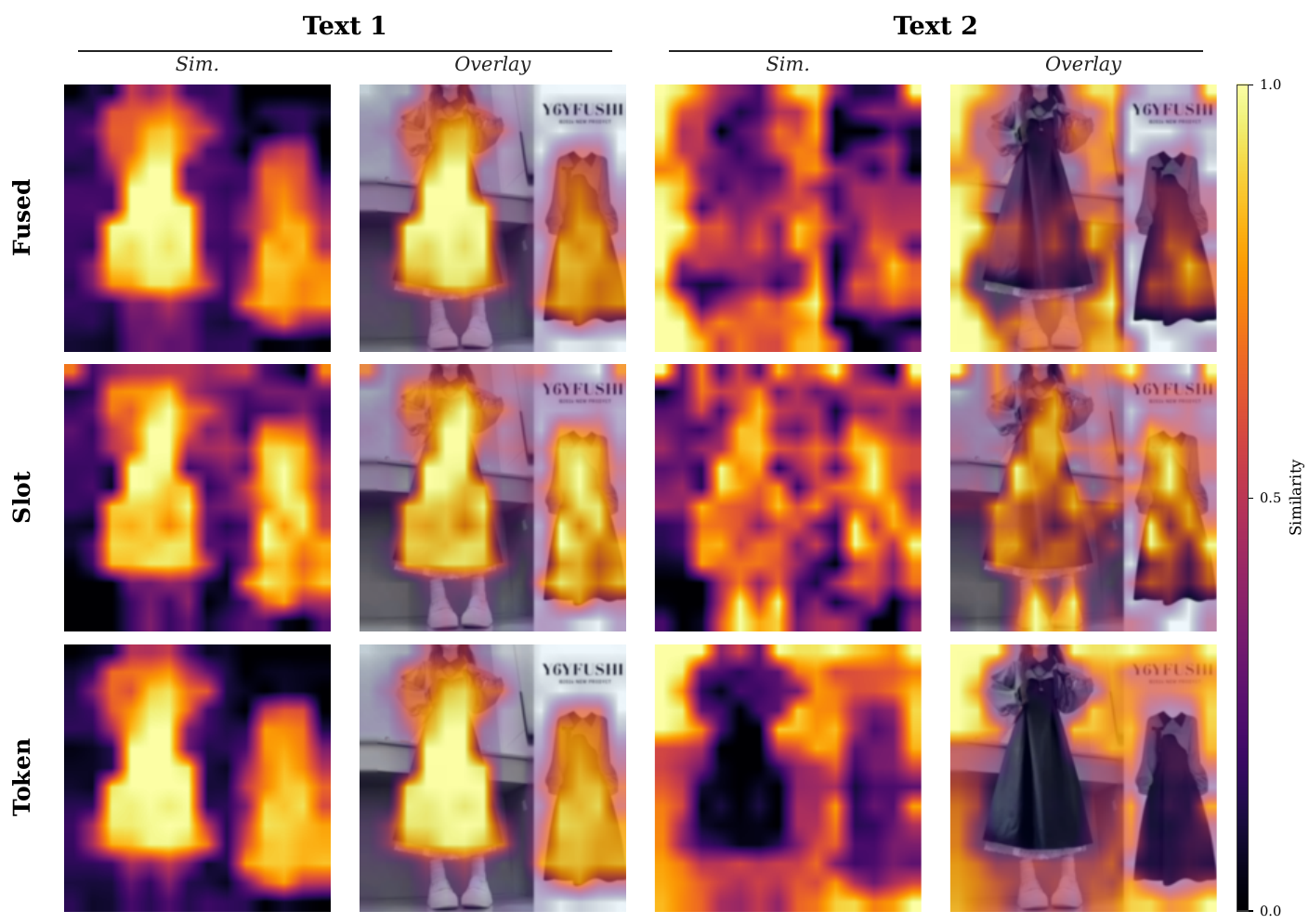}
        \caption{+\,D}
        \label{fig:case1_h}
    \end{subfigure}
    \hfill
    \begin{subfigure}[b]{0.48\linewidth}
        \centering
        \includegraphics[width=\linewidth]{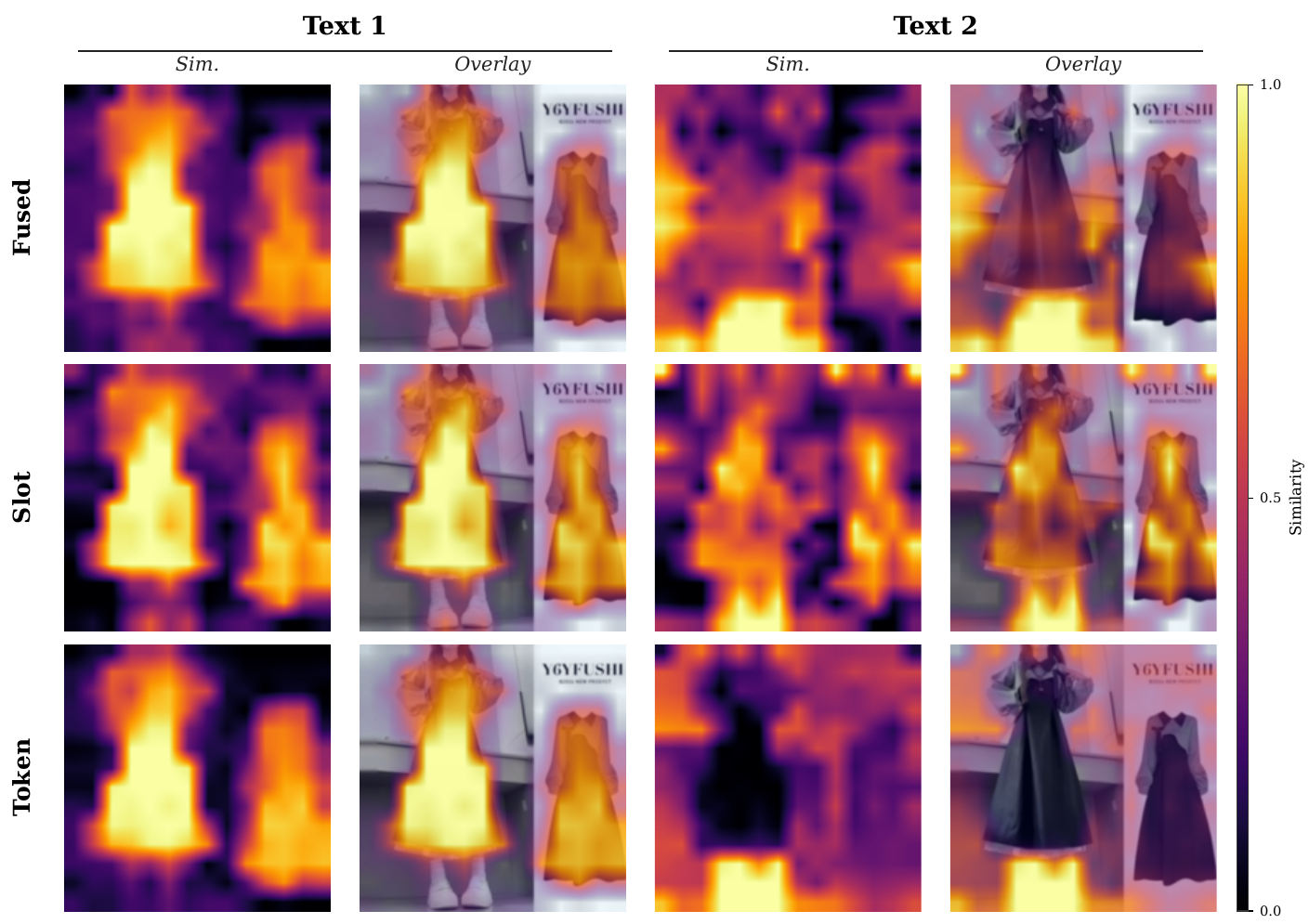}
        \caption{+\,T \textbf{(TIGER-FG)}}
        \label{fig:case1_i}
    \end{subfigure}
    \end{minipage}
 
    \caption{\textbf{Qualitative visualization of the additive ablation (Case 1/3: \emph{dress} and \emph{shoes} queries).} 
    Each panel corresponds one-to-one to a row in Table~\ref{tab:additive_ablation} on the same query--candidate pair; 
    from (c) onward ``$+\,\mathrm{X}$'' denotes adding component \textbf{X} on top of the previous panel. 
    Component abbreviations: \textbf{S}=slot- and CLS-guided cross-attention, 
    \textbf{B}=Target-anchored fused–region alignment, 
    \textbf{R}=Spatial-relational distillation, 
    \textbf{H}=Mismatched-title hard negatives, 
    \textbf{D}=Similarity-distribution distillation, 
    \textbf{T}=image--text contrastive regularizer; 
    ``Mosaic'' denotes switching training data from $1$ to $1{+}4$.}
    \label{fig:additive_ablation_case1}
\end{figure}

\begin{figure}[htbp]
    \centering
    \begin{minipage}{0.95\linewidth}
    \centering
    \captionsetup[subfigure]{font=small}
    \begin{minipage}[b]{0.48\linewidth}
        \centering
        \begin{subfigure}[b]{\linewidth}
            \centering
            \includegraphics[width=\linewidth]{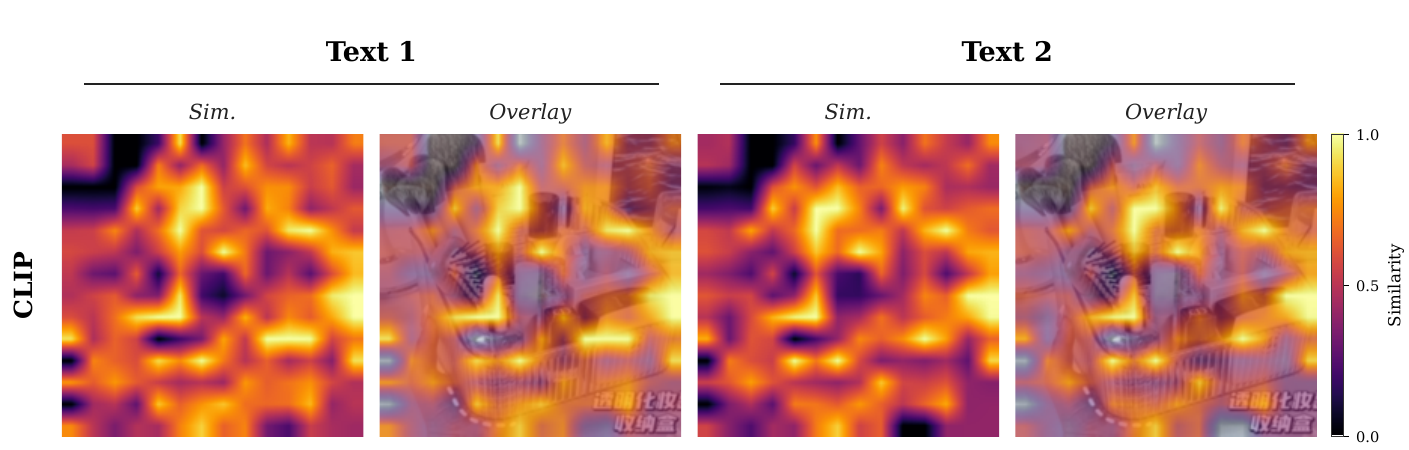}
            \caption{CLIP}
            \label{fig:case2_a}
        \end{subfigure}
 
        \vspace{1mm}
 
        \begin{subfigure}[b]{\linewidth}
            \centering
            \includegraphics[width=\linewidth]{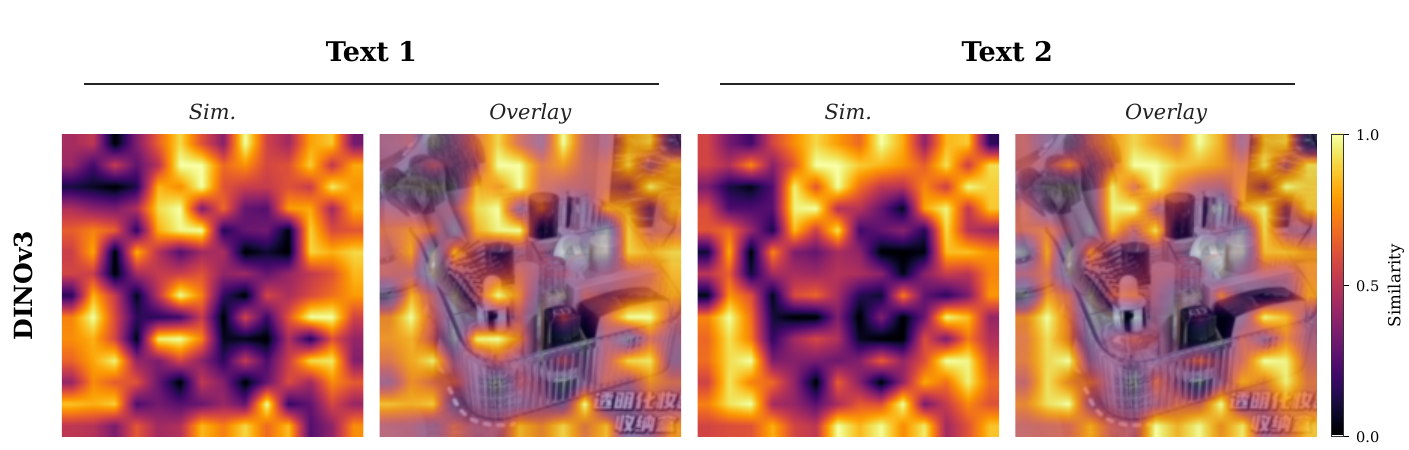}
            \caption{DINOv3}
            \label{fig:case2_b}
        \end{subfigure}
    \end{minipage}
    \hfill
    \begin{subfigure}[b]{0.48\linewidth}
        \centering
        \includegraphics[width=\linewidth]{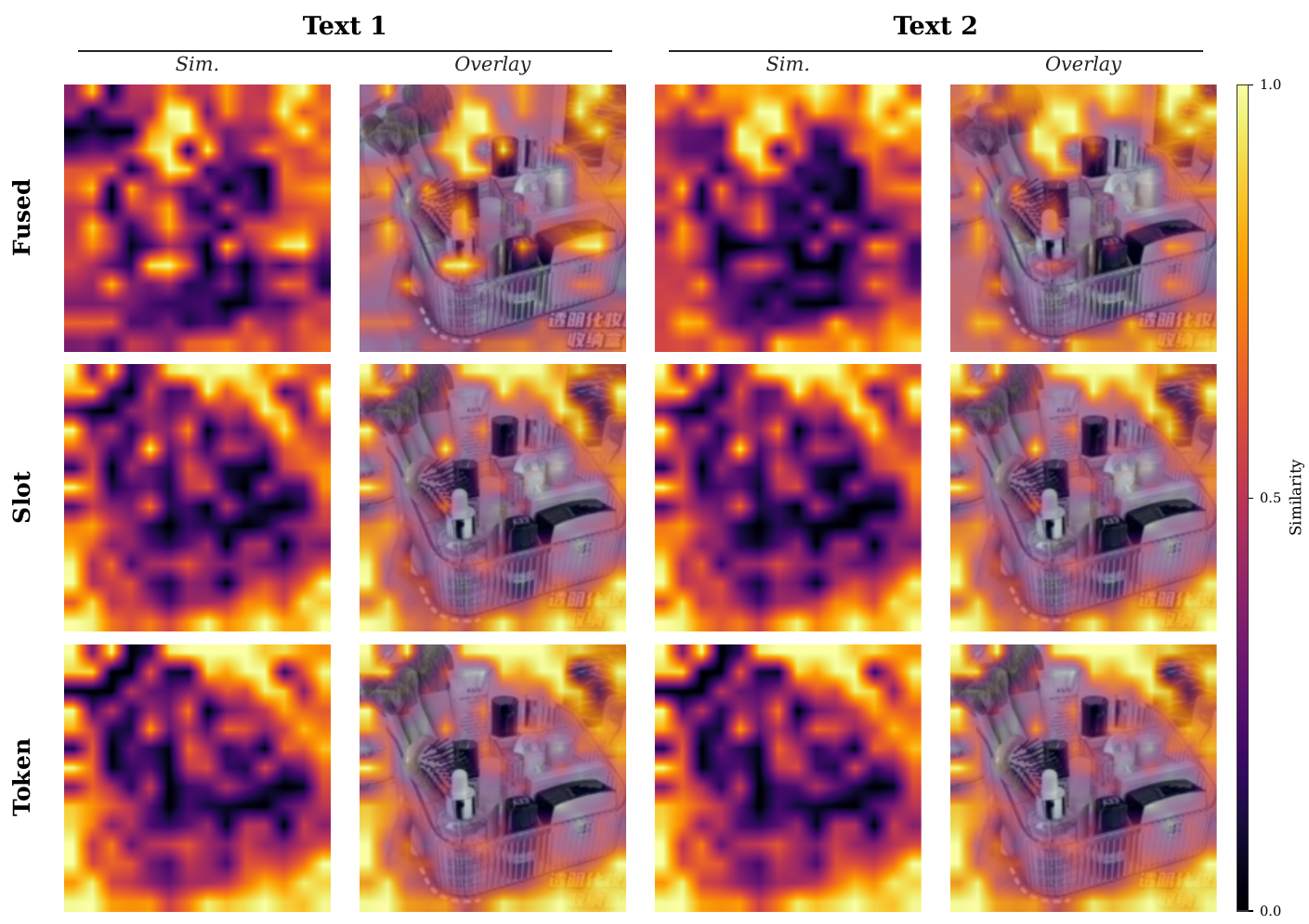}
        \caption{+\,S}
        \label{fig:case2_c}
    \end{subfigure}
 
    \vspace{1mm}
 
    \begin{subfigure}[b]{0.48\linewidth}
        \centering
        \includegraphics[width=\linewidth]{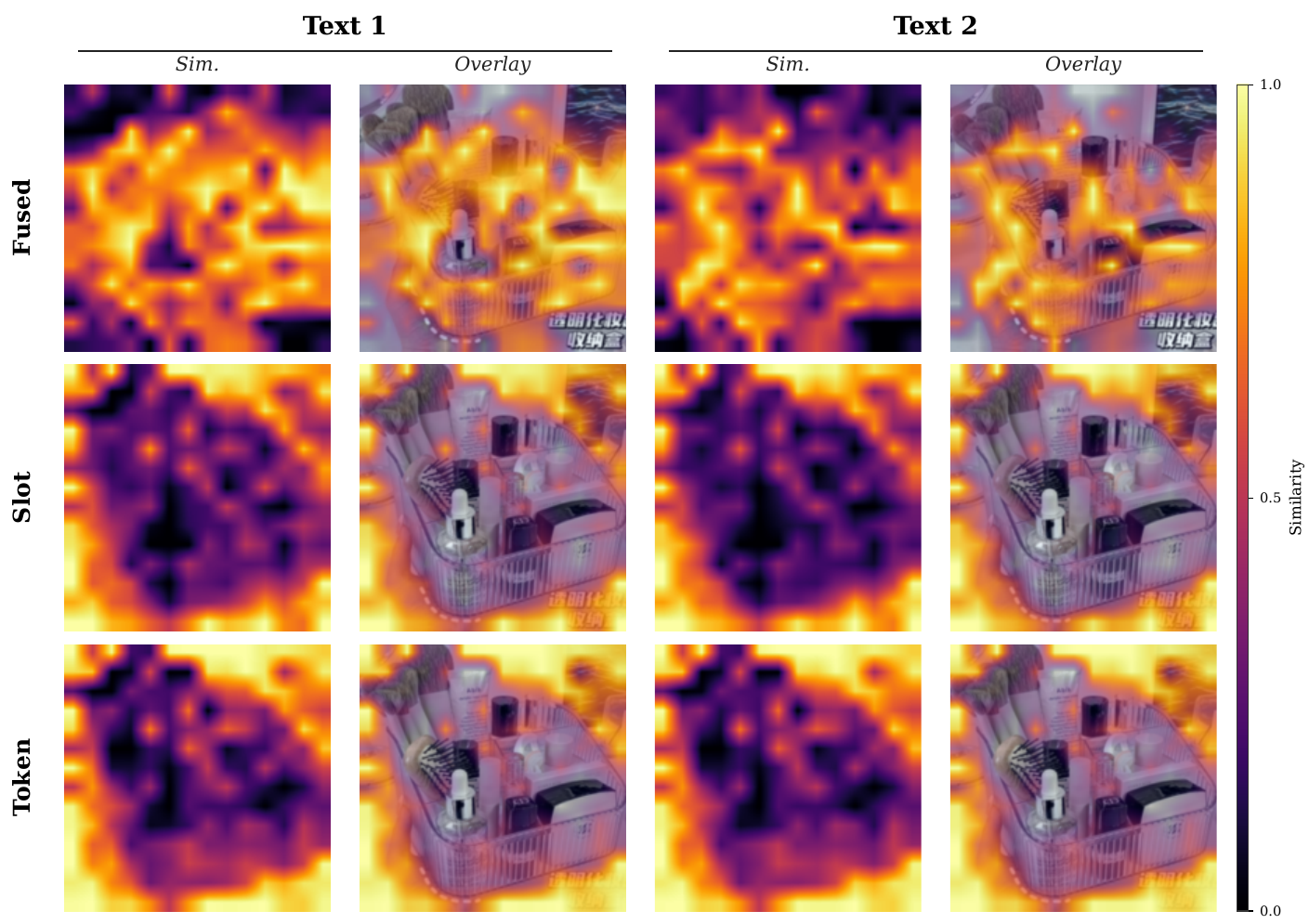}
        \caption{+\,B}
        \label{fig:case2_d}
    \end{subfigure}
    \hfill
    \begin{subfigure}[b]{0.48\linewidth}
        \centering
        \includegraphics[width=\linewidth]{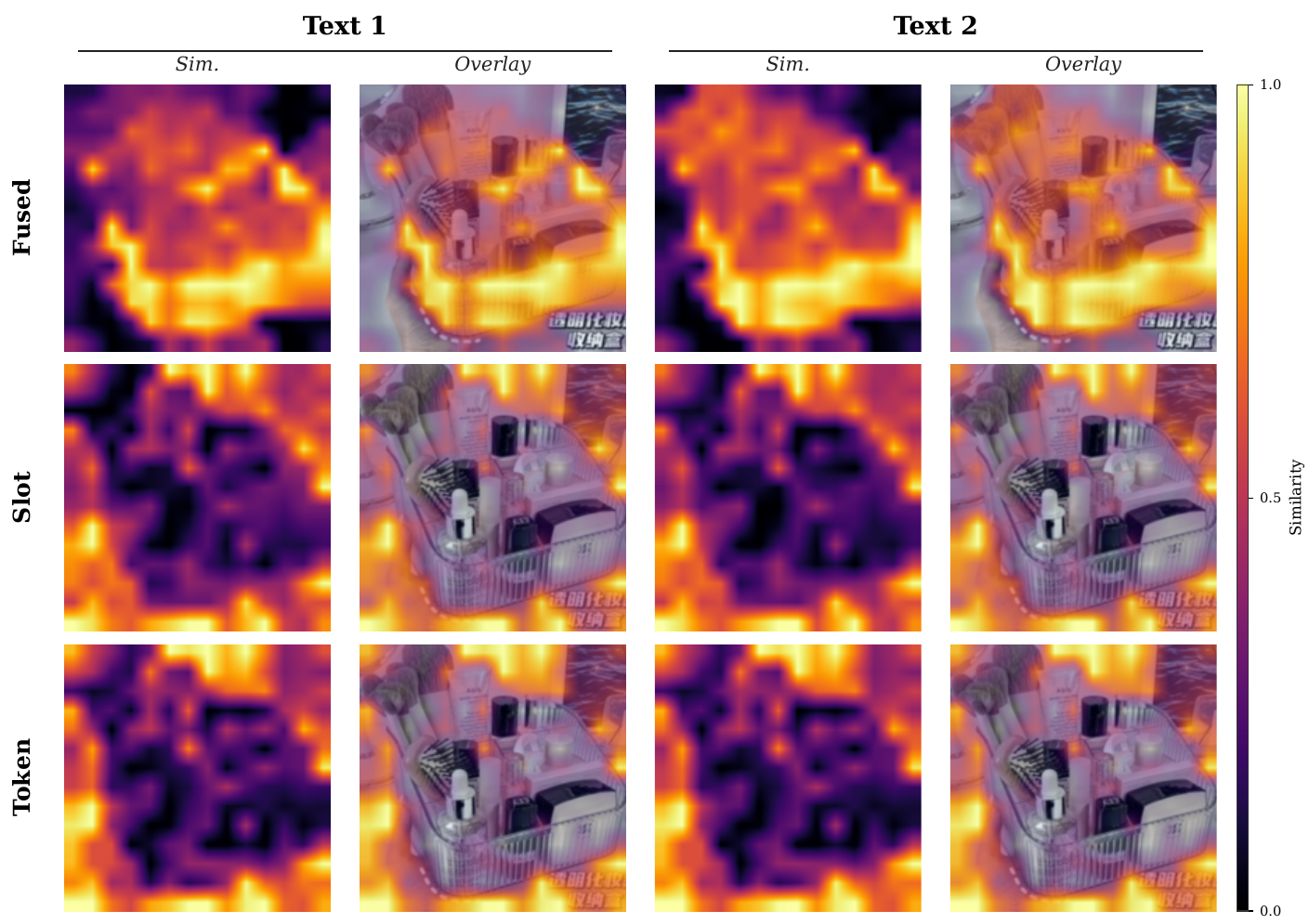}
        \caption{+\,R}
        \label{fig:case2_e}
    \end{subfigure}
 
    \vspace{1mm}
 
    \begin{subfigure}[b]{0.48\linewidth}
        \centering
        \includegraphics[width=\linewidth]{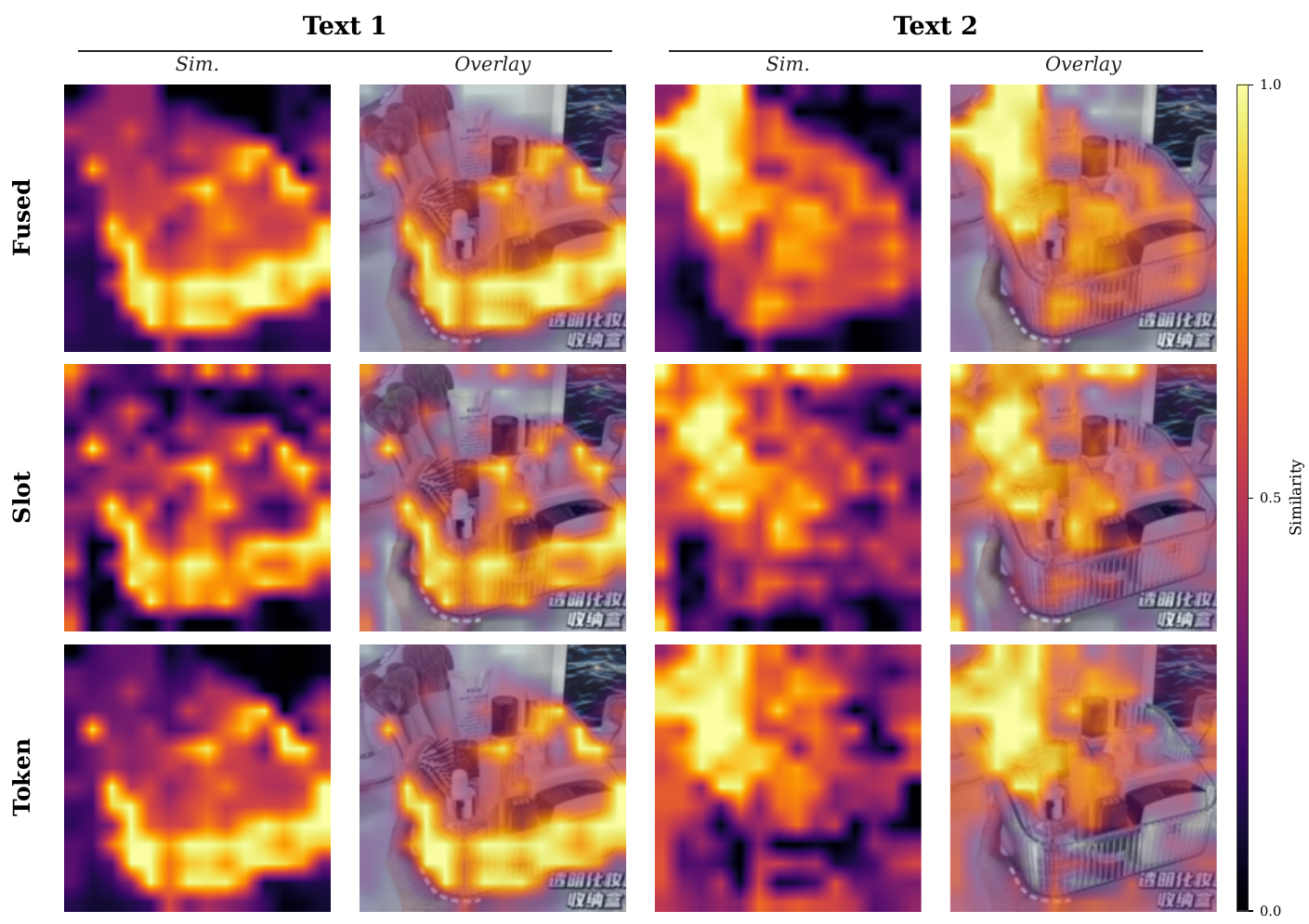}
        \caption{+\,H}
        \label{fig:case2_f}
    \end{subfigure}
    \hfill
    \begin{subfigure}[b]{0.48\linewidth}
        \centering
        \includegraphics[width=\linewidth]{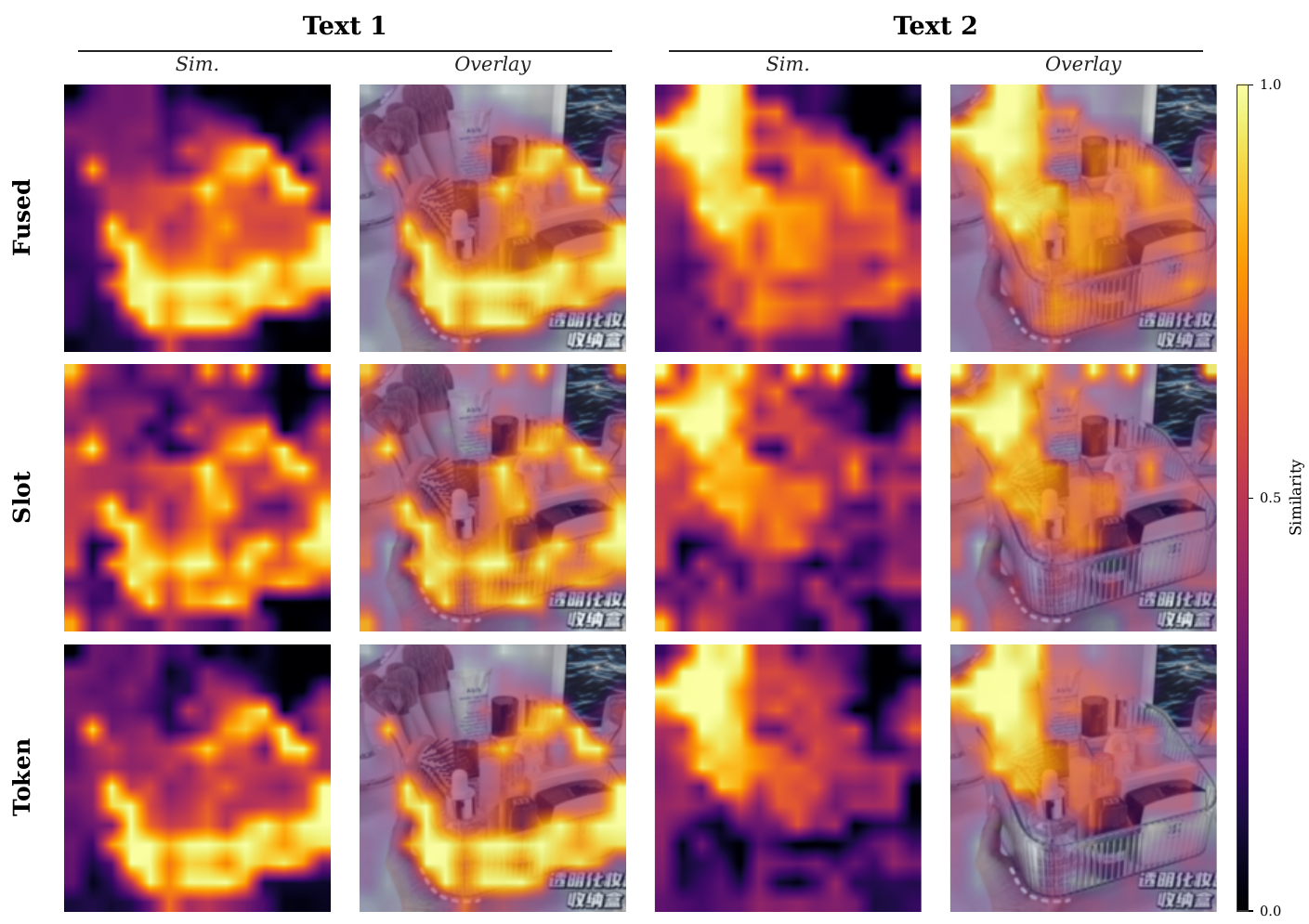}
        \caption{+\,Mosaic}
        \label{fig:case2_g}
    \end{subfigure}
 
    \vspace{1mm}
 
    \begin{subfigure}[b]{0.48\linewidth}
        \centering
        \includegraphics[width=\linewidth]{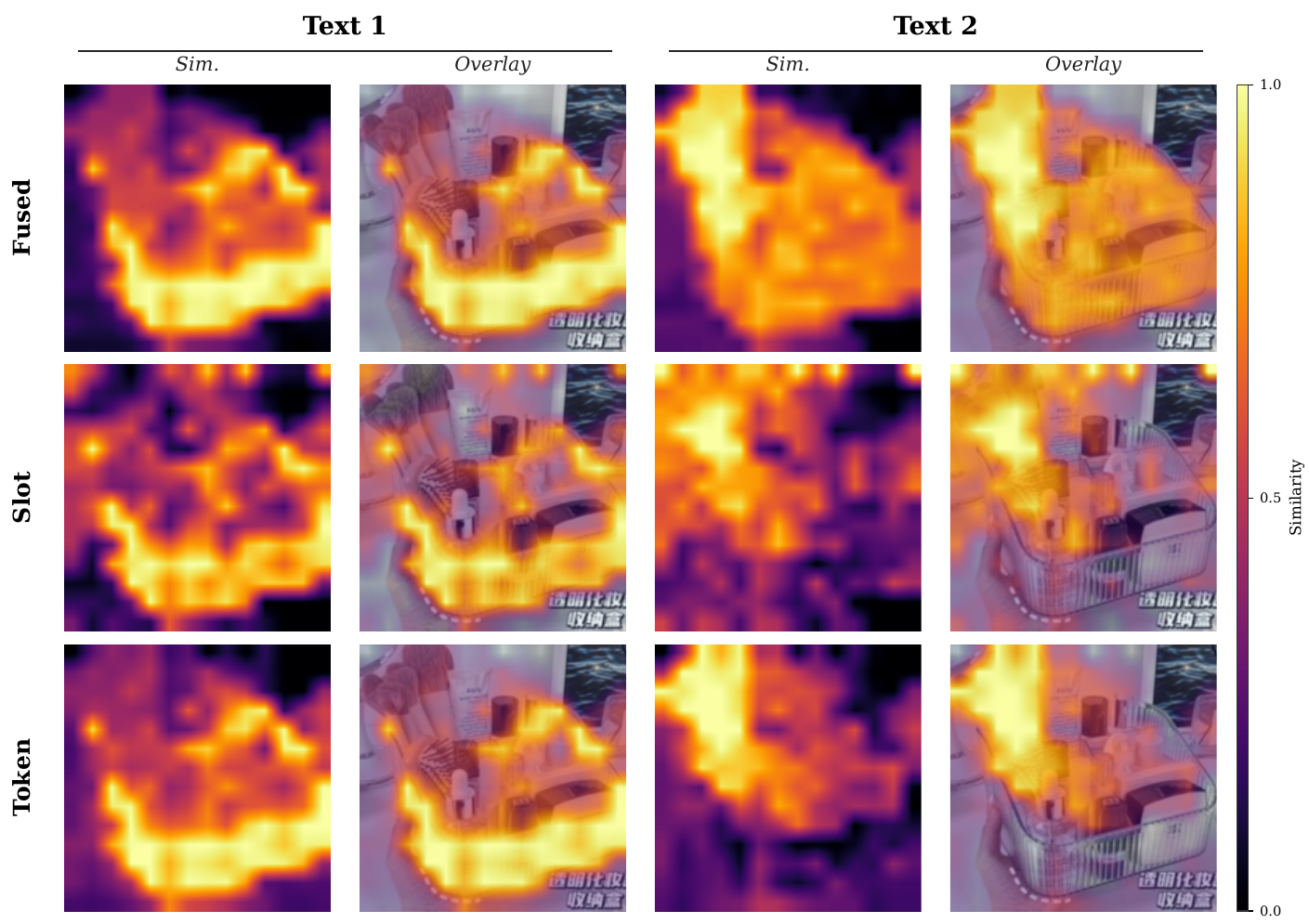}
        \caption{+\,D}
        \label{fig:case2_h}
    \end{subfigure}
    \hfill
    \begin{subfigure}[b]{0.48\linewidth}
        \centering
        \includegraphics[width=\linewidth]{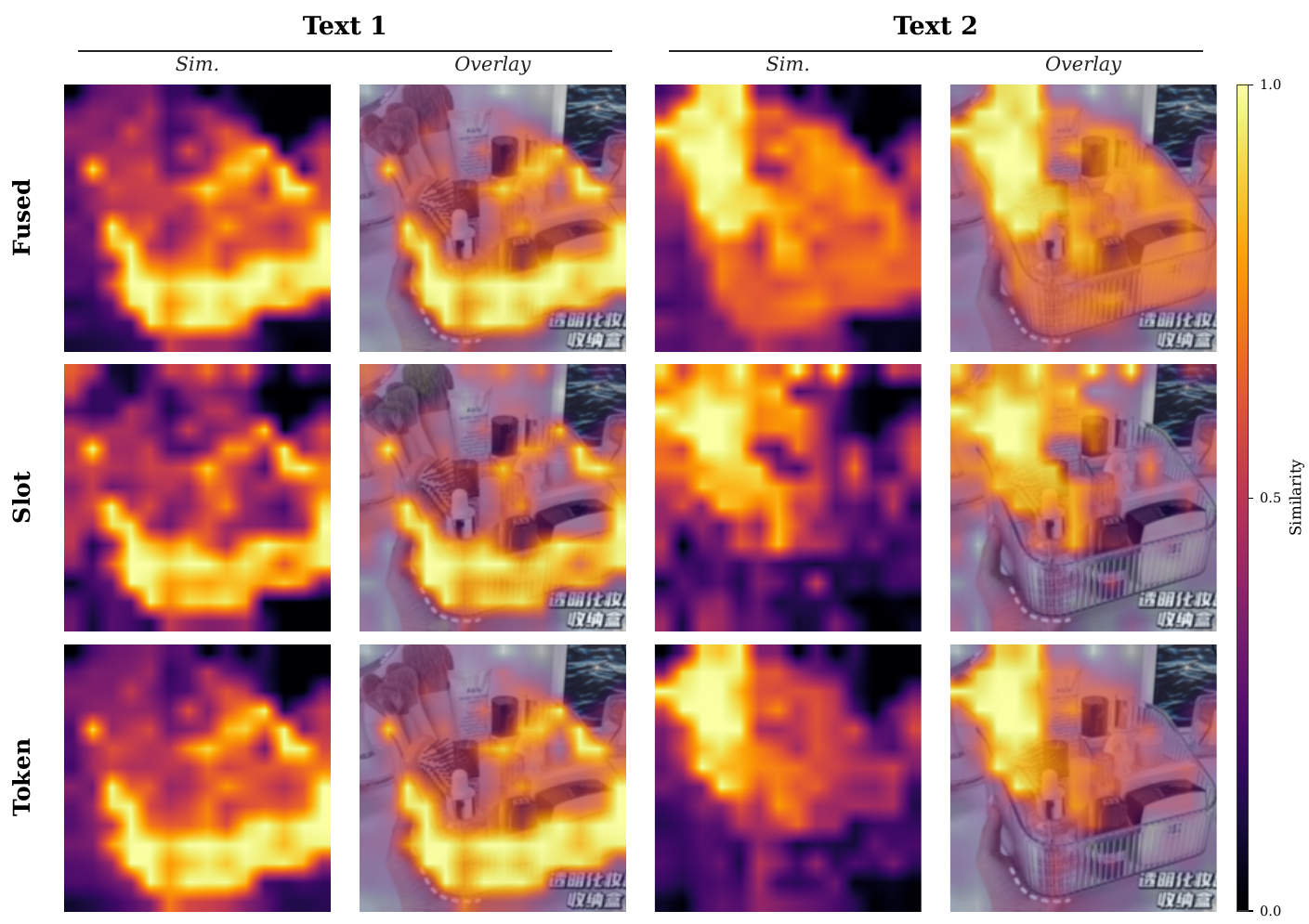}
        \caption{+\,T \textbf{(TIGER-FG)}}
        \label{fig:case2_i}
    \end{subfigure}
    \end{minipage}
 
    \caption{\textbf{Qualitative visualization of the additive ablation (Case 2/3: \emph{storage basket} and \emph{cosmetic brushes} queries).} 
    Panels follow the same additive layout as Figure~\ref{fig:additive_ablation_case1}; see its caption for component definitions. 
    This case emphasizes small-object queries in a cluttered home-goods scene, where the over-sharpening effect of \textbf{R}/\textbf{H} under raw-data training is particularly pronounced.}
    \label{fig:additive_ablation_case2}
\end{figure}

\begin{figure}[htbp]
    \centering
    \begin{minipage}{0.95\linewidth}
    \centering
    \captionsetup[subfigure]{font=small}
    \begin{minipage}[b]{0.48\linewidth}
        \centering
        \begin{subfigure}[b]{\linewidth}
            \centering
            \includegraphics[width=\linewidth]{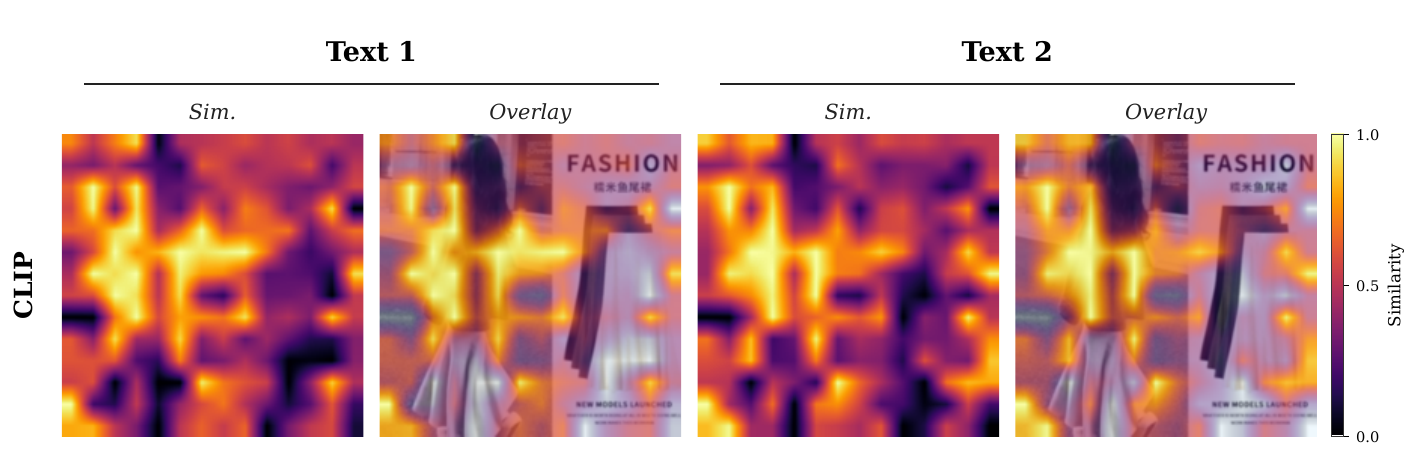}
            \caption{CLIP}
            \label{fig:case3_a}
        \end{subfigure}
 
        \vspace{1mm}
 
        \begin{subfigure}[b]{\linewidth}
            \centering
            \includegraphics[width=\linewidth]{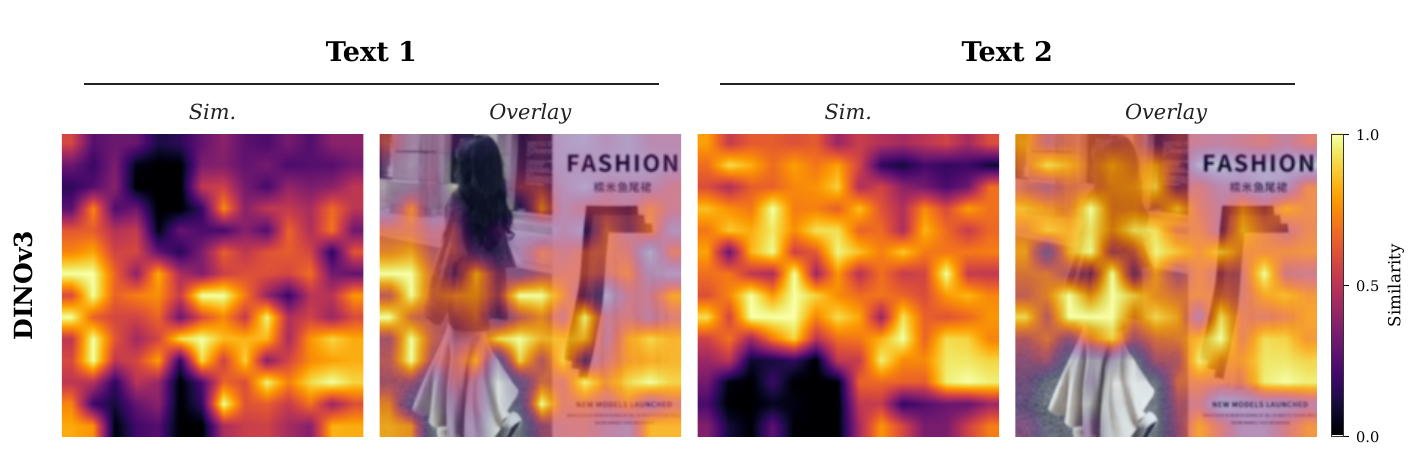}
            \caption{DINOv3}
            \label{fig:case3_b}
        \end{subfigure}
    \end{minipage}
    \hfill
    \begin{subfigure}[b]{0.48\linewidth}
        \centering
        \includegraphics[width=\linewidth]{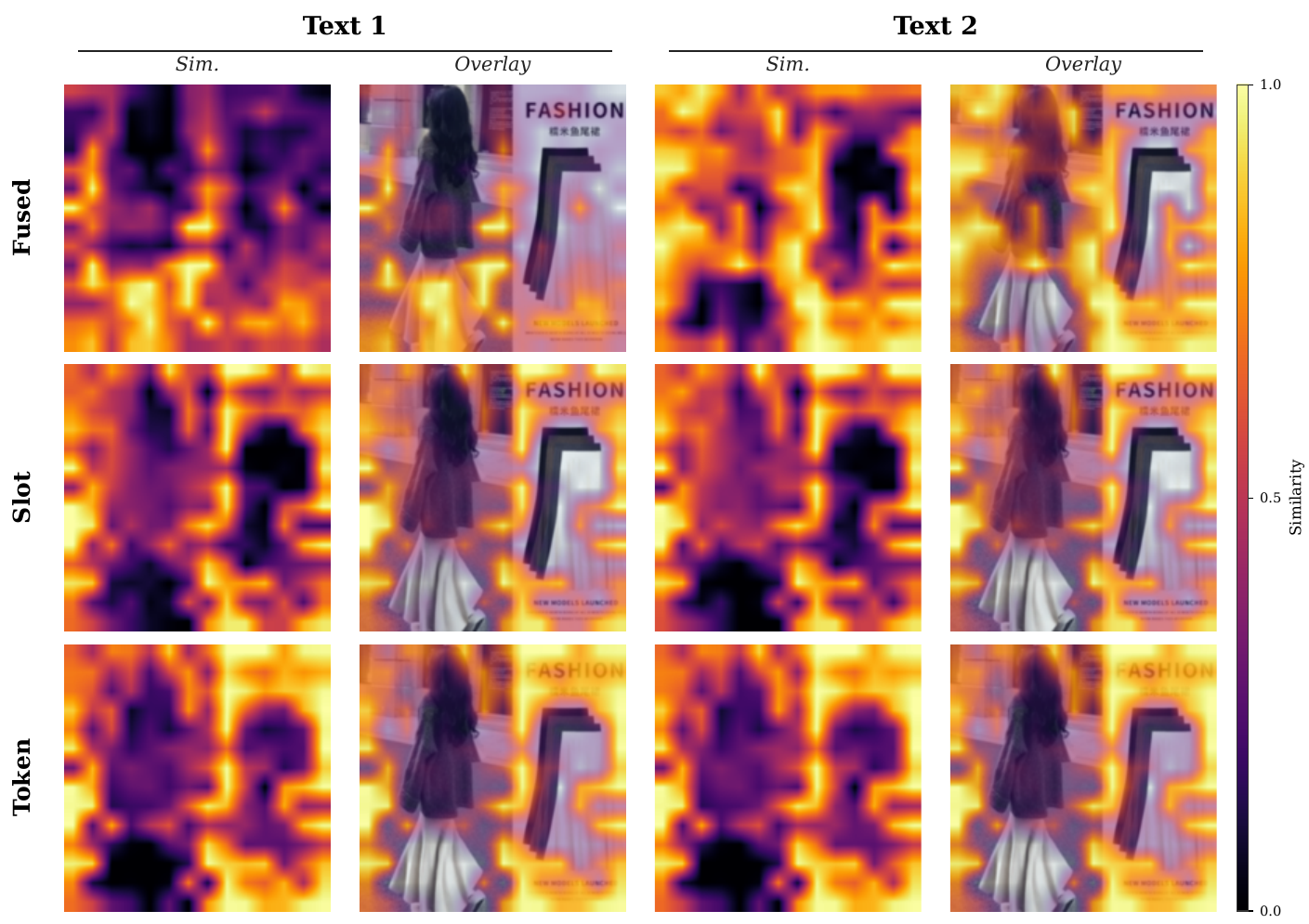}
        \caption{+\,S}
        \label{fig:case3_c}
    \end{subfigure}
 
    \vspace{1mm}
 
    \begin{subfigure}[b]{0.48\linewidth}
        \centering
        \includegraphics[width=\linewidth]{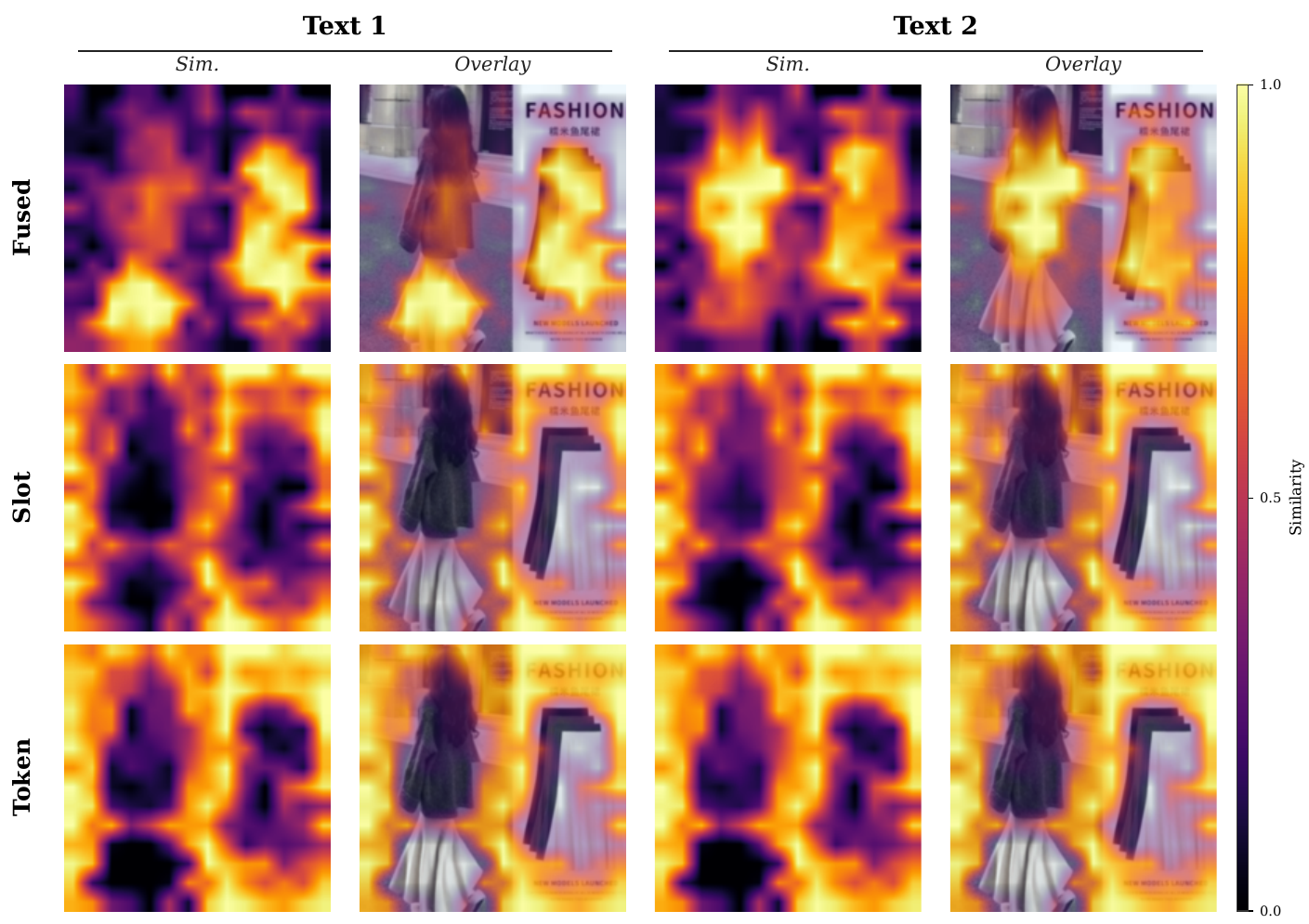}
        \caption{+\,B}
        \label{fig:case3_d}
    \end{subfigure}
    \hfill
    \begin{subfigure}[b]{0.48\linewidth}
        \centering
        \includegraphics[width=\linewidth]{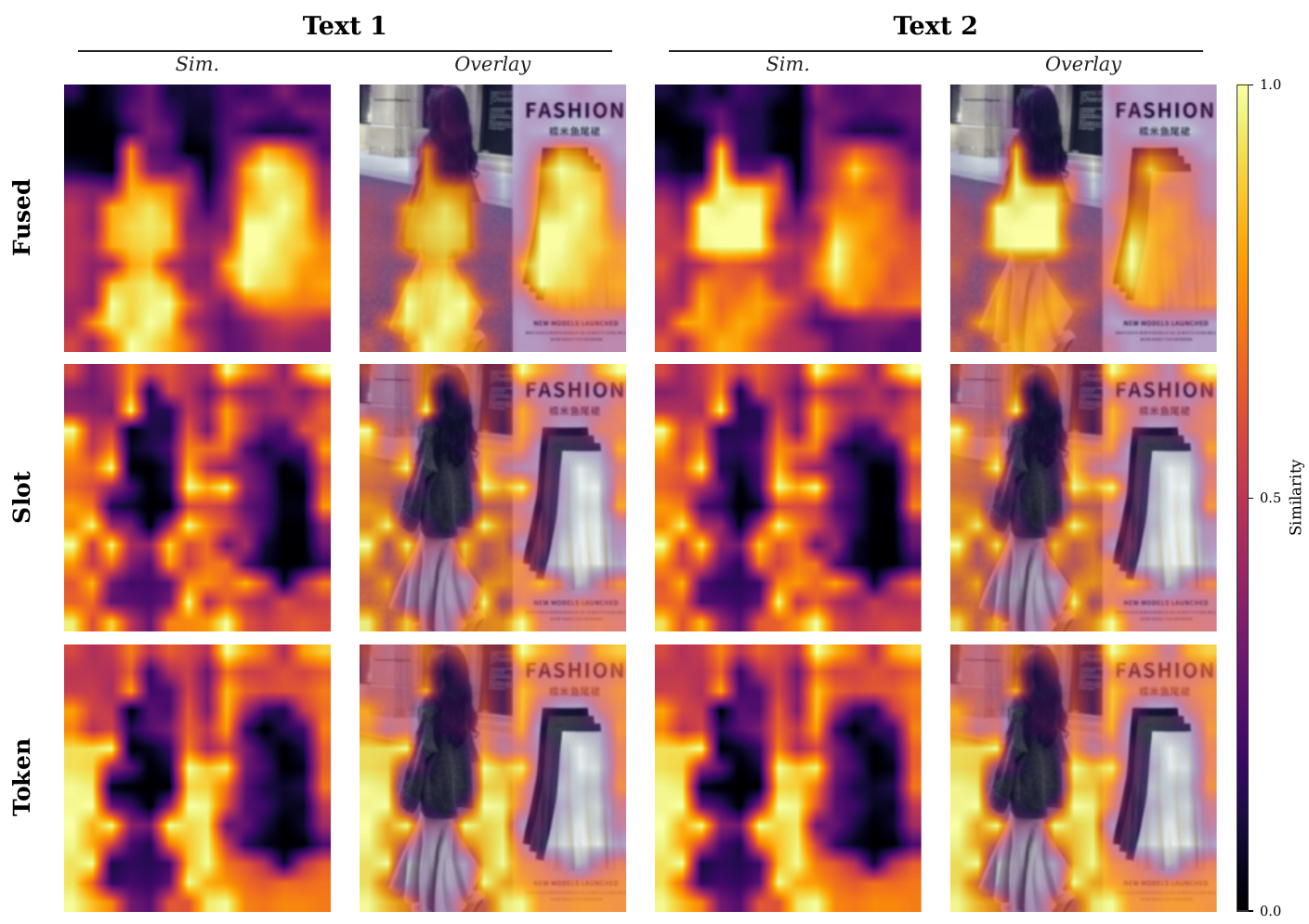}
        \caption{+\,R}
        \label{fig:case3_e}
    \end{subfigure}
 
    \vspace{1mm}
 
    \begin{subfigure}[b]{0.48\linewidth}
        \centering
        \includegraphics[width=\linewidth]{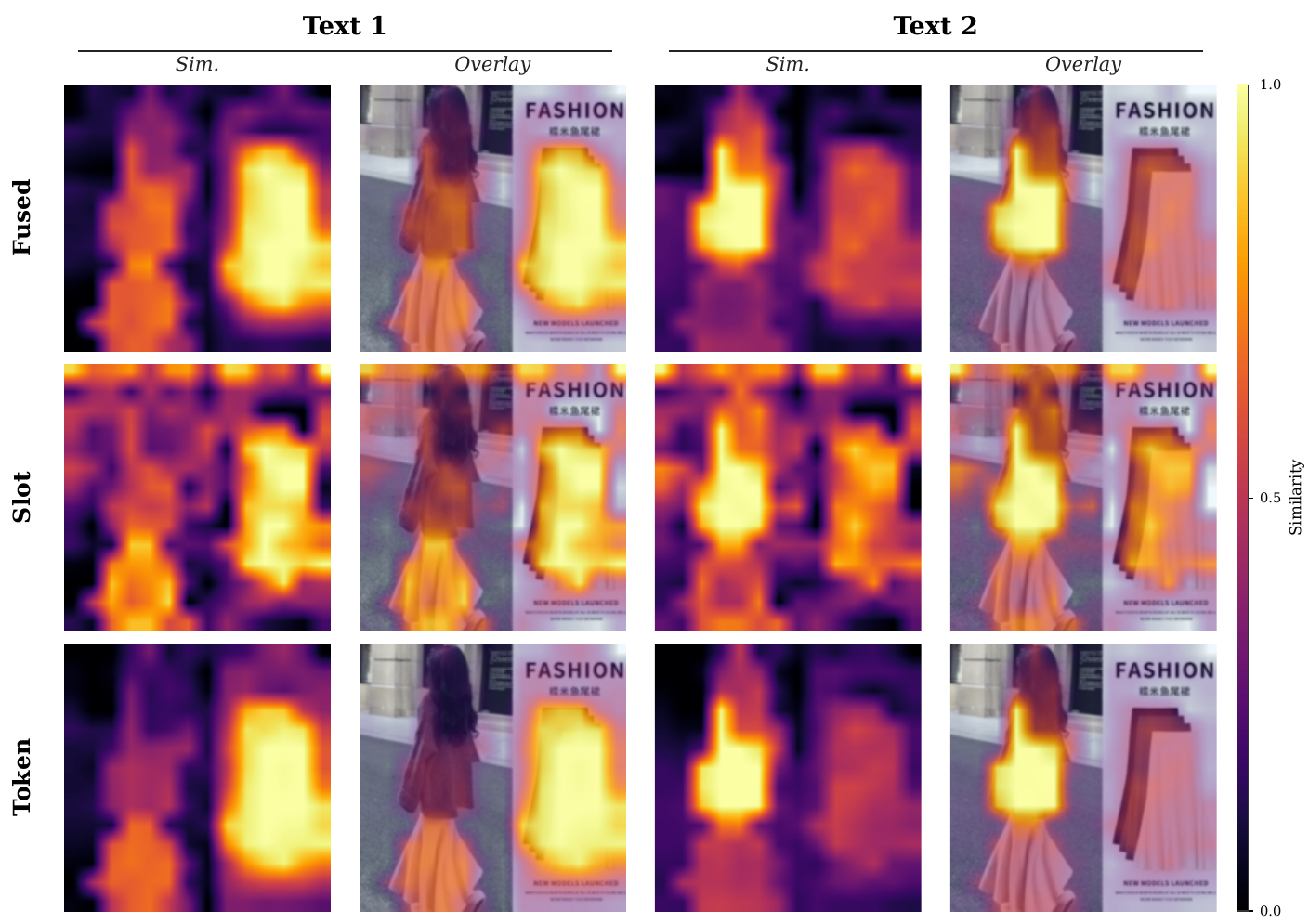}
        \caption{+\,H}
        \label{fig:case3_f}
    \end{subfigure}
    \hfill
    \begin{subfigure}[b]{0.48\linewidth}
        \centering
        \includegraphics[width=\linewidth]{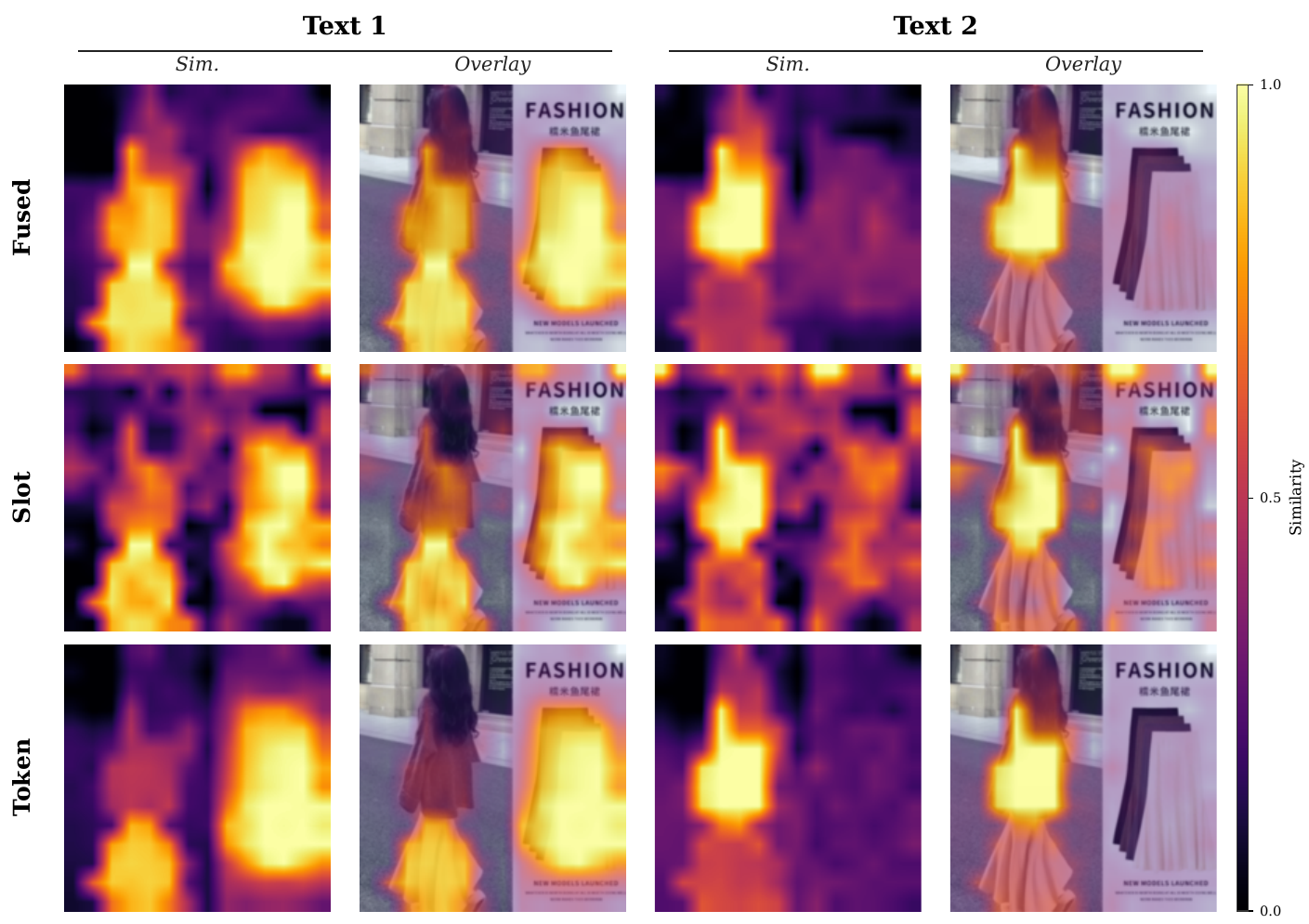}
        \caption{+\,Mosaic}
        \label{fig:case3_g}
    \end{subfigure}
 
    \vspace{1mm}
 
    \begin{subfigure}[b]{0.48\linewidth}
        \centering
        \includegraphics[width=\linewidth]{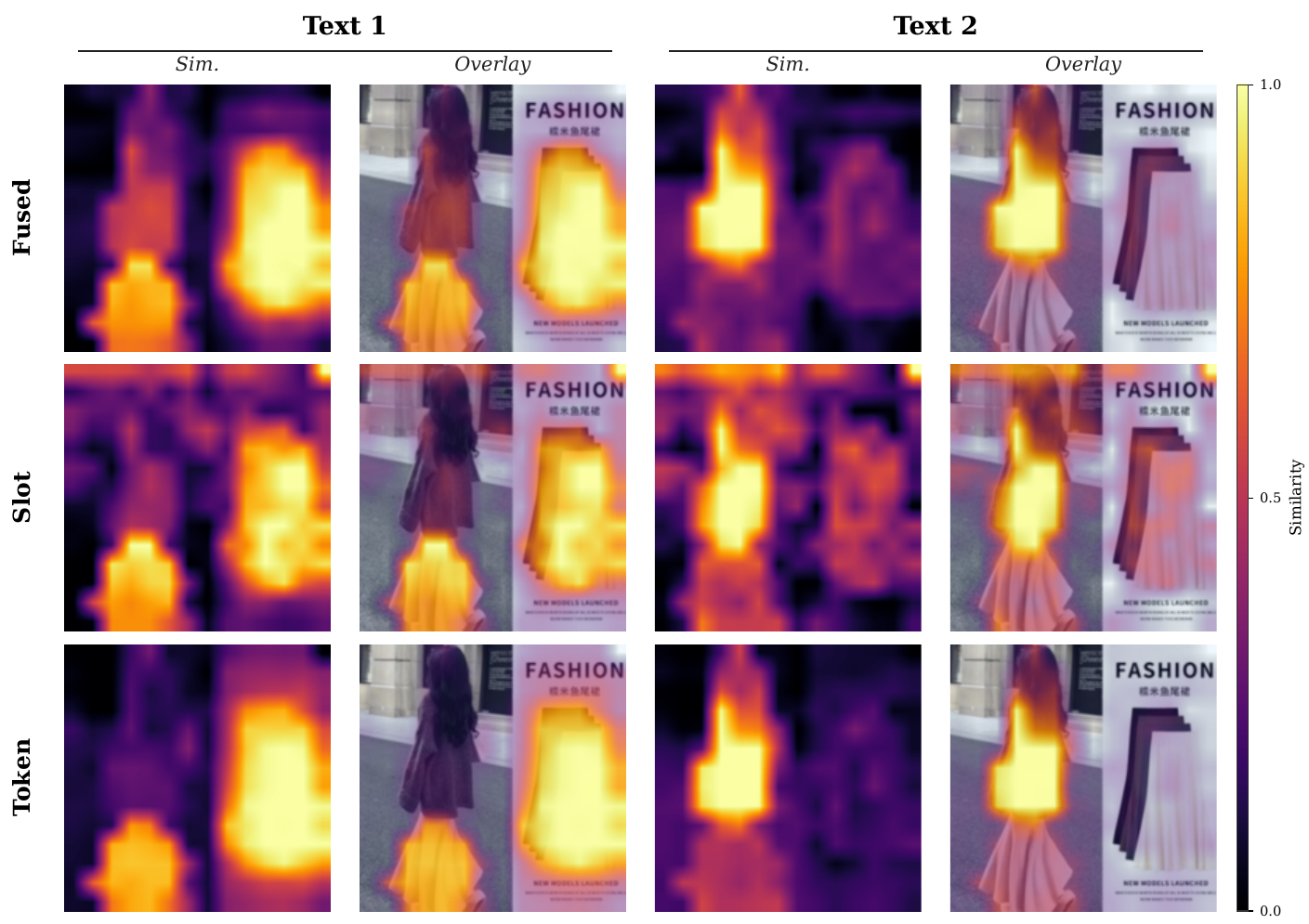}
        \caption{+\,D}
        \label{fig:case3_h}
    \end{subfigure}
    \hfill
    \begin{subfigure}[b]{0.48\linewidth}
        \centering
        \includegraphics[width=\linewidth]{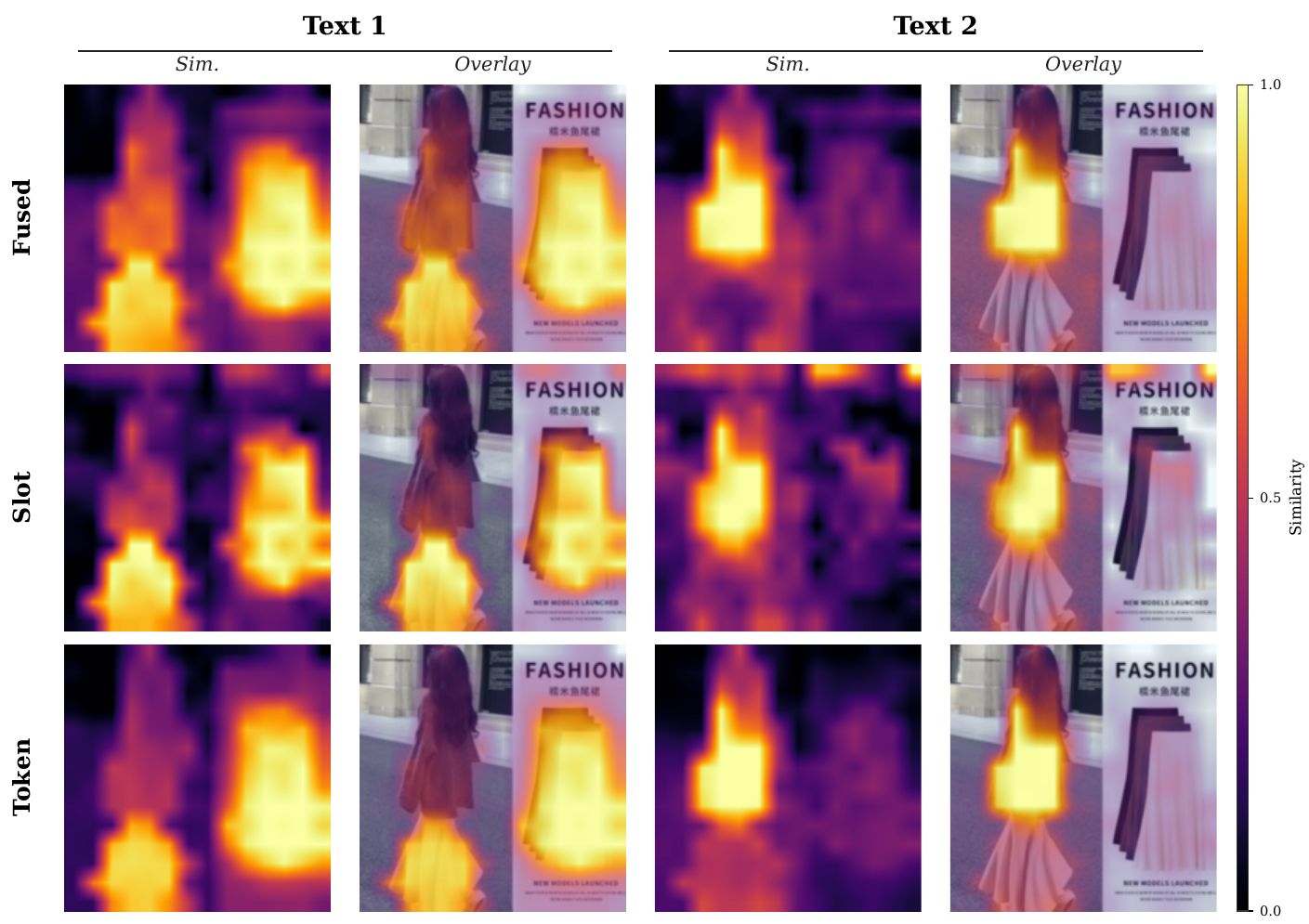}
        \caption{+\,T \textbf{(TIGER-FG)}}
        \label{fig:case3_i}
    \end{subfigure}
    \end{minipage}
 
    \caption{\textbf{Qualitative visualization of the additive ablation (Case 3/3: \emph{knitwear} and \emph{dress} queries).} 
    Panels follow the same additive layout as Figure~\ref{fig:additive_ablation_case1}; see its caption for component definitions. 
    This case contains two co-present garments in the candidate, and thus directly tests whether the model can route each text query to the correct garment---a capability that only fully emerges after Mosaic augmentation.}
    \label{fig:additive_ablation_case3}
\end{figure}

\paragraph{Early item-side changes bring modest gains.}
In block \emph{(a)}, replacing the CLIP visual backbone with DINOv3 improves Recall@1 from 46.80 to 49.74, with consistent gains across Recall, MRR, and NDCG. Further introducing the fusion design \textbf{S} yields only a marginal gain in the metrics, increasing Recall@1 from 49.74 to 49.94. The qualitative cases in Figures~\ref{fig:additive_ablation_case1}--\ref{fig:additive_ablation_case3} nevertheless show a clearer change from panels (a) to (c). The highlighted regions become more complete and more concentrated on product entities in the candidate image, suggesting improved entity-level localization. At the same time, the responses remain broad and are not yet reliably routed to the text-specified target, which is consistent with the limited metric gain at this stage.
 
\paragraph{Region supervision improves localization but hurts retrieval on raw data.}
Block \emph{(b)} shows a clear mismatch between qualitative and quantitative behavior. Adding box supervision \textbf{B} on top of \textbf{S} slightly improves the retrieval metrics, with Recall@1 increasing from 49.94 to 50.11. In the visualizations, panels (d) in Figures~\ref{fig:additive_ablation_case1}--\ref{fig:additive_ablation_case3} show that \textbf{B} further strengthens entity localization in the candidate image. Adding spatial-relational distillation \textbf{R} then makes the responses cleaner and more compact, with substantially reduced background activation, while adding hard negatives \textbf{H} begins to induce query-dependent behavior, so that different titles lead the model to attend to different objects. Despite these qualitative improvements, the retrieval metrics drop after \textbf{R} and remain low after \textbf{H}.
 
\paragraph{The metric drop reflects a distribution mismatch.}
This behavior is best understood as a mismatch between the raw-data training distribution and the Mosaic evaluation distribution. Under raw-data training, the item images are relatively clean, so stronger localization constraints encourage the model to rely on highly concentrated local evidence. This makes the responses visually sharper, but it also reduces reliance on broader cues that can still be useful for retrieval. Moreover, after \textbf{R} and especially \textbf{H}, the model begins to show weak query-dependent behavior, but this text-guided ability is still not reliable. When the guidance is correct, it can help the model focus on the relevant product entity; when it is incorrect, it can instead drive the model toward a mismatched entity that is visually salient but inconsistent with the title. In ECom-RF-IMMR-Mosaic, where candidate images contain more distractors and cross-object interference, such over-concentrated and occasionally misdirected matching can become brittle. The qualitative gains in block \emph{(b)} therefore indicate improved localization and emerging text guidance under the training distribution, but they do not yet translate into better retrieval under the Mosaic distribution.
 
\paragraph{Mosaic augmentation aligns training with the retrieval setting.}
Switching the training data from raw samples only to the raw+Mosaic setting yields the largest gain in Table~\ref{tab:additive_ablation}. Recall@1 jumps from 46.70 to 71.30, with similarly large improvements across Recall, MRR, and NDCG. The main reason is that Mosaic augmentation better matches the retrieval setting, where candidate images often contain multiple multiple products and stronger distractors. Under this training mixture, the model is no longer optimized only for clean single-object item images, but is instead exposed to the kind of multi-object interference that appears at retrieval time. This substantially reduces the train--test distribution disparity. More importantly, Mosaic augmentation helps the model learn which product entity in a multi-object item image should be matched to the title, rather than only sharpening attention on visually salient regions. As a result, the localization ability introduced by \textbf{R} and \textbf{H} becomes effective under realistic retrieval conditions, which leads to the large improvement in block \emph{(c)}.
 
\paragraph{Distillation and contrastive learning make title guidance more reliable.}
Building on the raw+Mosaic setting, adding spatial-relational and similarity-distribution distillation \textbf{D} further raises Recall@1 from 71.30 to 74.93. Unlike standard distillation, this objective does not simply match two models under identical inputs. Instead, it uses an e-commerce-trained visual teacher to provide box-level visual targets, and encourages the full-image multimodal item representation to align with the corresponding product region. This makes the title-guided image response more explicit and reduces failures in which the model attends to the wrong entity under a given title. Adding the image--text contrastive objective \textbf{T} then further improves Recall@1 to 75.22. Although the metric gain is smaller, the qualitative cases in Figures~\ref{fig:additive_ablation_case1}--\ref{fig:additive_ablation_case3} show a clearer benefit from panels (g) to (i): the response becomes more consistently routed to the title-matched object, especially in the challenging cases with competing products. This suggests that \textbf{T} helps preserve the text-guided role of the item-side text encoder and prevents the item representation from drifting toward a purely visual matching space. Together, \textbf{D} and \textbf{T} make title-guided retrieval substantially more reliable in the final TIGER-FG model.

\paragraph{Qualitative visualization.}
Figures~\ref{fig:additive_ablation_case1}--\ref{fig:additive_ablation_case3} visualize how the item-side response evolves as components are added for three representative query--candidate pairs with naturally multiple product entities. For each case, we show two text queries (\emph{Text\,1} and \emph{Text\,2}), together with the similarity map (\emph{Sim.}) and its overlay on the candidate image (\emph{Overlay}). For models with slot fusion, we further visualize the \emph{Fused}, \emph{Slot}, and \emph{Token} responses. The three cases cover different retrieval difficulties, namely a cluttered fashion scene with a black dress (Case~1), a home-goods scene with small-object queries (storage basket and cosmetic brushes, Case~2), and an apparel scene with two co-present garments (knitwear and dress, Case~3). Overall, the visualizations are consistent with the quantitative trends in Table~\ref{tab:additive_ablation}. Block \emph{(a)} mainly improves entity localization. Block \emph{(b)} yields sharper and more localized responses, but title-guided routing remains unstable under raw-data training. After introducing Mosaic-augmented training in block \emph{(c)}, the model better identifies which entity in the item image corresponds to the title. Adding \textbf{D} and \textbf{T} further strengthens this title-guided routing, making the response more consistently aligned with the title-matched object, especially in natural scenes with competing product entities.
 
%% file: tex_files/E.train_appendix.tex
\section{Additional Experimental Details}
\label{app:exp_details}

\subsection{Implementation Details}
\label{app:impl}

The query and item visual branches are both initialized from the original DINOv3 ViT-B/16 weights~\citep{simeoni2025dinov3} at $224\!\times\!224$ resolution and trained as two independent copies, without additional vision pretraining on our data.
The text encoder is initialized from the text branch of Chinese-CLIP ViT-B/16~\citep{chinese-clip}, which adopts a 12-layer RoBERTa-base architecture.
Before dual-encoder training, we further pretrain the text encoder with CLIP-style image--text contrastive learning on \textbf{ECom-RF-IMMR-10M}.

The unified embedding dimension is $C_u=256$.
The item branch uses $N_q=8$ learnable query tokens and $K=1$ mismatched-text hard negative per sample.
In the complementary visual branch, we set $\lambda_m=\lambda_c=0.5$.
All contrastive and distillation temperatures are fixed to $0.07$.
For the item-side objective, we set
$(\lambda_{v2t}, \lambda_{i2v}, \lambda_{\mathrm{SRD}})=(0.5,0.1,1.0)$
and $\lambda_h=0.1$.
For the dual-encoder objective, we set
$(\lambda_{q2i}, \lambda_{\mathrm{SDD}})=(1.0,1.0)$.
The final objective uses $\lambda_{\mathrm{dual}}=\lambda_{\mathrm{item}}=1.0$.

The $\mathcal{L}_{\mathrm{SRD}}$ teacher is a frozen original DINOv3 ViT-B/16.
The $\mathcal{L}_{\mathrm{SDD}}$ teacher is our in-house MoCo-pretrained image-to-image retrieval ViT-B/16, which also appears as a baseline in Table~\ref{tab:4_4_modalities}.
Both teachers remain frozen throughout training.

We train for 10 epochs on 8 NVIDIA H800 GPUs with batch size 256 and bf16 mixed precision.
Optimization uses AdamW with learning rate $2\!\times\!10^{-5}$, weight decay $0.01$, and a cosine schedule with 5\% warmup.
A full run takes about 14 GPU-hours.
Unless otherwise specified, TIGER-FG is trained with a $1\!:\!1$ mixture of original and Mosaic-augmented samples.
We also report TIGER-FG-RAW, which uses the same architecture and objectives but is trained only on the original samples, to isolate the effect of clutter-aware training.
For fair comparison, all trainable baselines are fine-tuned on \textbf{ECom-RF-IMMR-10M} under the same training settings, while off-the-shelf embedders are evaluated with their released weights.

\subsection{Public Benchmark Adaptation}
\label{app:public_benchmarks}

We additionally evaluate on two public e-commerce benchmarks, \textbf{eSSPR} and \textbf{LookBench}, to assess cross-dataset generalization.
Both benchmarks were originally designed for multimodal-to-multimodal retrieval, while IMMR uses a cropped visual query to retrieve image--text item candidates.
We therefore adapt their query and candidate formats to our evaluation protocol.

\textbf{eSSPR.}
eSSPR mainly contains clean item images, where each image usually depicts a single foreground item.
To adapt eSSPR to IMMR, we convert each query into a cropped visual region and keep the gallery as image--text item candidates.
Since ground-truth query boxes are not directly provided in the original benchmark, we first extract region proposals with an off-the-shelf object detector and then use a visual matching module to select the region that best matches the item target.
Samples that cannot be reliably parsed are removed.
The remaining samples are evaluated with the same retrieval protocol as our in-domain benchmarks.

\textbf{LookBench.}
LookBench provides localized query regions and a noisy candidate pool, where a query may correspond to multiple relevant items.
We use the provided bounding boxes as cropped visual queries.
For the candidate side, we construct multimodal item entries by pairing each candidate image with structured item text.
Since item titles are not directly available, we generate textual descriptions from the provided category and attribute annotations using Qwen3-VL-32B.
We aggregate results across all LookBench subsets.
In addition to Recall@K, MRR@K, and NDCG@K, we report HitRate@K on LookBench because each query may have multiple valid matches.